%% file: main.tex
\documentclass{iopart}

\expandafter\let\csname equation*\endcsname\relax
\expandafter\let\csname endequation*\endcsname\relax

\usepackage{textgreek}
\usepackage[sort]{cite} %package for sorting the cites
\usepackage{hyperref} %package for hyperref
\usepackage{url}  % Used for lxcat database entries in .bib file
\usepackage{chemformula}  % Formatting of species and reactions
\usepackage{siunitx}  % Formatting of numbers and quantities

\AtBeginDocument{\RenewCommandCopy\qty\SI}  
\DeclareSIUnit\bar{bar}  % Define \bar after it was deprecated from siunitx
\DeclareSIUnit{\litre}{l}  % Use lower case l for litre instead of L

\newcommand{\dv}[2]{\frac{d #1}{d #2}}

\newcommand{\pdv}[2]{\frac{\partial #1}{\partial #2}}

\begin{document}
    \input{00_titlepage}
    \input{01_abstract}

    \ioptwocol

    \input{02_introduction}
    \input{03_experiment}
    \input{04_simulation_framework}

    \input{05_results}
    \input{06_conclusion}

    \section{Acknowledgments}

    This work was funded by the German Research Foundation (DFG) via CRC 1316 (project number 327886311), project B5.
    
    \section{References}

    % \bibliographystyle{ieeetr}
    % \bibliographystyle{iopart-num}
    % \bibliography{main}
% \bibliography{references}{}
% \bibliographystyle{unsrt}
\providecommand{\newblock}{}

    \onecolumn
    \input{07_addendum}

\end{document}

%% file: 00_titlepage.tex
\title{Electron and gas temperature-driven chemistry during microdischarges formed in water vapour bubbles}

\author{Florens Grimm$^{1}$, Jan-Luca Gembus$^{1}$, Jana Schöne$^{1}$,  Peter Awakowicz$^{1}$, Lars Schücke$^{1}$, Andrew R. Gibson$^{1, 2}$}

\address{
    $^{1}$Chair of Applied Electrodynamics and Plasma Technology, Faculty of Electrical Engineering and Information Technology, Ruhr University Bochum, Bochum, Germany \\
    $^{2}$York Plasma Institute, School of Physics, Engineering and Technology, University of York, York, United Kingdom
}

\ead{grimm@aept.ruhr-uni-bochum.de}

%% file: 01_abstract.tex
\begin{abstract}

Microdischarges formed in bubbles immersed in liquids are of interest for materials synthesis and chemical conversion applications in the frame of plasma-driven electrochemistry. 
A key challenge associated with controlling such processes is the limited understanding of the gas-phase chemical kinetics in these microdischarges. 
Due to their large electron densities, and high gas temperatures, both electron and gas temperature driven chemistry are likely to be important. 
Here, a 0-D modelling approach, informed by experimental measurements, is used to study the chemical kinetics in these systems. 
A new reaction scheme is developed for microdischarges in water vapour, including reactions for both high electron density, and high gas temperature regimes. 
Microdischarges formed during plasma electrolytic oxidation are used as a test case, however, the key results are expected to be transferable to other plasma electrolysis systems with similar properties. 
Experimentally measured power densities are used as input to the 0-D model, together with estimates of temperatures and gas pressures within the gas bubble. 
Comparison of measured and simulated electron densities shows good agreement, given the limitations of both model and experiment. 
In the base case microdischarge, \ch{H2O} is found to be highly dissociated during the period of peak power density, with \ch{H} and \ch{O} making up the majority of the neutral gas in the bubble. 
The maximum ionization degree is around \qty{0.31}{\percent}, and the electronegativity during the period of peak electron density is found to be low. 
Species formation and reaction pathways are analysed under variation of the neutral gas temperature from \qty{2000}{K} to \qty{6000}{K}. 
At all temperatures, electron, ion, and neutral reactions with high threshold energies are found to be important for the overall chemical kinetics. 

\end{abstract}

\maketitle 

%% file: 02_introduction.tex
\section{Introduction}\label{sec:introduction}

Plasmas formed in direct contact with liquids are of interest for a broad range of applications, with key examples including materials synthesis and chemical conversion \cite{bruggeman_plasmaliquid_2016, bruggeman_plasma-driven_2021, graham_plasmas_2011, sen_gupta_contact_2017, yerokhin_plasma_1999, aliofkhazraei_review_2021, angelina_critical_2025}. 
In such applications, both the physical and chemical properties of the discharge are important drivers of application outcomes. 
These plasma systems come in a variety of forms, where the degree of coupling between the plasma and the liquid can vary strongly depending on the geometry of the electrodes with respect to the gas and the liquid \cite{bruggeman_plasmaliquid_2016}.
\\

One prominent geometrical arrangement is where the electrode system is fully immersed in a liquid. 
Such systems can be driven by various electrical signals ranging from short, high voltage pulses \cite{seepersad_investigation_2013, ceccato_time-resolved_2010, marinov_modes_2013, pongrac_spectroscopic_2018, grosse_nanosecond_2019, simeni_simeni_origins_2025}, to lower voltages applied over longer timescales \cite{schaper_plasma_2011, asimakoulas_fast_2020, clyne2015}. 
While discharges often occur in gas bubbles formed directly within the liquid in contact with the electrodes, studies of discharges formed in externally supplied gas bubbles have also been carried out \cite{babaeva_structure_2009, pillai_plasma_2022}. 
The physical properties of such discharges have been studied in detail by a number of authors, revealing insights into breakdown mechanisms \cite{seepersad_investigation_2013, ceccato_time-resolved_2010,schaper_plasma_2011,babaeva_structure_2009}, electron densities \cite{pongrac_spectroscopic_2018, grosse_nanosecond_2019, jovovicSpectroscopicStudyPlasma2012}, neutral gas temperatures \cite{wang_absolute_2021}, and bubble dynamics \cite{grosse_nanosecond_2019, clyne2015, gembusCharacterisationSingleMicrodischarges2025, forschner_statistical_2025, xiang_electric_2025}, for example. 
Surface modifications and particle formation induced by such discharges have also been studied \cite{curran_thermal_2005, troughton_effect_2016, krettek_creation_2024}.
\\

Discharges formed in conductive liquids are a major sub-category of in-liquid plasma sources. 
Such discharges include those used in contact glow discharge electrolysis \cite{sen_gupta_contact_2017} and plasma electrolytic oxidation~\cite{aliofkhazraei_review_2021}, for example. 
In these systems, gas formation at the electrodes is closely tied to the formation of short-lived microdischarges, which appear stochastically on the electrode surface. 
The close coupling between microdischarge and gas bubble, as well as the unpredictable nature of their formation, means that experimental measurements and simulations of microdischarge properties are highly challenging. 
Despite these challenges, a number of works have provided insights into bubble and microdischarge formation mechanisms in these systems~\cite{clyne2015,gembusCharacterisationSingleMicrodischarges2025,forschner_statistical_2025, troughton_effect_2016,clyne_review_2019}. 
Nevertheless, many of the key physical and chemical phenomena occurring in these discharges are poorly understood. 
This, in turn, limits the extent to which they can be optimised in applications. 
\\

As described above, quantification of the physical properties of bubble-based microdischarges has received significant attention in the literature, with a range of important insights obtained. 
However, quantitative studies of the chemical composition of such discharges are more limited. 
For many in-liquid discharges, the temperature of the neutral gas is typically in the range of several thousand Kelvin~\cite{wang_absolute_2021}. 
Mededovic and Locke proposed a model where the chemical kinetics in the discharge were driven by these high gas temperatures, and electron-driven reactions were neglected~\cite{mededovic_primary_2007}. 
Results of this model were found to be in good agreement with the experimentally measured production rates and concentrations of several species. 
Wang~\emph{et~al.} found that a similar model was able to reproduce the absolute densities and temporal variations of OH radicals, measured by laser induced fluorescence, within the timeframe of in-bubble microdischarge formation~\cite{wang_absolute_2021}. 
These studies support the importance of gas temperature driven chemistry in these systems. 
Zheng~\emph{et~al.}, applied a different numerical model and reaction scheme to study plasma and species formation in a cathodic plasma electrolysis system \cite{zheng_understanding_2019}. 
In that work, the chemistry was largely electron driven, with relatively low gas temperatures of between \qty{400}{\kelvin} and \qty{600}{\kelvin} used in the model. 
\\

In this work, we aim to build on insights from these previous studies, and develop a greater understanding of the relative importance of electron and gas temperature driven chemistry in in-bubble microdischarges. 
To do this, a new reaction scheme for the gas phase chemistry of microdischarges formed in water vapour is developed, focusing on the treatment of reactions that are expected to be important under regimes of high electron density and high gas temperature. 
Inclusion of both classes of chemistry within the same model allows for their relative importance to be quantified. 
This reaction scheme is implemented in a 0-D plasma-chemical kinetics model. 
Experimental measurements of the electrical characteristics and bubble dynamics related to microdischarges formed during plasma electrolytic oxidation of aluminium are used to define the time-resolved power density during an individual microdischarge, which is used as an input to the model. 
The experimental setup is described in Sec.~\ref{sec:experiment}, followed by a description of the model in Sec.~\ref{sec:simulation-framework}. 
In Sec.~\ref{sec:results}, experimental validation of the model is described, followed by a detailed analysis of charged and neutral particle formation during the microdischarge lifetime, as well as their most important production and consumption reactions. 
Lastly, the role of varying gas temperature on the production and consumption reactions of \ch{H2O} is studied, and the relative importance of electron and gas temperature driven chemistry is discussed.

%% file: 03_experiment.tex
\section{Experiment}\label{sec:experiment}
\begin{figure}[h]
    \centering
    \includegraphics[width=1\linewidth]{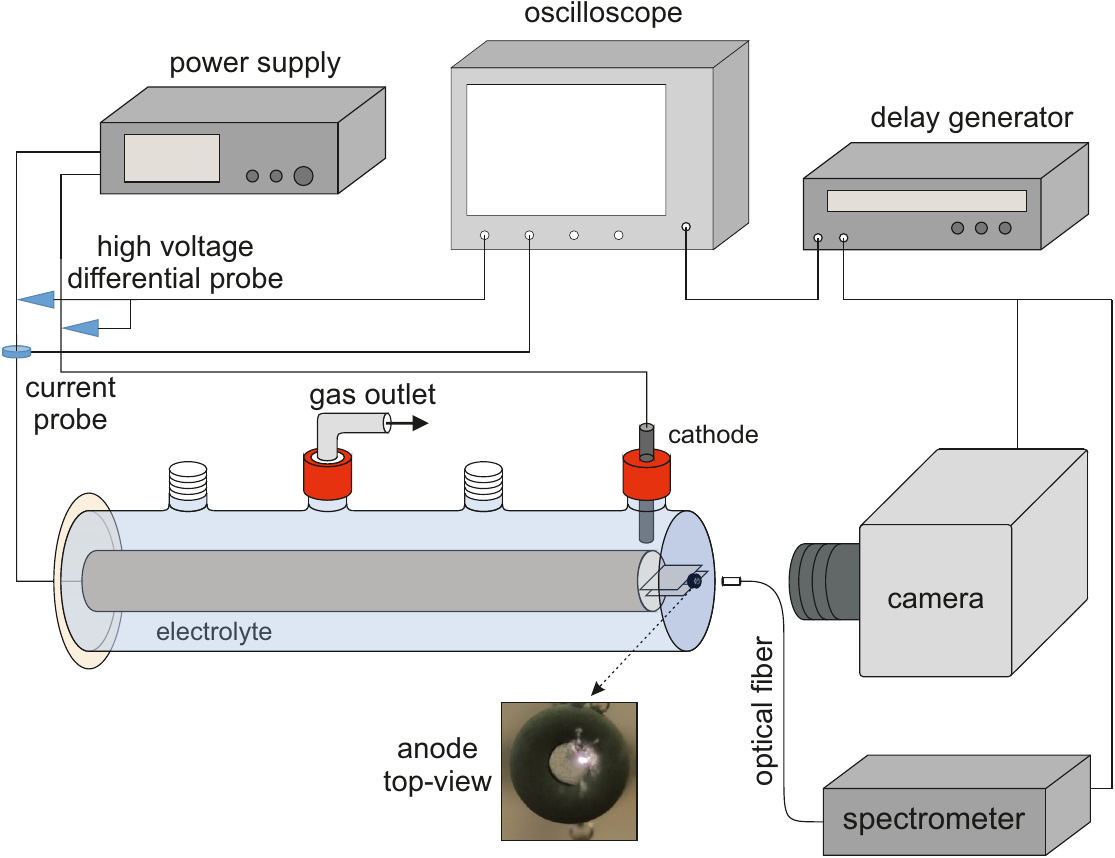}
    \caption{Experimental setup for single microdischarge studies. Image reproduced from \cite{gembusCharacterisationSingleMicrodischarges2025}.}
    \label{fig:smd-setup}
\end{figure}

To study the behavior of single microdischarges during a plasma electrolytic oxidation (PEO) process, a
specialized setup (SMD: \textbf{S}ingle\ \textbf{M}icrodischarge\ \textbf{S}etup) was developed, as shown in Fig.~\ref{fig:smd-setup}. 
This setup is described in detail in \cite{Bracht_diss, gembusCharacterisationSingleMicrodischarges2025}.
\\

The experimental setup consists of a quartz glass cylinder containing a stainless steel cathode and an aluminium wire anode.
The cylinder contains a solution of distilled water with an electrolyte, typically potassium hydroxide (KOH), at concentrations ranging from \qtyrange[per-mode=symbol]{0.5}{4}{\gram\per\litre}. 
For simplicity, only KOH at a concentration of \qty[per-mode=symbol]{1}{\gram\per\litre} is used for comparison between experiment and simulation in this work. 
The influence of KOH concentration on microdischarge properties is discussed in~\cite{gembusCharacterisationSingleMicrodischarges2025}.
\\

In order to minimize the number of microdischarges (in an ideal case to one), the anode wire is restricted to a diameter of \qty{1}{\milli\metre} and is insulated with a shrinking tube. 
The wire is mounted inside the glass tube with a polyether ether ketone (PEEK) holder. 
In addition, a heat-resistant fluorine rubber ring is placed at the tip to protect the insulation from damage during the PEO process. 
The insulation, combined with the small wire diameter, confines the ignition of individual microdischarges to the tip of the wire.
This design enables the study of individual discharges with lifetimes ranging from a few \unit{\micro\second} to several hundred \unit{\micro\second}.
\\

Optical access is provided on one side of the glass cylinder by a quartz glass window. 
This allows for the investigation of individual microdischarges using diagnostic tools such as high-speed cameras or spectrometers, which are aligned to the substrate tip. 
In this study, only a brief overview of the methods used is given.
A more detailed description of the setup and the instruments used, as well as additional experimental studies done on the setup, can be found in the
works of Bracht~\cite{Bracht_diss} and Gembus~\emph{et al.}~\cite{gembusCharacterisationSingleMicrodischarges2025}.
In this work, the experimental measurements are used for two purposes. 
Firstly, experimental measurements are used to inform simulation inputs, such as the power density, gas temperature and gas pressure inside gas bubbles formed during PEO. 
Secondly, electron densities measured in the experiment are used to provide a basic validation of the simulation results.

\subsection{Spectroscopic analysis}\label{subsec:spectroscopic-analysis}

During PEO, characteristic light emission results from atomic and molecular transitions of species such as hydrogen, oxygen, metallic ions, and electrolyte residues \cite{dunleavy2009, jovovicSpectroscopicStudyPlasma2012}. 
Among these, the hydrogen Balmer lines are of particular interest in this work due to their high sensitivity to the electron density, caused by the linear Stark-effect~\cite{konjevicHydrogenBalmerLines2012}. 
A relatively calibrated, high-resolution echelle spectrometer (ARYELLE Butterfly, LTB Lasertechnik Berlin GmbH) enables the observation of the H$\alpha$ (\qty{656}{\nano\metre}) and H$\beta$-lines (\qty{486}{\nano\metre}) during the PEO process. 
The H$\alpha$ line has a much higher intensity and therefore a better signal-to-noise ratio than the H$\beta$ line and is thus primarily used for determining electron densities. 
For this purpose, the measured H$\alpha$ profile is fitted with a Voigt function, where the Gaussian component accounts for Doppler and instrumental broadening, while the Lorentzian component reflects the effects of Van der Waals and Stark broadening \cite{Bracht_diss, hillebrandDeterminationPlasmaParameters2020}. 
The instrumental broadening was first characterized using low-pressure argon plasma, allowing for the deconvolution of the Voigt profile. 
Due to the highly transient microdischarges and a comparably long exposure time of the spectrometer of \qty{10}{\second}, both single- and two-density Voigt fits were determined. 
The two-density approach provided a better fit, especially at the line center. 
Physically, the improved fit using two densities suggests significant variations in electron density during the measurement period. 
This is consistent with the work of Dorval~\emph{et al.} \cite{dorval_bayesian_2025}, who found that accounting for the variation in electron density over the discharge time during nanosecond pulsed discharges in liquids allowed for a better fitting to the H$\alpha$ profile compared to a single density fit.
From our two-density fitting analysis, electron densities were estimated to be in the range of approximately \qtyrange{5e20}{5e22}{\per\cubic\metre}.
\\

In contrast, a relatively calibrated, low spectral resolution spectrometer (QE65000, Ocean Optics) is used to measure the continuum radiation during the PEO process in the range between 200 and \qty{1100}{\nano\metre}. 
This radiation results from the interaction between free electrons and neutral or ionized species, which can be broadly categorized into free-free and free-bound processes \cite{Bilek_2021}. 
These interactions result in Bremsstrahlung and recombination radiation, respectively. 
Only the electron-neutral Bremsstrahlung is considered in this work due to much higher neutral than ion density during the PEO process \cite{gembusCharacterisationSingleMicrodischarges2025}. 
Additionally, thermal radiation emitted from the heated anode surface is approximated by black-body radiation. 
By fitting the measured spectrum with models of Bremsstrahlung and black-body radiation, both electron and surface temperatures can be estimated. 
For aluminium substrates, surface temperatures typically range from \qtyrange{2600}{3750}{\kelvin}, while electron temperatures are between \qty{5000}{\kelvin} and \qty{12200}{\kelvin} \cite{gembusCharacterisationSingleMicrodischarges2025}. 
More details on the fitting of the continuum radiation are given in \cite{gembusCharacterisationSingleMicrodischarges2025}.
For the simulations used in this work, we assume that the neutral gas temperature during a microdischarge is in a similar range to the measured surface temperature. 

\subsection{High-speed imaging}\label{subsec:high-speed-imaging}

A high-speed imaging system synchronized with current-voltage measurements is utilized to investigate the evolution of gas bubbles during the lifetime of individual microdischarges.
The high-speed camera (VEO 410L IMP, Vision Research) is operated at a resolution of $128 \times 128$ pixels and a frame-rate of \qty{150000}{fps}. 
With this, a temporal resolution of \qty{6.66}{\micro\second} and, given that the wire is magnified so that it fills the entire camera sensor, a spatial resolution of \qty{8}{\micro\metre} can be achieved. 
Individual microdischarge ignitions generate approximately rectangularly shaped current pulses \cite{gembusCharacterisationSingleMicrodischarges2025, clyne2015}. 
Triggering on the rising edge of these current pulses, a series of pictures is taken, spanning a timeframe of around \qty{1}{\milli\second}.
Depending on the conditions, multiple bubble formations and collapses can occur during this time. 
Synchronized current-voltage measurements give insight into the electrical parameters during the lifetime of each microdischarge.
\begin{figure}
    \centering
    \includegraphics[width=1\linewidth]{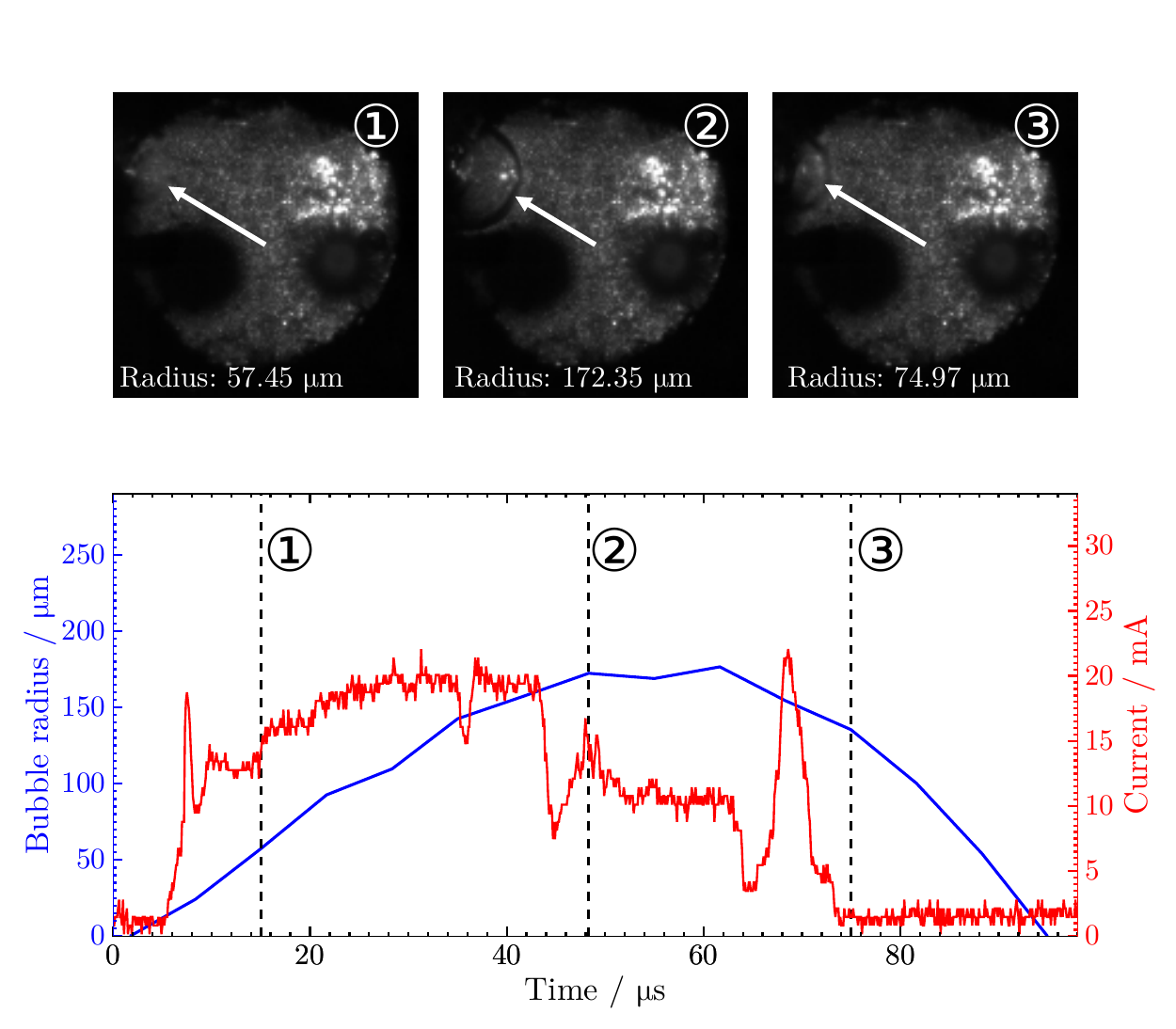}
    \caption{Current associated with an individual microdischarge event with corresponding high-speed camera images and bubble radii for an electrolyte composition of \qty[per-mode=symbol]{1}{\gram\per\litre} KOH at a constant ignition voltage of \qty{481}{\volt}. The arrows in each image indicate the position of the bubble.}
    \label{fig:bubble_images_current_voltage_measurement}
\end{figure}
Fig.~\ref{fig:bubble_images_current_voltage_measurement} shows a typical current profile associated with an individual microdischarge, along with corresponding bubble images and radii. 
During this time period, the voltage remains approximately constant at \qty{481}{\volt}. 
Further information on the analysis of current, voltage and bubble radius to calculate the power density profile of the microdischarge is given in Sec.~\ref{sec:simulation-framework}.
\\

A more detailed description of the setup, operational parameters and experimental studies of electrical parameters and bubble dynamics can be found in
the works of Bracht~\emph{et al.}~\cite{Bracht_diss} and Gembus~\emph{et al.}~\cite{gembusCharacterisationSingleMicrodischarges2025}. 
In these works, calculations of the neutral gas pressure inside the bubbles using the Rayleigh-Plesset equation were conducted.
Time-resolved radius measurements served as input to derive the corresponding pressure evolution.
In general, the neutral gas pressure is found to be dependent on time, as an individual bubble expands and collapses. 
While the extent of the variation in pressure differs from bubble to bubble, it is generally found that these pressures reach up to a few bar at the start of the bubble, decrease by around a factor of ten when the bubble reaches its maximum size, and then increase again as the bubble collapses. 
On average, the bubble pressure is in the range of \qty{1}{\bar}. 
For a more detailed discussion of these measurements, see the works of Bracht~\cite{Bracht_diss} and Gembus~\emph{et al.}~\cite{gembusCharacterisationSingleMicrodischarges2025}.

%% file: 04_simulation_framework.tex
\section{Simulation framework}\label{sec:simulation-framework}
A 0-D plasma-chemical kinetics simulation framework was developed to simulate the evolution of neutral and charged species densities, as well as the electron temperature, during the lifetime of typical PEO microdischarges.
For the implementation of the underlying equations, the Julia programming language is used \cite{Julia-2017}.
The model solves the mass-balance equation for each included species, which governs the temporal evolution of species densities in the plasma, taking into account production and loss processes.
Additionally, the energy-balance equation is solved for the electron energy density to determine the electron temperature.
\\

The progression of a simulation starts with the definition of the initial densities, temperatures, plasma working conditions, and the reaction scheme, which are loaded from input files.
In some cases, there are dependencies between these values, which are calculated afterwards.
For example, the initial density for \ch{H2O} is determined using the ideal gas law, which depends on the pre-defined gas temperature and pressure.
A start time $t_{\mathrm{start}}$ and an end time $t_{\mathrm{end}}$ are also chosen to define the simulation time frame. 
The simulation then follows these steps for each time step:
\begin{enumerate}
    \item The reaction rate coefficients are calculated based on the current electron and gas temperatures.
    \item These reaction rate coefficients are used in the mass-balance and energy-balance equations to compute the temporal
    changes in species densities and the electron temperature.
    \item The calculated changes are applied to update the densities and the electron temperature.
    \item The time is advanced by a time step $\Delta t$ such that $t_{\mathrm{new}}=t_{\mathrm{old}} + \Delta t$.
\end{enumerate}
This process is repeated until $t_{\mathrm{new}} \geq t_{\mathrm{end}}$.
The progression of the simulation is handled by \textit{DifferentialEquations.jl} \cite{rackauckas2017differentialequations} using the stiff ordinary differential equation solver \textit{TR-BDF2} \cite{hosea1996analysis}.
Because of the solver used, the number and size of the time steps $\Delta t$ vary between individual simulations.

\subsection{Mass-balance equation}\label{subsec:mass-balance-equation}

The mass-balance equation is derived from the continuity equation, neglecting spatial derivatives:
\begin{equation}
    \pdv{n_{i}}{t} = R^{\mathrm{prod}}_{i}-R^{\mathrm{loss}}_{i}
    \label{eq:continuity-equation}
\end{equation}
\noindent with the number density\ $n_{i}$ for the \textit{i}-th species. 
$R^{\mathrm{prod}}_{i}$\ and\ $R^{\mathrm{loss}}_{i}$\ represent particle production or consumption due to collisions, respectively. 
Because the key focus of this work is the gas phase chemistry, we focus solely on gas-phase production and loss processes, neglecting production and losses due to surface interactions.
The net reaction rate $R^{\mathrm{total}}_i$ for the \textit{i}-th species is defined as
\begin{equation}
    R^{\mathrm{total}}_{i} = R^{\mathrm{prod}}_{i} - R^{\mathrm{loss}}_{i} = \sum\limits_{k} \nu_{i,k}k_{k}(T)\prod\limits_{j}n_{j}^{\nu_{j,k}}  
    \label{eq:reaction-rate}
\end{equation}
\noindent with the stoichiometric coefficient\ $\nu_{i,k}$\ of the \textit{i}-th species in reaction \textit{k}, the reaction rate coefficient\ $k_{k}(T)$\ for the \textit{k}-th reaction and the number density $n_{j}$ of species \textit{j}.
The reaction rate coefficients $k_{k}$ depend on either the electron temperature $T_{\mathrm{e}}$ or the gas temperature $T_{\mathrm{g}}$, depending on the nature of the reaction. 
In this work, three different variants of rate coefficients are used.

\begin{enumerate}
    \item \textbf{Calculated from cross-sections} \\
    Rate coefficients for electron impact reactions can be calculated from cross-section data using 
    \begin{equation}
        k(T_{\mathrm{e}}) = \int_0^\infty \sigma(\epsilon) \epsilon f(\epsilon, T_{\mathrm{e}}) \, \mathrm{d}\epsilon,
        \label{eq:rate:cross-section-equation}
    \end{equation}
    \noindent with the reaction cross-section $\sigma$ evaluated at the mean energy $\epsilon$, and the electron energy distribution function (EEDF) $f(\epsilon,T_{\mathrm{e}})$.
    For this, information about the EEDF is necessary, which, in this work, is assumed to be Maxwellian, following the form 
    \begin{equation}
        f(\epsilon) = \frac{2}{\sqrt{\pi}} \frac{\epsilon^{1/2}}{(k_{\mathrm{B}} T_{\mathrm{e}})^{3/2}} \exp\left(-\frac{\epsilon}{k_{\mathrm{B}} T_{\mathrm{e}}}\right),
        \label{eq:maxwellian-eedf}
    \end{equation}

    \item \textbf{Arrhenius rate coefficient} \\    
    Gas-temperature-dependent reactions are expressed in a modified Arrhenius form:
    \begin{equation}
        k(T) = A \cdot \left(\frac{T}{T_{0}}\right)^n \cdot \exp\left(-\frac{E_{a}}{k_{B}T}\right),
        \label{eq:rate:arrhenius-equation}
    \end{equation}
    \noindent with the pre-exponential factor $A$. 
    $T_{0}$ is a normalisation temperature usually taken to be \qty{300}{\kelvin}, and $E_{a}$ is the activation energy of the reaction and $k_{\mathrm{B}}$ the Boltzmann constant. 
    The factor $(T/T_0)^{n}$ describes the temperature dependence of the rate coefficient, in addition to the exponential term. 
    A subset of electron impact reactions are included using a similar form.
    
    \item \textbf{Pressure dependent three-body rate coefficient} \\    
    While most three-body rate coefficients are considered using the modified Arrhenius form, an exception is the reaction \ch{OH + OH + M -> H2O2 + M} (R310), which follows the rate expression described in \cite{baulch.1992}.
    As described in \cite{baulch.1992}, the rate coefficient $k_{\mathrm{R}310}$ is calculated as 
    \begin{equation}
        k_{\mathrm{R}310} = \frac{k_{0} [M] k_{\infty}}{k_0[M] + k_{\infty}} \cdot F
    \end{equation}
    \noindent where $k_0$ and $k_{\infty}$ are the low- and high-pressure limiting rate coefficients, respectively. 
    $[M]$ is the density of the third body, which is considered to be any neutral species in the reaction scheme. 
    $F$ is a broadening factor given by 
    \begin{align}
        log_{10} F &\cong \frac{\log_{10}F_{c}}{\left[1 + \left(\frac{\log_{10}(k_{0} [M] / k_{\infty})}{N}\right)^2\right]}
    \end{align}
    \noindent with $N = 0.75 - 1.27 \log_{10} F_{c}$. 
    The values used for $k_0$, $k_{\infty}$ and $F_{c}$ in the simulation are listed in Tab.~\ref{tab:hydroxyl-ratecoeff}, following the preferred values given in \cite{baulch.1992}. 

    \begin{table}
        \centering
        \begin{tabular}{cl}
        \hline\hline
            $k_0$ & $\num{4.0e-36}\,\mathrm{m}^{3}\mathrm{molecule}^{-1}\mathrm{s}^{-1}$ \\
            $k_\infty$ & $\num{1.5e-17}\cdot(T/300)^{-0.37}\,\mathrm{m}^{3}\mathrm{molecule}^{-1}\mathrm{s}^{-1}$ \\
            $F_c$ & 0.5 \\
            \hline
        \end{tabular}
        \caption{Values used for the rate coefficient $k_{\mathrm{R}310}$ of reaction \ch{OH + OH + M -> H2O2 + M} following the preferred values given in \cite{baulch.1992}.}
        \label{tab:hydroxyl-ratecoeff}
    \end{table}
\end{enumerate}

\subsection{Energy-balance equation}\label{subsec:energy-balance-equation}

Electron-heavy particle reactions are generally important drivers of reaction kinetics and energy transfer processes in plasmas.
As the rate coefficients for these reactions depend directly on the electron temperature $T_{\mathrm{e}}$, it is necessary to solve the electron energy balance equation to determine the temporal evolution of the electron temperature. 
The key source of electrons to gain energy is via absorption of the electrical power input to the discharge, while they typically lose energy in collisional processes.
In the model, these processes are included in the energy-balance equation for electrons, which takes the following form:
\begin{align}
    \label{eq:theoretical-energybalance}
    \frac{3}{2}k_{\mathrm{B}}\dv{\left(T_{\mathrm{e}}n_{\mathrm{e}}\right)}{t} = P_{\mathrm{dens}} + S_{\mathrm{ela}} + S_{\mathrm{inela}}
\end{align}
with the Boltzmann constant $k_{\mathrm{B}}$, the electron temperature $T_{\mathrm{e}}$ and the electron density $n_{\mathrm{e}}$.
The term $P_{\mathrm{dens}}$ describes the power absorbed per unit volume defined as
\begin{equation}
    P_{\mathrm{dens}} = \frac{P_{\mathrm{in}}}{V}
    \label{eq:absorbed-power-term}
\end{equation}
with $P_{\mathrm{in}}$ the external power coupled into the plasma system and $V$ the volume in which the power is deposited.
A more detailed discussion of the absorbed power assumed in this work can be found in section\ \ref{subsec:power-density}.
The remaining terms $S_{\mathrm{ela}}$
\begin{equation}
    S_{\mathrm{ela}} = - \frac{3}{2} \left(k_{\mathrm{B}}T_{\mathrm{e}} - k_{\mathrm{B}}T_{\mathrm{g}}\right)
    \sum\limits_{k}^{N_{\mathrm{ela}}} 2\frac{m_{\mathrm{e}}}{m_{\mathrm{X}}} k_{k} n_{\mathrm{e}} n_{\mathrm{X}}
    \label{eq:elastic-collision-term}
\end{equation}
and $S_{\mathrm{inela}}$
\begin{equation}
    S_{\mathrm{inela}} = -\sum\limits_{k}^{N_{\mathrm{inela}}} \epsilon_{k}k_{k} \prod\limits_{j} n_{j}^{\nu_{j,k}}
    \label{eq:inelastic-collision-term}
\end{equation}
account for energy changes due to elastic and inelastic electron-neutral collisions, respectively.
Here, $N_{\mathrm{ela}}$ and $N_{\mathrm{inela}}$ are the number of elastic and inelastic reactions, $m_{\mathrm{e},\mathrm{X}}$ and $n_{\mathrm{e},\mathrm{X}}$ are the masses and the densities of the electron and the heavy collision partner, respectively, and $\epsilon_{k}$ is the energy threshold for the \textit{k}-th reaction,
describing the electron energy lost or gained due to the reaction.

\subsection{Power density}\label{subsec:power-density}

A fully self-consistent simulation of microdischarges contained within bubbles, over all relevant timescales, presents a
significant challenge due to the dynamic and complex nature of these systems.
Here, a simplified approach employs several experimentally measured quantities as inputs to the 0-D model.
In the frame of the 0-D model used in this work, spatial variations are neglected.
The temporal variation of the system features an initial expansion of a gas bubble driven by the microdischarge, which
subsequently begins collapsing after the microdischarge extinguishes.
\\

To capture the temporal variation of the system in this work, instead of a fully self-consistent approach, a temporally varying power density calculated from experimentally measured microdischarge properties serves as input.
To define this power density, several assumptions are made.
The electrical power measured in the system is assumed to be initially deposited into electrons, representing a source term in the electron energy equation discussed above.
This power is further assumed to be deposited homogeneously within the volume of the gas bubble.
While these assumptions enable the use of experimentally measured quantities as input to the model, they represent significant simplifications of the overall system.
For instance, the microdischarges themselves are significantly smaller than the gas bubbles in which they form.
As a result, the power is likely deposited inhomogeneously within the bubble volume, with larger power density in the microdischarge region and smaller power density in the region surrounding the visible microdischarge.
Further discussion of limitations is given in Sec.~\ref{sec:results}, where simulated electron density is compared to those measured in the experiment.
\\

\begin{figure}
    \centering
    \includegraphics[width=1\linewidth]{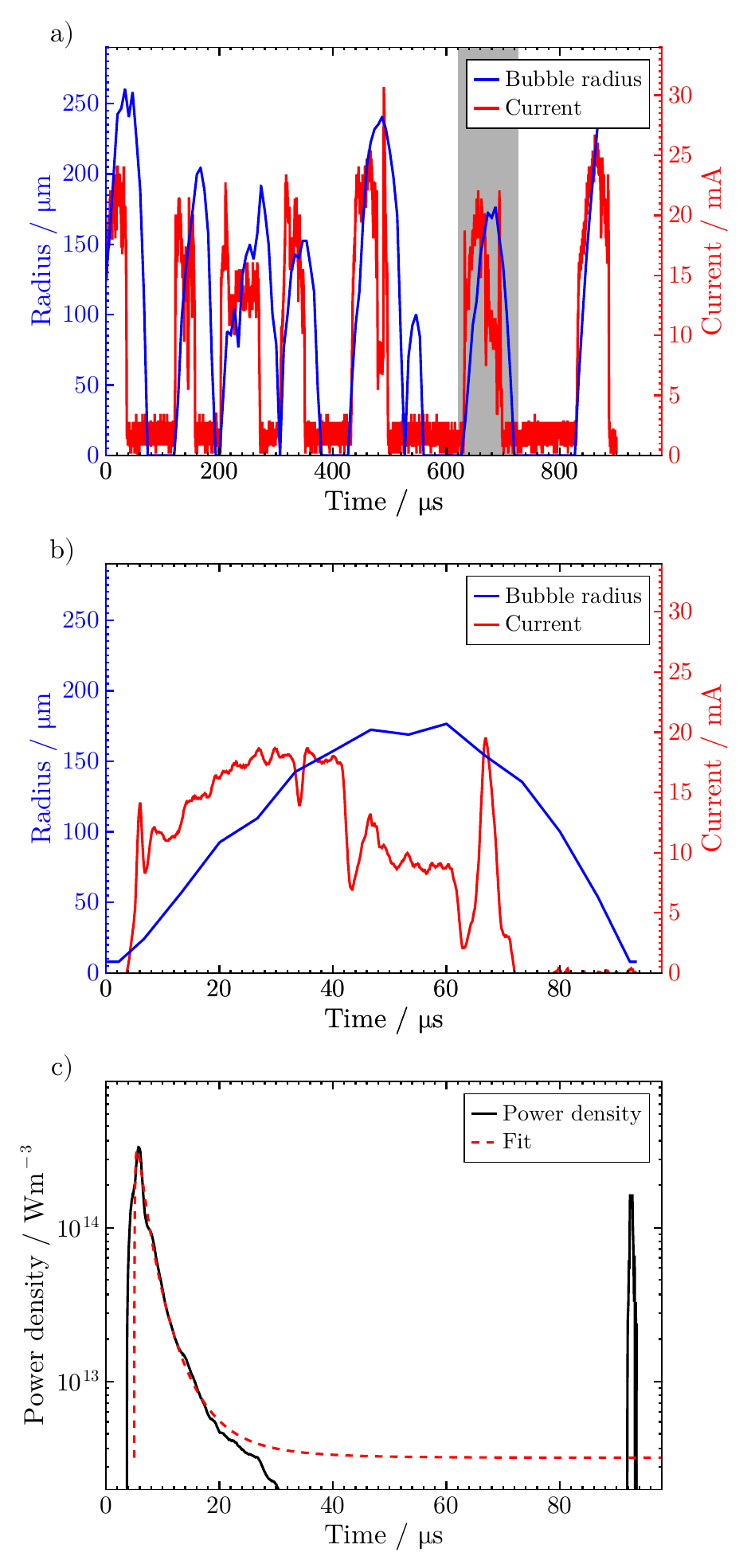}
    \caption{a) Synchronised current and bubble radius measurements as functions of time, including a series of consecutive microdischarge ignitions. b) Detailed view of the current and radius associated with the microdischarge event marked by the grey shared area in a), processed according to the assumptions described in the text. c) Power density calculated from the microdischarge in b), and subsequent fit used as input to the simulation.}
    \label{fig:current-voltage-power-density}
\end{figure}

The power density calculation utilizes the high-speed imaging and synchronized current-voltage measurements described in Sec.~\ref{sec:experiment}. 
The results of one such measurement are shown in Fig.~\ref{fig:current-voltage-power-density}~a). 
During this time period, the voltage remains approximately constant around \qty{481}{\volt}. 
Here, a series of current pulses associated with individual microdischarge events can be observed. 
Each current pulse is accompanied with the synchronised expansion of a gas bubble. 
Further discussion of these dynamics in this experimental system are given in~\cite{gembusCharacterisationSingleMicrodischarges2025}. 
From these measurements, the data for individual microdischarges are selected and processed.
The power density is then calculated for each extracted data series, starting with evaluating the deposited power
as the product of voltage and current, as expressed in:
\begin{equation}
    \label{eq:power}
    P_{in}(t) = U(t) \cdot I(t),
\end{equation}
Assuming a hemispherical bubble on the electrode, the bubble volume $V$ is computed as:
\begin{equation}
    \label{eq:bubble-volume}
    V(t) = \frac{2\pi}{3} \left(R(t)\right)^3
\end{equation}
The power density is then calculated as:
\begin{equation}
    \label{eq:power-density}
    P_{dens}(t) = \frac{P_{in}(t)}{V(t)}
\end{equation}

Due to limitations stemming from the experimental setup, a number of additional assumptions need to be made about the bubble radii and current-voltage data, to enable the calculation of power densities suitable for the simulation. 
Current-voltage measurements are inherently overlaid by noise, which, if used as is, would lead to sharp temporal variations in the power density profile used in the model, making the equation system more difficult to solve.
A Savitzky-Golay filter is used to smooth both current and voltage signals. 
A short filter window size is used for the current signal to preserve its temporal profile.
Since the voltage is mostly constant during an individual measurement, the filter window size is less important.
\\

The current measurement shows an additional background current, which is likely associated with electrolysis and unrelated to the microdischarge.
This background electrolysis current is determined from a timeframe where no bubble or microdischarge is observed. 
This is identified by a stable behaviour of the current. 
This current is then subtracted from the measured current before the power is calculated. 
Through testing of different values, the \qty{75}{\percent} quantile of the electrolysis current was found to be most suitable for use as a background value, rather than the mean, as it more effectively reduced the influence of transient fluctuations or noise spikes. 
After subtraction, negative current values are clipped to zero to prevent unphysical negative power densities.
\\

Lastly, an additional constraint for the bubble radius needs to be made.
Sec.~\ref{sec:experiment} explains the acquisition process of bubble radii by the high-speed camera, where individual frames are captured with a temporal resolution of $\Delta t=\qty{6.66}{\micro\second}$ for a total duration of around \qty{1}{\milli\second}.
This results in only \numrange{2}{3} datapoints being available for the beginning of a bubble formation.
In general, these data points do not exactly align with the rising edge of the current, therefore making it challenging to ascertain when the precise start of the bubble occurs.
To address this misalignment, interpolation between the last image frame with no visible bubble radius and the first image frame with a non-zero radius is typically employed to estimate the bubble radius at this time. 
However, this approach is not ideal as the alignment of the zero bubble radius with the current measurement remains uncertain, leading to inaccuracies in the early-stage bubble dynamics. 
Consequently, this uncertainty is a significant challenge for the calculation of the power density, as the small radius values near zero can inflate the power density unrealistically.
To mitigate this issue, the bubble radius is constrained to a minimum of \qty{8}{\micro\metre}, which corresponds to the spatial resolution of the high-speed camera. This provides a practical limit to minimise unphysical results in the power density calculation.
In Fig.~\ref{fig:current-voltage-power-density}~b) data for a single bubble is presented, with these additional assumptions applied.
The resulting calculated power density, following equations~\ref{eq:power}, \ref{eq:bubble-volume}, and \ref{eq:power-density}, is shown in Fig.~\ref{fig:current-voltage-power-density}~c).
\\

To integrate the experimental power density into the 0-D model, a fitting function captures the general trend of $P_{dens}(t)$, prioritizing peak height over exact reproduction. 
The fitting function combines a log-normal component (equation \ref{eq:lognormal-fit-part}) for the initial peak, an exponential decay for the tail (equation \ref{eq:exponential-fit-part}), and a transition function to enforce $P_{dens}(t=0) = 0$ (equation \ref{eq:transition-func}):
\begin{equation}
    f_{lognorm}(t) = \frac{A_1}{t \sigma \sqrt{2\pi}} \cdot \exp
    \left\{ -\frac{\left( \log_{10}(t) - \mu \right)^2}{2\sigma^2} \right\}
    \label{eq:lognormal-fit-part}
\end{equation}
\begin{equation}
    f_{exp}(t) = A_2 \cdot \exp\left\{ -\frac{t}{\tau} \right\}
    \label{eq:exponential-fit-part}
\end{equation}
\begin{equation}
    f_{tf}(t) = 0.5 \cdot \left( 1 + \tanh(k \cdot t) \right)
    \label{eq:transition-func}
\end{equation}
\noindent with $A_1$, $\mu$, $\sigma$, $A_2$, and $\tau$ being fitting parameters, while across all power density profiles fitted in this study, the parameter $k$ is fixed at $10^{10}$.
\\

The transition function ensures a steep transition from zero to the fitted curve within the first few nanoseconds. 
This minimizes distortion of the log-normal and exponential components and ensures that the power density function remains smooth, so that the equation system can be solved more easily.
Since the log-normal term is undefined at $t=0$, $f_{full}(0)$ is manually set to zero, to align with the physical expectation of no power delivery before bubble formation. 
Outliers in the power density data, arising from residual electrolysis current combined with small bubble radii and incompatible with microdischarge dynamics, are excluded during fitting to ensure that the power density profiles remain as realistic as possible.
The full fitting function is:

\begin{equation}
    f_{full}(t) = \left( f_{lognorm}(t) + f_{exp}(t) \right) \cdot f_{tf}(t)
    \label{eq:full-fit}
\end{equation}
with the result for the previously presented power density data being shown in Fig.~\ref{fig:current-voltage-power-density}~c).

\subsection{Plasma-chemical reaction scheme}\label{subsec:chemistry-set}

The reaction scheme developed in this work consists of 310 reactions, including 21 species.
The included species can be found in Tab.~\ref{tab:species}, while the complete list of reactions can be found in the appendix. 
The ionization potentials and electron affinities corresponding to the positive and negative ions associated with each species are given in Tab.~\ref{tab:ion-attributes}, for reference.
\\

\begin{table}[h!]
    \caption{Species Included in the Model}
    \begin{tabular}{l l}
        \hline\hline
        Category          & Species                                                        \\
        \hline
        Electrons         & \ch{e}                                                         \\
        Neutral atoms     & \ch{H}, \ch{O} \\
        Neutral molecules & \ch{H2}, \ch{H2O}, \ch{H2O2}, \ch{HO2} \\
                          & \ch{O2}, \ch{O3}, \ch{OH}  \\
        Positive ions     & \ch{H+}, \ch{H2+}, \ch{H3+} \\
                          & \ch{O+}, \ch{O2+} \\
                          & \ch{OH+}, \ch{H2O+} \\ 

        Negative ions     & \ch{H-},\ch{O-},\ch{O2-},\ch{OH-} \\
        Sum of neutral species    & \ch{M}                                                         \\
        \hline\hline
    \end{tabular}
    \label{tab:species}
\end{table}

\begin{table}[h!]
    \caption{Electron affinities and ionization potentials for species considered in this work. Values are taken from the NIST Chemistry WebBook~\cite{nist_webbook}.}
    \begin{tabular}{l l l}
        \hline\hline
        Species & Electron affinity & Ionization potential  \\
        \hline
        \ch{H} & 0.754 & 13.59                              \\
        \ch{O} & 1.46 & 13.61                               \\
        \ch{H2} & - & 15.43                                 \\
        \ch{H2O} & - & 12.61                                \\
        \ch{O2} & 0.45 & 12.07                              \\
        \ch{OH}& 1.83 & 13.1                                \\
        \ch{H3} & - & -                                     \\
        \hline\hline
    \end{tabular}
    \label{tab:ion-attributes}
\end{table}

The reaction scheme has been constructed with the aim of including a comprehensive treatment of both electron and high gas temperature driven reactions. 
To this end, a range of electron impact processes are included for all neutral species that are expected to be present at high densities, i.e. \ch{H2O}, \ch{H2}, \ch{O2}, \ch{H}, \ch{O}, and \ch{OH}. 
Electron-ion recombination, ion-ion recombination, and electron-impact detachment from negative ions are included to describe interactions between charged particles.
Charge exchange, ion association, and detachment of negative ions during collisions with neutrals are included to describe ion-neutral reactions. 
For reactions among neutrals, both reactions with low and high temperature threshold are included, so that the model should be applicable over a wide range of gas temperatures. 
These include purely thermal dissociation reactions of all neutral molecules. 
As far as possible, neutral reaction rates that have been measured or that arise from reviews where they are recommended over a wide gas temperature range have been included.
\\

A key simplification that has been made in the reaction scheme is that only ground-state neutral and charged species are explicitly simulated, while excited-state species are grouped to their corresponding ground-state.
However, reactions that would produce these excited states are retained in the reaction network.
This approach ensures that, for example, electron energy loss reactions are captured in the energy-balance equation, even though the densities of excited-states formed from them are not individually tracked. 
A weakness in this approach is that an explicit role of excited atoms or molecules in the reaction scheme is lost, such as where these excited states can increase the rate coefficients for individual reactions. 
While these can certainly be important processes that should be included in future reaction schemes, their absence is not expected to affect the overall trends and conclusions presented in this work. 
In the appendix, reactions where an excited state would otherwise be formed, but is not included explicitly in the model, are marked with an additional line of text indicating the excited state involved. 
The same has been done in instances where the rate constant from the source publication is for a slightly different reaction to that which it has been used for here.
Additionally, reactions producing complex ions, such as \ch{H3O+} or \ch{HO2+}, are modified to align with the species defined in Tab.~\ref{tab:species}.
For example, reactions such as:
\begin{align*}
    \ch{H2+} + \ch{H2O} &\rightarrow \ch{H} + \ch{H3O+}&& (\mathrm{R}219) \\
    \ch{H2+} + \ch{O2} &\rightarrow \ch{H} + \ch{HO2+}&& (\mathrm{R}220) \\
    \ch{OH+} + \ch{H2O} &\rightarrow \ch{H3O+} + \ch{O}&& (\mathrm{R}223) \\
    \ch{H2O} + \ch{H3+} &\rightarrow \ch{H2} + \ch{H3O+}&& (\mathrm{R}225) \\
    \ch{H3+} + \ch{O2} &\rightarrow \ch{H2} + \ch{HO2+}&& (\mathrm{R}226)
\end{align*}
are not included in their original form. 
Instead, these reactions are reformulated to produce products consistent with the tracked species, such as \ch{H2O+} or \ch{O2+}, to avoid introducing additional ion species like \ch{H3O+}, \ch{HO2+}, or positive water cluster ions. 
This approach simplifies the reaction network while preserving the essential pathways that are expected to be most relevant to the overall chemical kinetics. 
A similar approach is applied for negative ions, where negative water clusters are also neglected. 
Reactions that have been modified in this way are also marked in the reaction table in the appendix.

\subsection{Base case simulation conditions}\label{subsec:base-case}

\begin{figure}
    \centering
    \includegraphics[width=1\linewidth]{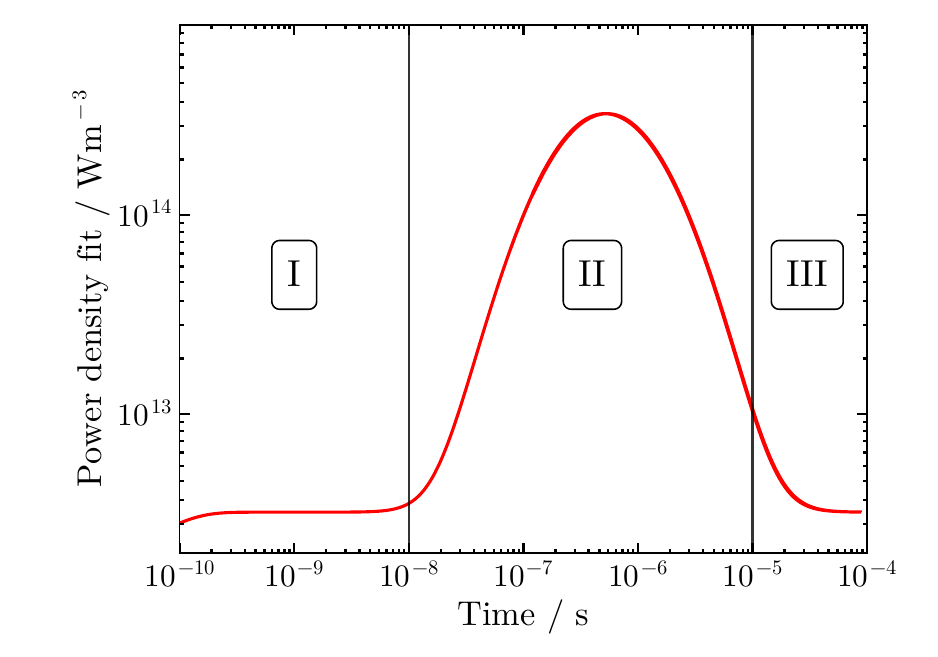}
    \caption{Power density fit function used in the simulation. The simulation timeframe is divided into three phases based on the behaviour of the fitted power density. These phases will be used in the discussion of the results. This is the same power density fit profile shown in Fig.~\ref{fig:current-voltage-power-density}~c).}
    \label{fig:power-density-fit}
\end{figure}

\begin{table}[h!]
    \centering
    \caption{Simulation inputs for the base case}
    \begin{tabular}{lll}
        \hline\hline
        & Value & Description \\
        \hline
        $t_{\mathrm{start}}$ & \qty{0}{\second} & Simulation start time \\
        $t_{\mathrm{end}}$ & \qty{8.85e-5}{\second} & Simulation end time \\
        $p$  & \qty{1}{\bar} & Bubble pressure\\
        $T_{\mathrm{g}}$  & \qty{2000}{\kelvin} &  Gas temperature \\
        $T_{\mathrm{e}}$  & \qty{1}{eV} & Initial electron temperature \\
        $n_{\mathrm{e}}$ & \qty{e10}{\metre^{-3}} & Initial electron density\\
        $n_{\mathrm{H2O}}$ & \qty{3.7e24}{\metre^{-3}} & Initial $\ch{H2O}$ density\\
        $n_{\mathrm{X}}$ & \qty{e10}{\metre^{-3}} & Remaining species densities\\
        \hline\hline
    \end{tabular}

    \label{tab:base-case_parameters}
\end{table}

Tab.~\ref{tab:base-case_parameters} shows the conditions used for the base case simulation.
A gas pressure of \qty{1}{\bar} is assumed as the average pressure within the bubble, roughly consistent with time-resolved measurements of bubble radii analyzed using the Rayleigh-Plesset equation, as detailed in Sec.~\ref{subsec:high-speed-imaging} and shown in more detail in \cite{gembusCharacterisationSingleMicrodischarges2025}.
For the base case simulation, the gas temperature is set to \qty{2000}{\kelvin} as a conservative estimate, assuming the neutral gas temperature is in the same range as the substrate surface temperature. 
The gas temperature in the simulation is kept constant during the microdischarge lifetime. 
Experimentally, the substrate surface temperature has been measured to be in the range \qtyrange{2600}{3750}{\kelvin}, as discussed in Sec.~\ref{subsec:spectroscopic-analysis}.
Since the gas temperature is approximated by the surface temperature, it is a relatively poorly defined input to the model. 
Based on this, results for simulations performed at $T_{\mathrm{g}} = \qty{2000}{\kelvin}, \qty{4000}{\kelvin}$, and \qty{6000}{\kelvin} are discussed in Sec.~\ref{sec:results} to better understand how the gas temperate affects the results of the model and the overall chemical kinetics.

At $T_{\mathrm{g}}=\qty{2000}{\kelvin}$ and $p=\qty{1}{\bar}$ the gas density for \ch{H2O}, $n_{\ch{H2O}}$ is \qty{3.7e24}{\per\cubic\metre}, using the ideal gas law.
The gas bubble is assumed to consist of essentially pure \ch{H2O} at the start of the simulation. 
This neglects processes such as the formation of molecules due to electrolysis at the electrode, as well as the presence of components of the electrolyte and the substrate in the gas phase, such as potassium and aluminium.
The initial electron density and densities of other species are set to \qty{1e10}{\metre^{-3}}, representing trace concentrations before discharge ignition. 
The simulation runtime is set based on the measured bubble lifetime of \qty{8.85e-5}{\second}.
Multiple initial electron temperatures were tested in the scope of this work with no detectable impact on the resulting densities and reaction pathways, as presented in Sec.~\ref{sec:results}.
Therefore, the initial value is set to $T_{e} = \qty{1}{\electronvolt}$.
The calculation of the power density used for the base case simulation is discussed in Sec.~\ref{subsec:power-density}.
\\

Since the simulation results exhibit important features at early times of the bubble lifetime the results are generally presented on logarithmic axes scales.
Similarly, the power-density fit used in the simulation is presented on logarithmic axis scales in Fig.~\ref{fig:power-density-fit}.
For an easier discussion of the results in the following sections, the timescale of the microdischarge is divided into three phases, which are roughly correlated with different regimes of the power density profile:
\begin{itemize}
    \item \textbf{Phase I (\qtyrange{e-10}{e-8}{\second})}: discharge onset
    \item \textbf{Phase II (\qtyrange{e-8}{e-5}{\second})}: power density peak
    \item \textbf{Phase III (\qtyrange{e-5}{e-4}{\second})}: power density decline
\end{itemize}

%% file: 05_results.tex
\section{Results}\label{sec:results}

\subsection{Electron densities and temperatures}\label{subsec:electron-properties}

\begin{figure}
    \centering
    \includegraphics[width=\linewidth]{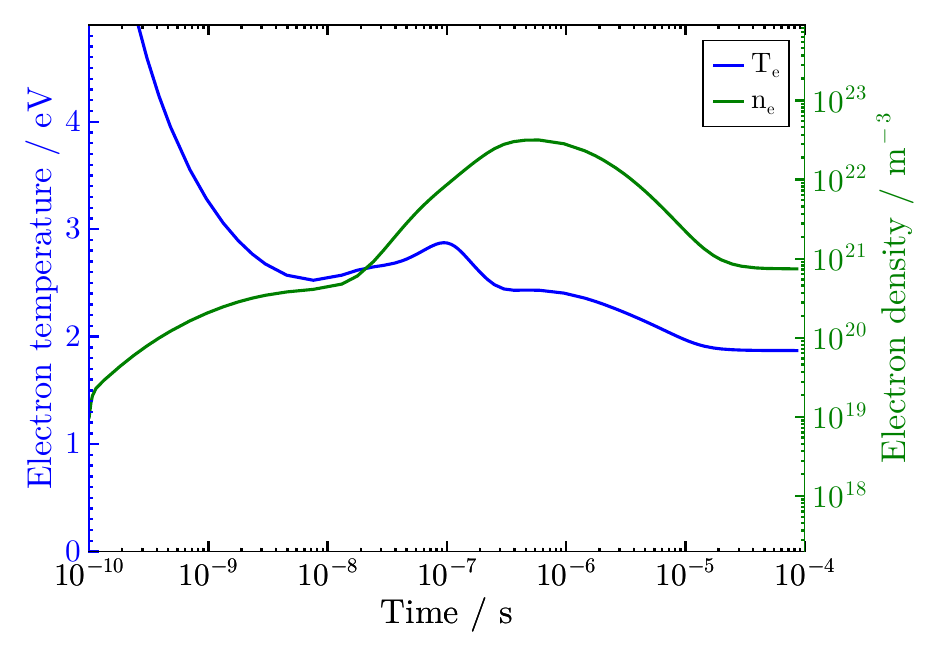}
    \caption{Simulated electron densities and temperatures for the base case simulation. The base case uses the power density profile shown in Fig.~\ref{fig:power-density-fit} and assumes a gas temperature of \qty{2000}{\kelvin}. The full simulation inputs for this case are shown in Tab.~\ref{tab:base-case_parameters}.}
    \label{fig:base-case-ne_te_film00_peak06}
\end{figure}

The electron density and temperature as a function of time throughout an individual microdischarge event are shown in Fig.~\ref{fig:base-case-ne_te_film00_peak06}, for the base case simulation.
The base case uses the power density profile described in Sec.~\ref{sec:simulation-framework} and shown in Fig.~\ref{fig:current-voltage-power-density} and \ref{fig:power-density-fit}, and assumes a gas temperature of \qty{2000}{\kelvin}.
The electron density is initialised at a low value, and the power density is \qty{0}{\watt\per\cubic\metre} at the start of the simulation.
As the power density increases, and the electron density is still low, the electron energy density in the energy equation is balanced by artificially large electron temperatures.
As electron impact ionization begins to occur, the electron density increases such that it contributes towards the electron energy density and the electron temperature decreases towards the expected range of a few eV between \qty{e-10}{\second} and \qty{e-9}{\second}.
For the remaining duration of the discharge lifetime, the electron temperature shows a decreasing behaviour, dropping to values of around \qtyrange{2}{2.5}{\electronvolt} in Phase II between \qtyrange{e-8}{e-5}{\second} and reaching its steady-state value of around \qty{1.7}{\electronvolt} by \qty{e-5}{\second} up until bubble collapse at \qty{e-4}{\second}.
The initial increase in electron density occurs after around \qty{e-10}{\second}, and continues to increase until the peak of the power density occurs after around \qty{e-6}{\second}.
As the power density begins to decrease, the electron density also decreases, reaching a relatively constant plateau value after around \qty{e-5}{\second}, until the end of the simulation, when the bubble collapses in the experiment.
\\

A detailed experimental validation of such temporally varying density profiles in transient microdischarges produced in bubbles is challenging.
Here, we aim to provide an indication of the accuracy of the simulations by comparing simulated electron densities with those measured by emission spectroscopy.
As described in Sec.~\ref{sec:experiment}, electron densities are measured by Stark broadening of the H$\alpha$ emission line.
An individual measurement accounts for emission over a large number of microdischarge events.
Fitting of the emission lines is achieved through a two-density fit, where the two densities are assumed to represent a combination of variations between individual discharges over the measurement time, as well as variations of electron density throughout an individual microdischarge event.
Measurements are also carried out at different time points throughout a PEO treatment.
As discussed in detail in \cite{gembusCharacterisationSingleMicrodischarges2025}, the properties of microdischarges vary substantially over the treatment time as the thickness of the oxide layer on the treated anode increases.
\\

\begin{figure}[h!]
    \centering
    \includegraphics[width=1\linewidth]{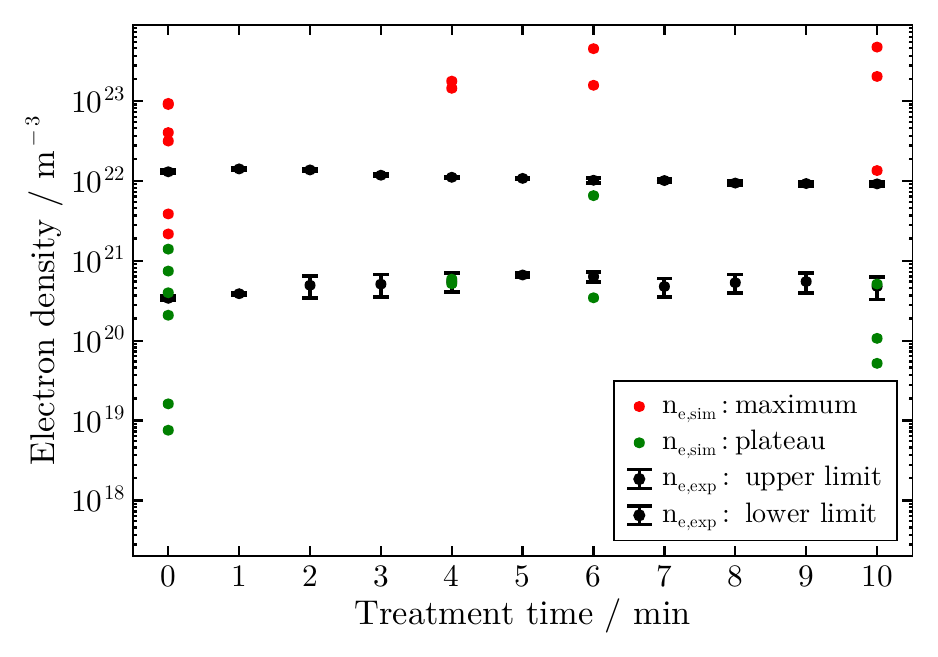}
    \caption{Comparison of electron densities from the simulations at their maximum, and during the plateau region after the main power pulse, with the two densities derived from Stark broadening of the measured H$\alpha$ emission line.}
    \label{fig:base-case_sim_vs_exp_2000K}
\end{figure}

To include these effects in the simulations, a series of electrical characteristics and time-resolved bubble radius measurements have been carried out at different treatment times, and power density profiles calculated from these.
Overall, six power density profiles at \qty{0}{minutes} treatment time, two at \qty{4}{minutes}, two at \qty{6}{minutes}, and three at \qty{10}{minutes} were found to be usable as simulation inputs.
Simulations are carried out for each power density profile, the maximum and plateau electron densities are extracted from each and used for comparison with the experimentally measured electron densities.
The results of these comparisons are shown in Fig.~\ref{fig:base-case_sim_vs_exp_2000K}.
The two electron densities derived from fitting of the H$\alpha$ profile differ by one to two orders of magnitude and do not vary strongly with treatment time.
The largest number of simulations are carried out at a time of around \qty{0}{minutes}, as this was the time point that the largest number of usable power density profiles could be determined from the available experimental data.
Here, simulations of individual microdischarges differ significantly from each other, due to large variations in the calculated power density profiles.
Between different microdischarges, the maximum and plateau values of the electron densities vary by one to two orders of magnitude.
For an individual microdischarge, the maximum electron densities are also around one to two orders of magnitude higher than the densities at the plateau.
When compared with the experiment, some microdischarge simulations have densities that sit between the two densities derived from the experiment, while others have higher maximum densities than the highest density derived from the experiment.
While this is a relatively coarse comparison that has significant limitations, the electron densities predicted by the simulations are generally consistent with the order of magnitude of the experimentally measured densities.
More broadly, this analysis demonstrates that there is significant variation both between and within individual microdischarges, which is consistent with the fact that the fitting of the experimentally measured H$\alpha$ emission can be improved assuming two densities, rather than one.
\\

In contrast to the experimental measurements, both the maximum and the plateau densities from the simulation tend to increase with increasing treatment time.
This is related to the larger power densities that are derived from the experimental measurements at later treatment times.
The fact that this behaviour is not consistent with the experimental measurements is likely related to the assumption made here that the experimentally measured power is all deposited into the electrons.
The increasing deviation between experiment and simulation as treatment time increases may be an indication that this assumption is better fulfilled at earlier treatment times.
This may be because more energy is dissipated into the breakdown of the thicker oxide layer that is present at later treatment times, a process that is not included in the model.
Overall, the power input to the electrons is likely to be overestimated in all cases, however, for early treatment times, this still allows for a reasonable approximation of the experimentally measured electron densities.
As a result, we focus on further analysis of the base case power profile, which is taken at a treatment time of around \qty{0}{minutes}.

\subsection{Base case: species densities}\label{subsec:base-case-species}

Having discussed the electron properties of the base case and how they compare with experimental measurements, the temporal variation of charged and neutral species in the base case microdischarge will be outlined in this section.
\\

\begin{figure}
    \centering
    \includegraphics[width=\linewidth]{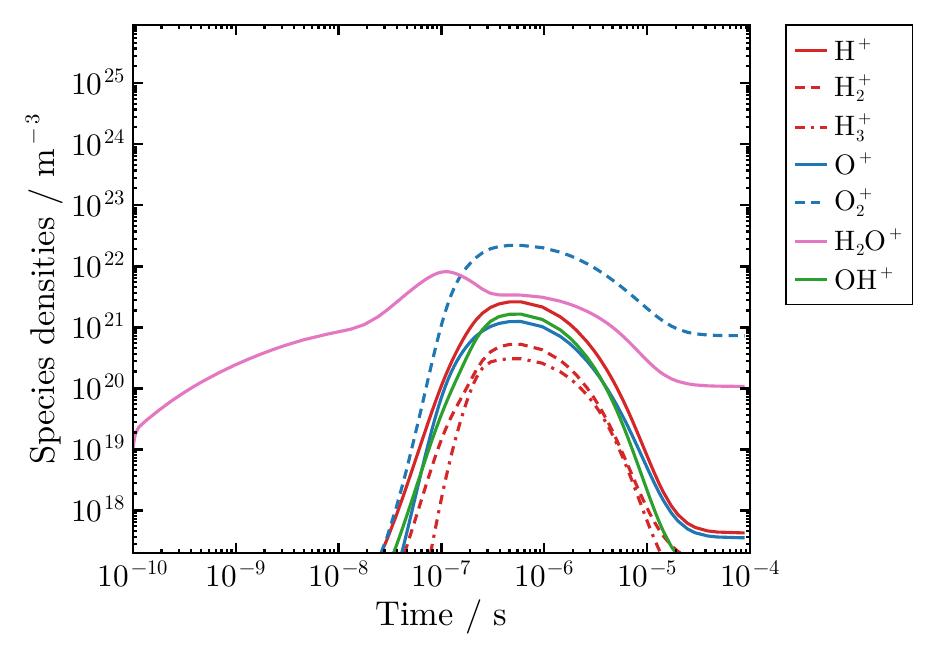}
    \caption{Simulated positive species densities for the base case simulation.}
    \label{fig:base-case-positive_species_film00_peak06}
\end{figure}

The temporal variations of the positively charged species are shown in Figure \ref{fig:base-case-positive_species_film00_peak06}.
Here, the \ch{H2O^+} density increases first, following the trend in electron density, as it is directly formed from electron impact ionization of \ch{H2O}.
At later times, \ch{O2^+} begins to increase, and becomes the dominant positive ion after around \qty{e-7}{\second}.
Other ions are formed at lower densities and do not contribute significantly to the total positive ion density at any time during the microdischarge.
It should be emphasised here that \ch{H2O^+} is used as the terminal \ch{H2O} related positive ion, as discussed in Section \ref{subsec:chemistry-set}.
Because of this, its density can be viewed as inclusive of other species such as \ch{H3O^+} and larger water cluster ions that would form in a more extensive reaction scheme.
\\

\begin{figure}
    \centering
    \includegraphics[width=\linewidth]{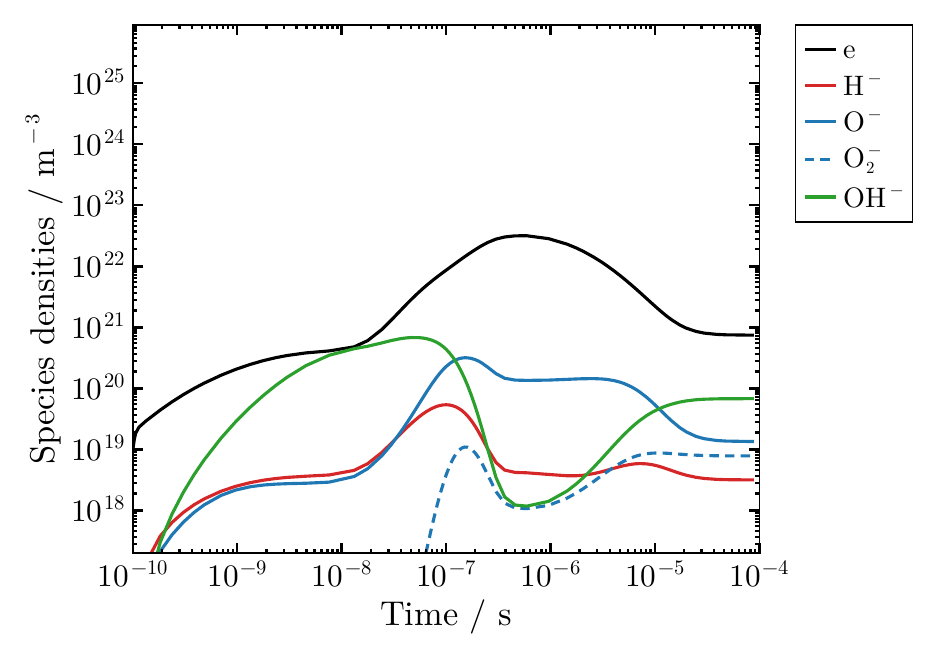}
    \caption{Simulated negative species densities for the base case simulation.}
    \label{fig:base-case-negative_species_film00_peak06}
\end{figure}

The temporal variations of the negatively charged species are shown in Figure \ref{fig:base-case-negative_species_film00_peak06}.
As discussed earlier, the electron density increases rapidly at early times.
This is followed by an increase in the density of \ch{OH^-} which reaches a similar value to that of the electrons shortly after \qty{e-8}{\second}.
This is the point in time where the electronegativity, i.e. the total negative ion density over the electron density, reaches its largest value.
After this time, the electron density continues to increase, while the density of \ch{OH^-} decreases rapidly, with \ch{O^-} becoming the main negative ion, albeit with a significantly lower density than that of the electrons.
At later times, between \qty{e-5}{\second} and \qty{e-4}{\second}, when the electron density has decreased, \ch{OH^-} again becomes the dominant negative ion.
Similarly to the case of the positive ions, the negative ion chemistry is kept purposefully simplified by neglecting negative ion clusters.
\\

\begin{figure}
    \centering
    \includegraphics[width=\linewidth]{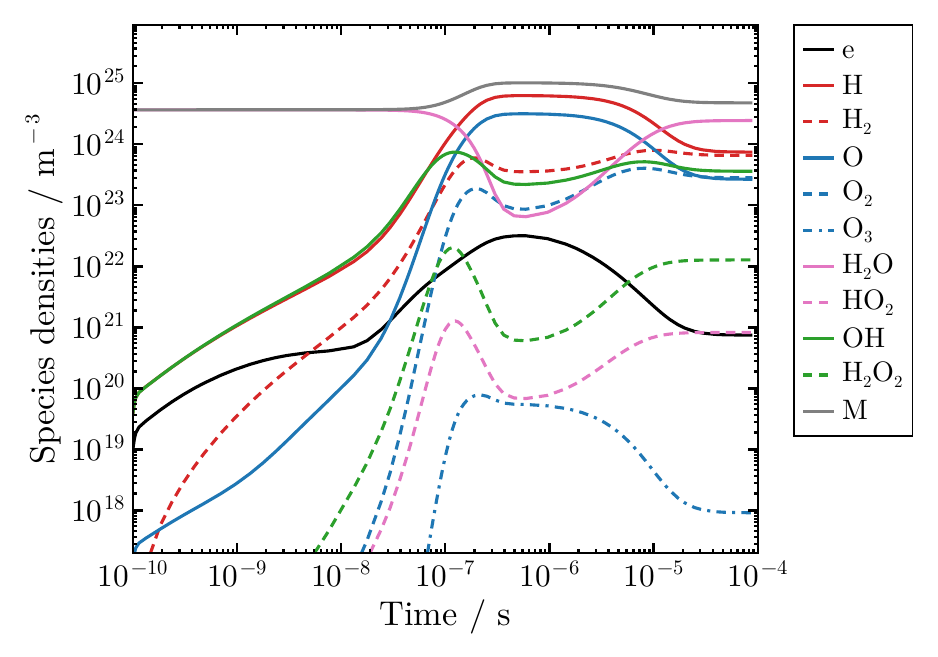}
    \caption{Simulated neutral species densities for the base case simulation. The electron density is also shown for reference.}
    \label{fig:base-case-neutral_species_film00_peak06}
\end{figure}

The temporal variations of the neutral species, with the electron density added for reference, are shown in Figure \ref{fig:base-case-neutral_species_film00_peak06}.
As the electron density increases at earlier times, the densities of \ch{H} and \ch{OH}, which are formed directly from the dissociation of \ch{H2O}, begin to increase.
After around \qty{e-7}{\second}, the dissociation of \ch{H2O} begins to become significant, such that its density decreases substantially.
Between \qty{e-7}{\second} and \qty{e-5}{\second}, \ch{H2O}, and the \ch{OH} that has formed from it, are significantly dissociated and the main constituents of the gas are \ch{H} and \ch{O}.
Towards the end of the bubble lifetime, when the electron density starts to decrease, \ch{H} and \ch{O} recombine so that the densities of \ch{H2O}, \ch{OH}, \ch{H2}, and \ch{O2} increase and they contribute significantly to the overall gas composition.
Other species such as \ch{HO2}, \ch{H2O2}, and \ch{O3} are minor constituents of the gas mixture at all times.
\\

\subsection{Base case: species production and consumption pathways}\label{subsec:base-case-species-pathways}

In the previous section, the temporal variations in species densities were shown and discussed.
In the following, the pathways leading to the production and consumption of some of the most important species in the gas mixture are discussed.
This discussion is separated out into different time periods during the microdischarge, as introduced in Sec.~\ref{subsec:base-case}.

\subsubsection{Phase I: discharge onset}\label{subsubsec:base-case-early-phase}

\begin{figure}
    \centering
    \includegraphics[width=\linewidth]{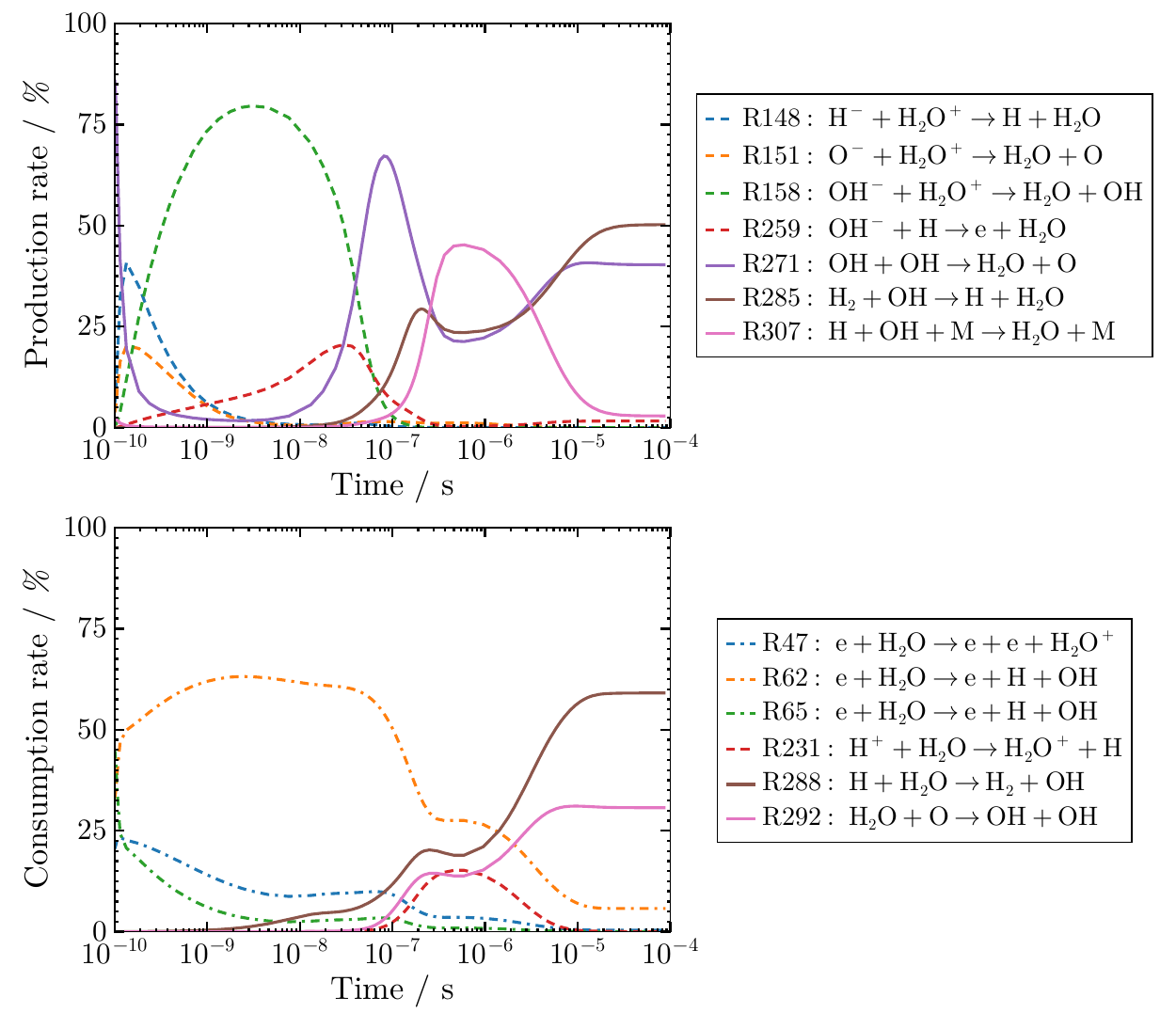}
    \caption{Production and consumption pathways of \ch{H2O} for the base case simulation.}
    \label{fig:base-case-prod_cons_H2O_film00_peak06_scaled}
\end{figure}

\begin{figure}
    \centering
    \includegraphics[width=\linewidth]{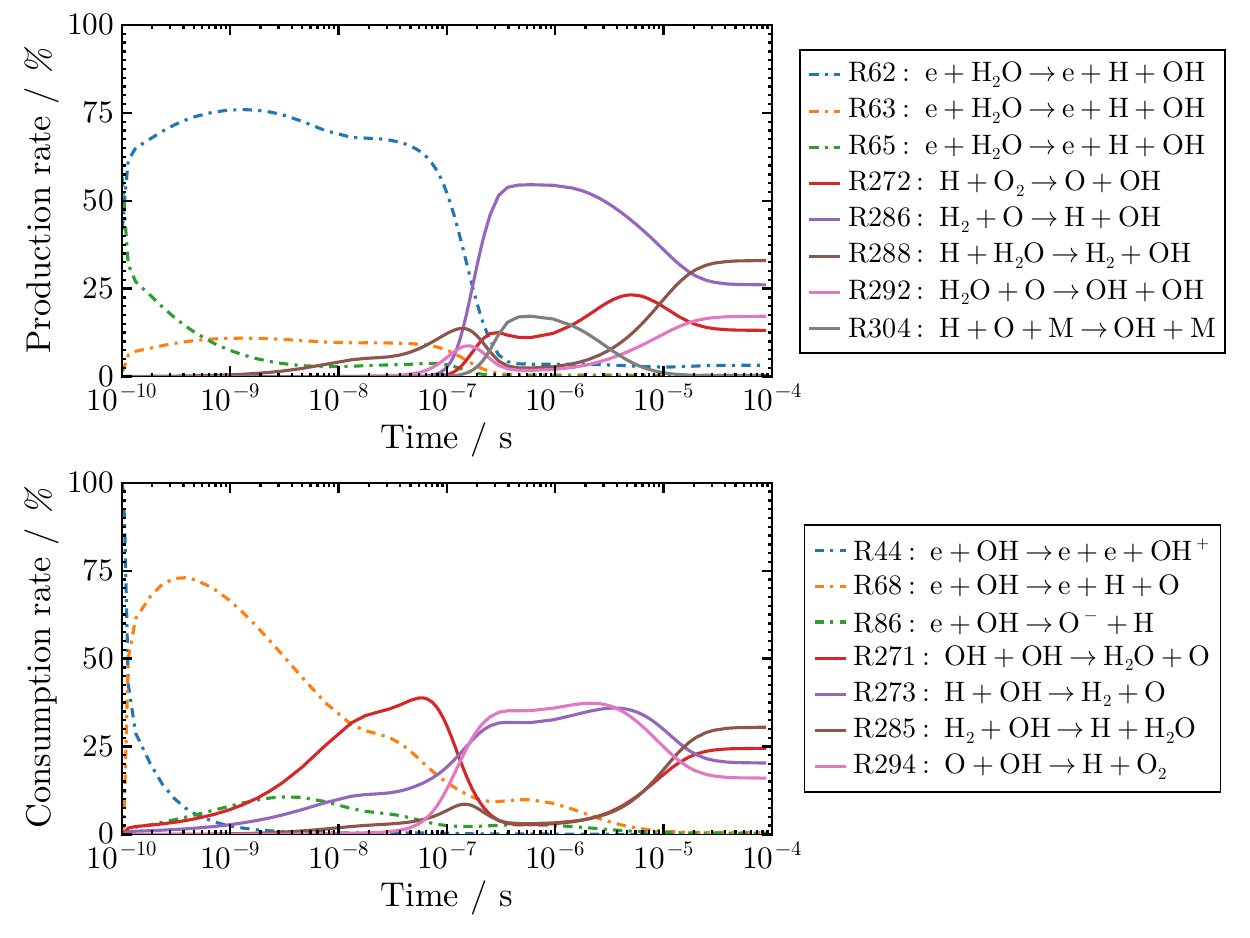}
    \caption{Production and consumption pathways of \ch{OH} for the base case simulation.}
    \label{fig:base-case-prod_cons_OH_film00_peak06_scaled}
\end{figure}

\begin{figure}
    \centering
    \includegraphics[width=\linewidth]{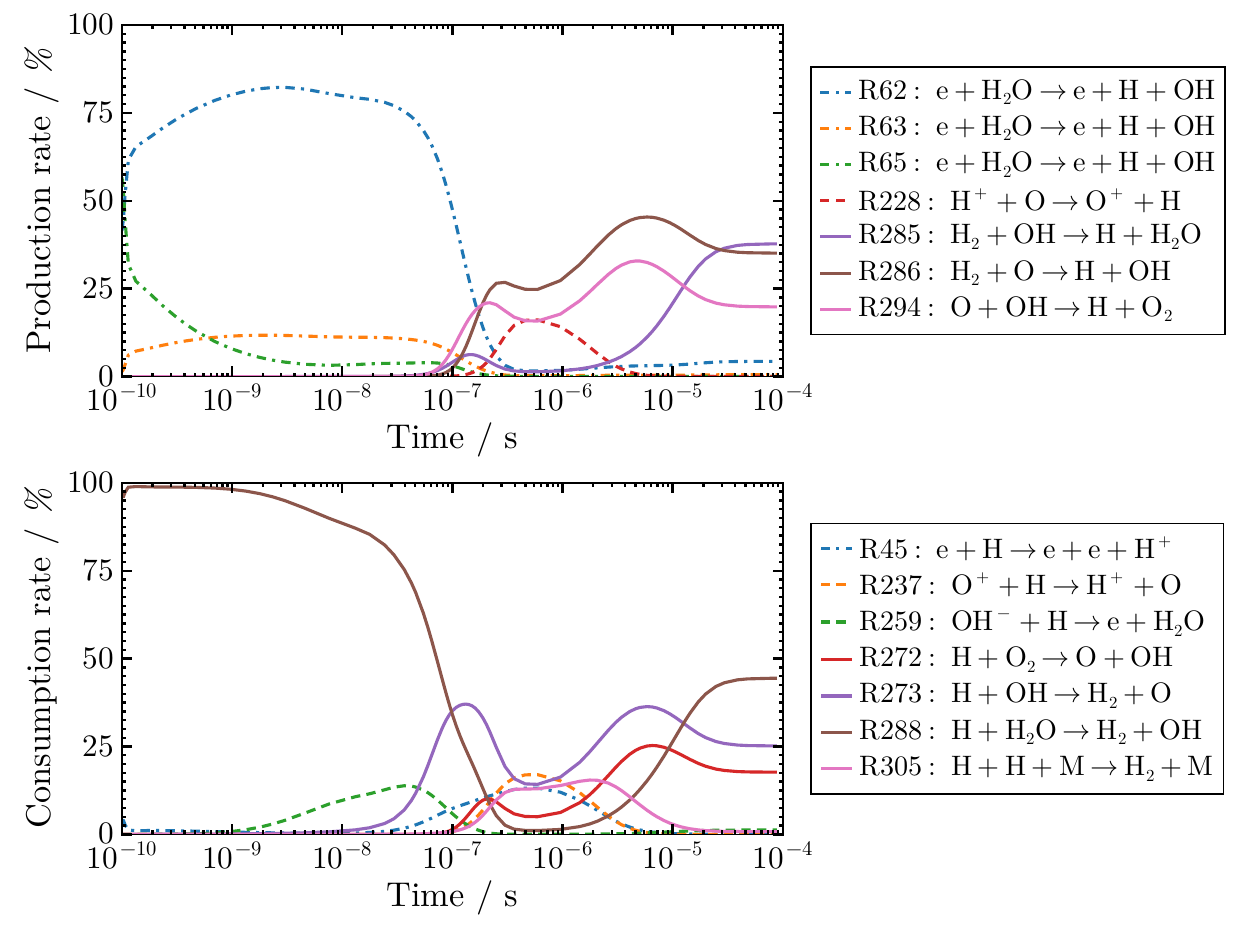}
    \caption{Production and consumption pathways of \ch{H} for the base case simulation.}
    \label{fig:base-case-prod_cons_H_film00_peak06_scaled}
\end{figure}

Firstly, the processes involved in production and consumption of the key neutral species in the discharge are shown and discussed.
The main production and consumption reactions of \ch{H2O}, \ch{OH} and \ch{H} are given in Fig.~\ref{fig:base-case-prod_cons_H2O_film00_peak06_scaled}, \ref{fig:base-case-prod_cons_OH_film00_peak06_scaled}, and \ref{fig:base-case-prod_cons_H_film00_peak06_scaled}, respectively.
\\

In Phase I (\qtyrange{e-10}{e-8}{\second}), high electron energy threshold processes dominate the consumption of \ch{H2O}, primarily electron-impact ionization (e.g., \ch{e + H2O -> e + e + H2O^+} (R47)) and dissociation (e.g., \ch{e + H2O -> e + H + OH} (R62), driven by a power input of around \qty{2e12}{\watt\per\cubic\metre}.
These reactions produce \ch{H} and \ch{OH}, with \ch{OH} further dissociating into atomic oxygen and hydrogen mainly via \ch{e + OH -> e + H + O} (R68), as illustrated in Fig.~\ref{fig:base-case-prod_cons_OH_film00_peak06_scaled}.
These reactions primarily lead to the continuous growth in the densities of \ch{H}, \ch{OH}, \ch{O}, and electrons shown in Fig.~\ref{fig:base-case-neutral_species_film00_peak06} during this phase of the microdischarge.
Due to the abundance of \ch{H2O}, its density remains largely unaffected during Phase I.
\\

\begin{figure}
    \centering
    \includegraphics[width=\linewidth]{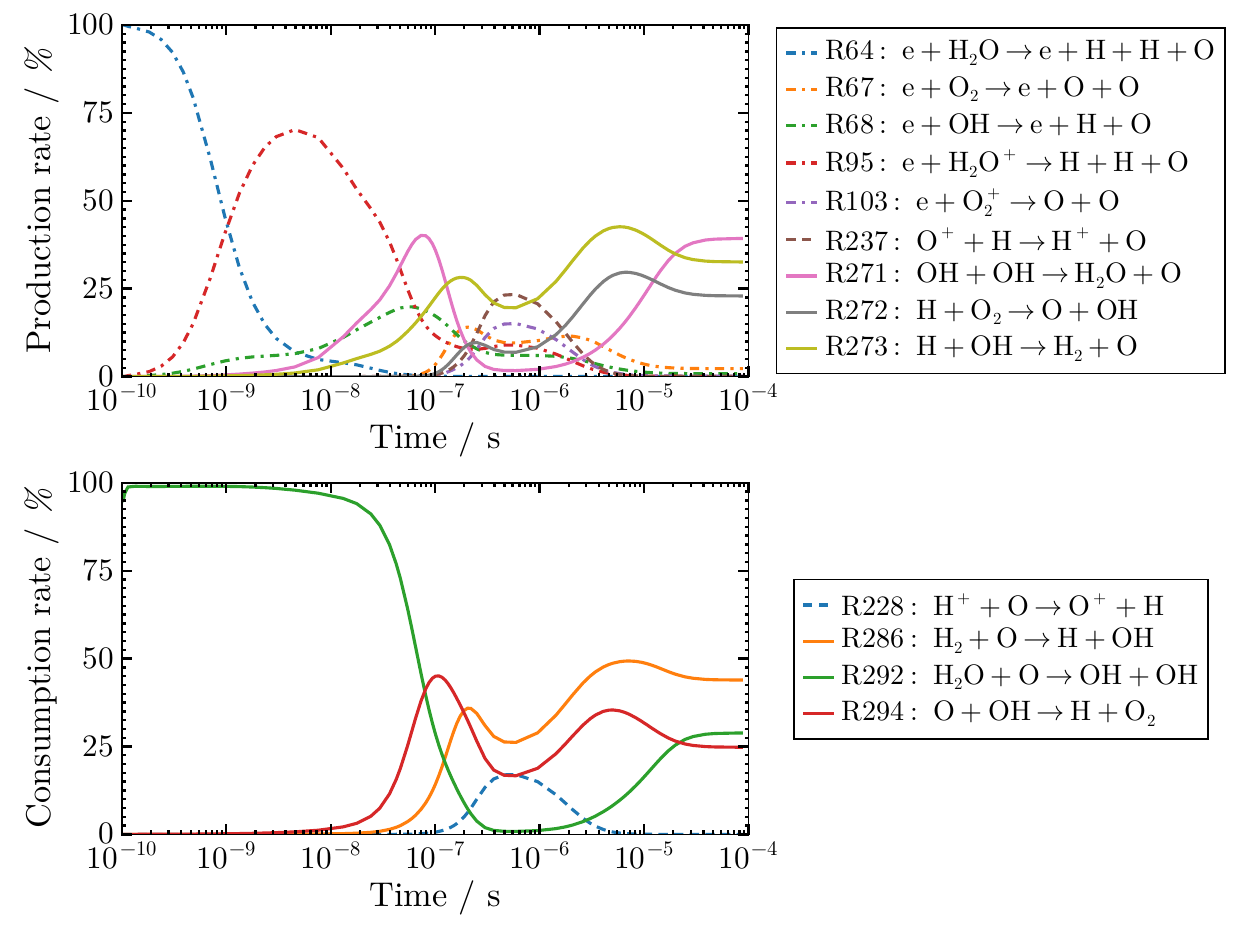}
    \caption{Production and consumption pathways of \ch{O} for the base case simulation.}
    \label{fig:base-case-prod_cons_O_film00_peak06_scaled}
\end{figure}

\begin{figure}
    \centering
    \includegraphics[width=\linewidth]{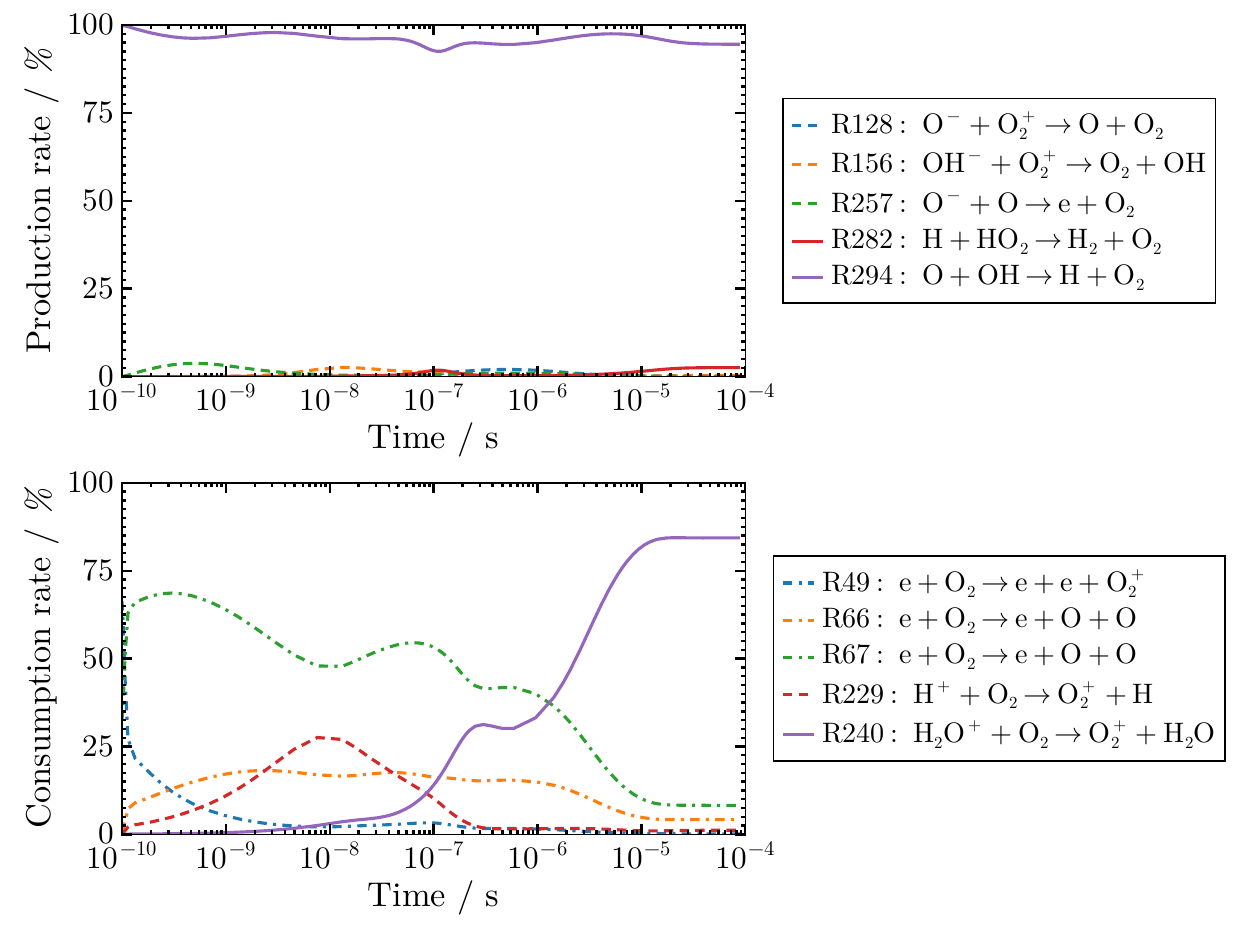}
    \caption{Production and consumption pathways of \ch{O2} for the base case simulation.}
    \label{fig:base-case-prod_cons_O2_film00_peak06_scaled}
\end{figure}

In Phase I, dissociation dominates over molecule formation, resulting in a significant imbalance between atomic and molecular species.
For oxygen, the \ch{O}/\ch{O2} ratio is extremely high, reaching approximately \num{2e7}:1 at around \qty{e-10}{\second}, due to the absence of direct \ch{O2} production pathways from \ch{H2O}.
Production and consumption pathways for \ch{O} and \ch{O2} are shown in Fig.~\ref{fig:base-case-prod_cons_O_film00_peak06_scaled} and~\ref{fig:base-case-prod_cons_O2_film00_peak06_scaled}.
\\


The primary source of \ch{O2} is the reaction \ch{O + OH -> H + O2} (R294), which depends on the \ch{OH} and \ch{O} densities.
This reaction is responsible for nearly \qty{100}{\percent} of \ch{O2} production throughout the whole simulation duration (\qtyrange{e-10}{e-4}{\second}).
Meanwhile, the consumption of \ch{O2} is governed by electron-impact processes, primarily dissociation via \ch{e + O2 -> e + O + O} (R66, R67) and, to a lesser extent, ionization via \ch{e + O2 -> e + e + O2+} (R49).
With the increase of \ch{H2O+} density in this phase, the rate of charge exchange reactions like reaction (R240) increases towards the end of the phase.
For atomic oxygen, production is driven by \ch{e + H2O -> e + H + H + O} (R64) at the beginning of the phase, which is surpassed by the contribution of \ch{e + H2O+ -> H + H + O} (R95) with increasing electron and \ch{H2O+} densities.
Conversely, \ch{O} consumption is dominated by \ch{H2O + O -> OH + OH} (R292), contributing nearly \qty{100}{\percent} during this early phase, with reactions \ch{O + OH -> H + O2} (R294) and \ch{H2 + O -> H + OH} (R286) only gaining relevance after around \qty{e-8}{\second}.
The initially large difference between atomic and molecular oxygen continually balances out between \qtyrange{e-10}{e-8}{\second}, with \ch{O2} production rates via (R294) increasing significantly by the beginning of the next phase at \qty{e-7}{\second}, resulting in a ratio of approximately 7:1 in favor of atomic oxygen.
\\

\begin{figure}
    \centering
    \includegraphics[width=\linewidth]{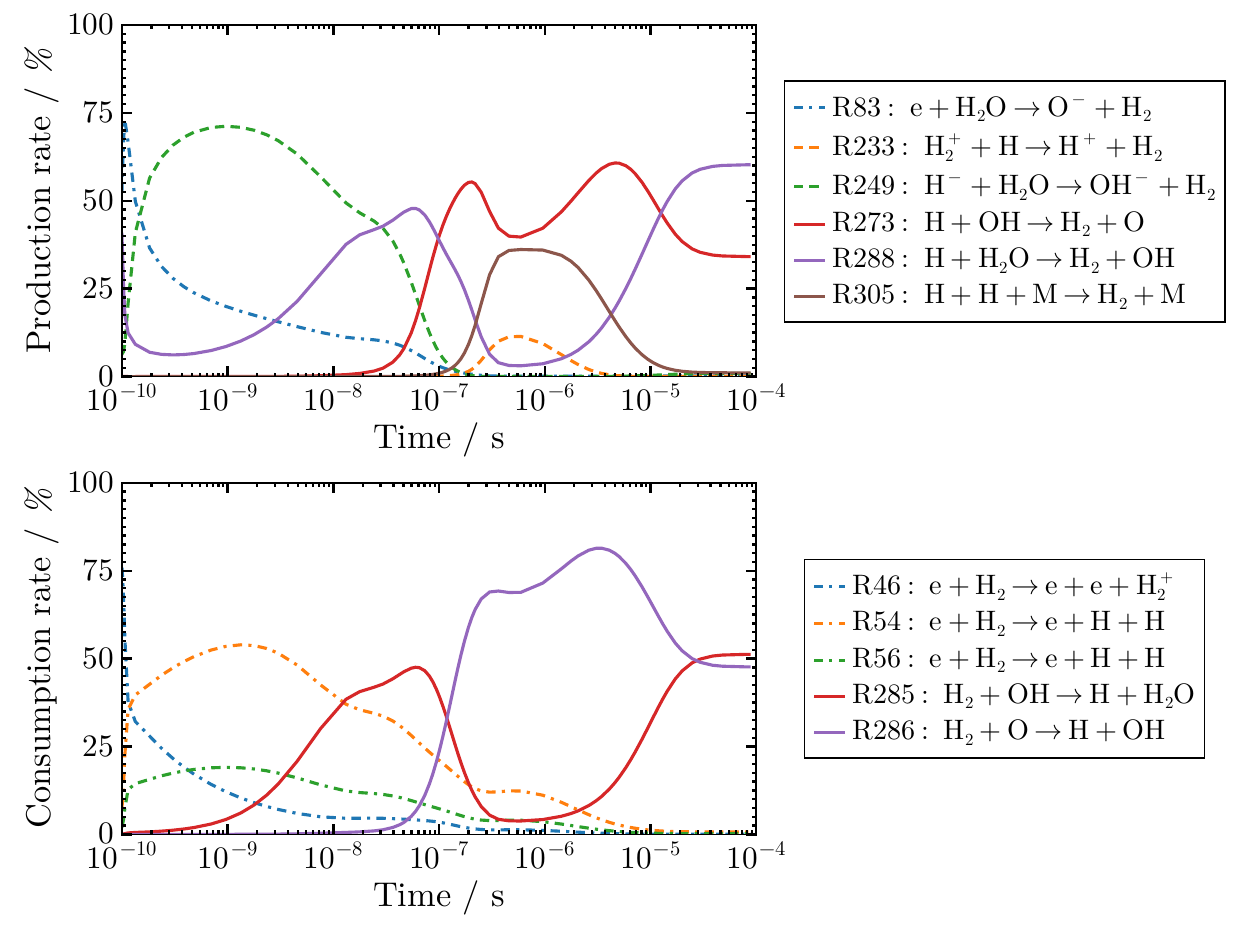}
    \caption{Production and consumption pathways of \ch{H2} for the base case simulation.}
    \label{fig:base-case-prod_cons_H2_film00_peak06_scaled}
\end{figure}
For the \ch{H}/\ch{H2} ratio, similar behavior to the \ch{O}/\ch{O2} ratio can be observed, with dissociation dominating at discharge onset, peaking at approximately \num{15000}:1 around \qty{e-10}{\second}.
However, unlike \ch{O2}, \ch{H2} benefits from direct production pathways from \ch{H2O}, such as \ch{H + H2O -> H2 + OH} (R288), \ch{e + H2O -> O- + H2} (R83) and \ch{H- + H2O -> OH- + H2} (R249), leading to higher \ch{H2} production.
These pathways are detailed in Fig.~\ref{fig:base-case-prod_cons_H2_film00_peak06_scaled}.
Here, the reactions (R249) and (R83) are responsible for around \qty{70}{\percent} of \ch{H2} production during the beginning and middle of the phase.
As the \ch{H} density grows (R288) gains importance, becoming a significant \ch{H2} production pathway by \qty{e-8}{\second}.
For \ch{H2} consumption, electron-impact dissociation (\ch{e + H2 -> e + H + H}, (R54), and to a lesser extent (R56)) and ionization (\ch{e + H2 -> e + e + H2+}, (R46)) dominate initially, but \ch{H2 + OH -> H + H2O} (R285) grows in relevance, contributing \qtyrange{40}{45}{\percent} by \qty{e-8}{\second}.
Complementing this, the production and consumption pathways for \ch{H} are illustrated in Fig.~\ref{fig:base-case-prod_cons_H_film00_peak06_scaled}.
The primary \ch{H} production pathway is electron-impact dissociation of \ch{H2O} via \ch{e + H2O -> e + H + OH} (R62), which dominates throughout this timeframe.
Reactions (R63) and (R65), which also play a non-negligible role, are two variants of the same reaction that would produce excited states of OH (R63) and H (R65).
The consumption of \ch{H} is mainly due to the reaction \ch{H + H2O -> H2 + OH} (R288).
\\

\begin{figure}
    \centering
    \includegraphics[width=\linewidth]{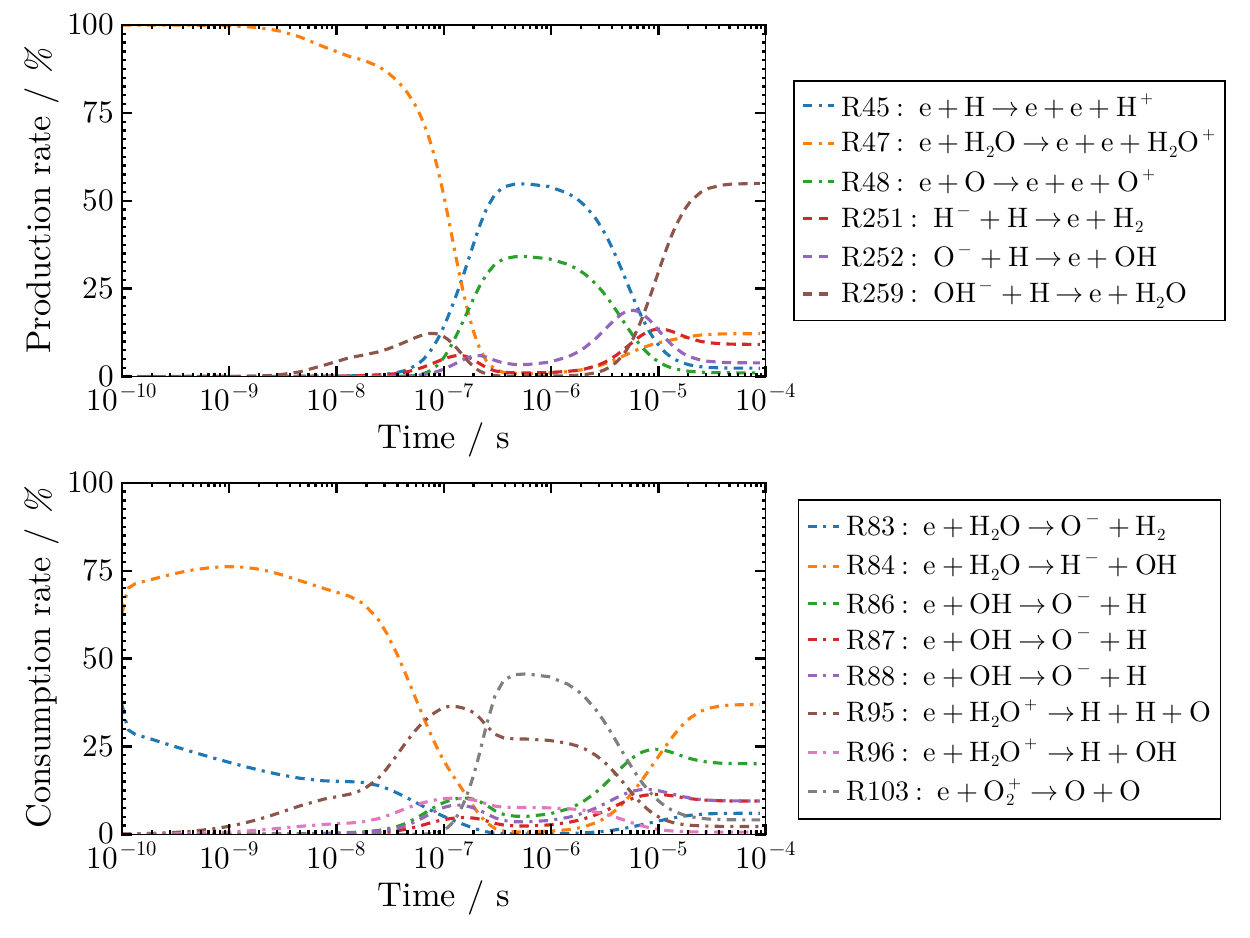}
    \caption{Production and consumption pathways of electrons for the base case simulation.}
    \label{fig:base-case-prod_cons_ne_film00_peak06_scaled}
\end{figure}

\begin{figure}
    \centering
    \includegraphics[width=\linewidth]{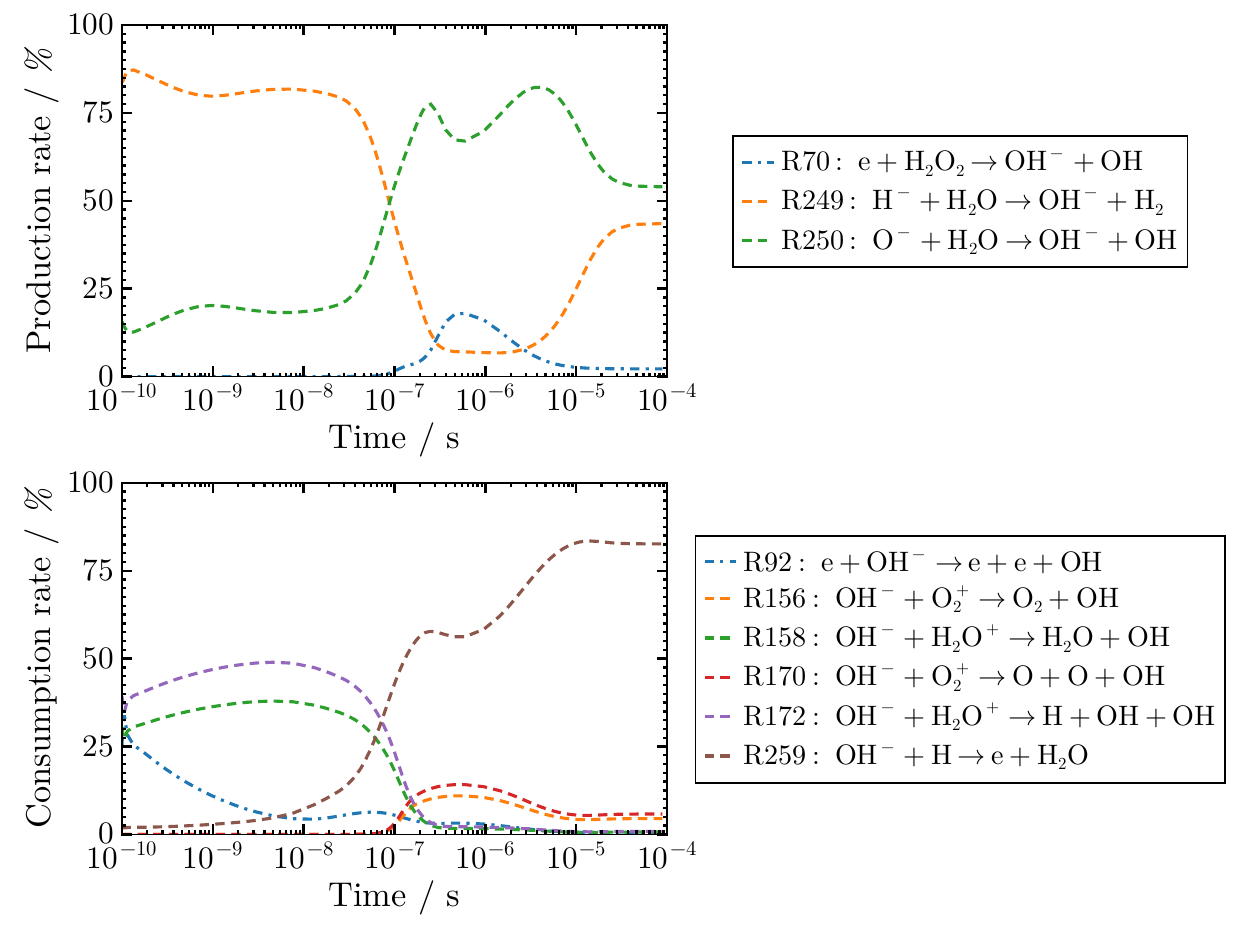}
    \caption{Production and consumption pathways of \ch{OH-} for the base case simulation.}
    \label{fig:base-case-prod_cons_OH-_film00_peak06_scaled}
\end{figure}

Next, the production and consumption pathways of charged particles will be considered.
Fig.~\ref{fig:base-case-prod_cons_ne_film00_peak06_scaled} shows the major production and consumption processes for electrons.
Here, the ionization of \ch{H2O} via (R47) is the major source of electron production during Phase I, generating \ch{H2O+}.
Electron consumption occurs primarily through dissociative attachment processes, such as \ch{e + H2O -> H- + OH} (R84) and \ch{e + H2O -> O- + H2} (R83), with (R84) accounting for up to \qty{75}{\percent} of the total electron consumption rate.
These reactions increase \ch{H-} and \ch{O-} densities to around \qty{3e18}{\per\cubic\metre}, which drive \ch{OH-} formation via \ch{H- + H2O -> OH- + H2} (R249) and \ch{O- + H2O -> OH- + OH} (R250).
The production and consumption processes of \ch{OH-} are presented in Fig.~\ref{fig:base-case-prod_cons_OH-_film00_peak06_scaled}.
Consequently, \ch{OH-} and electrons are the major negatively charged species, each reaching \qty{5e20}{\per\cubic\metre} by \qty{e-8}{\second}, as shown in Fig.~\ref{fig:base-case-negative_species_film00_peak06}, while \ch{H-} and \ch{O-} remain present at lower densities.
\\

\begin{figure}
    \centering
    \includegraphics[width=\linewidth]{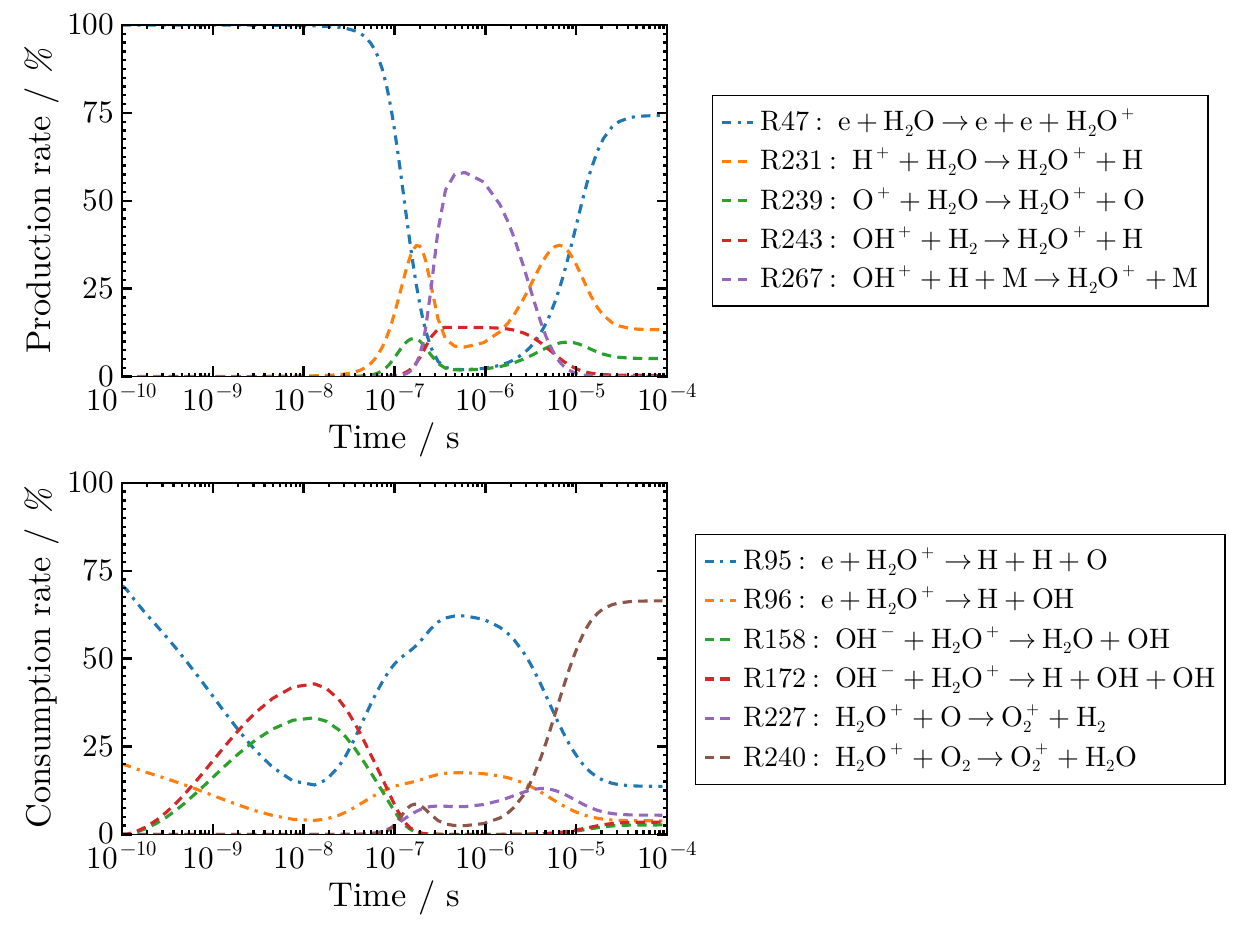}
    \caption{Production and consumption pathways of \ch{H2O+} for the base case simulation.}
    \label{fig:base-case-prod_cons_H2O+_film00_peak06_scaled}
\end{figure}

\begin{figure}
    \centering
    \includegraphics[width=\linewidth]{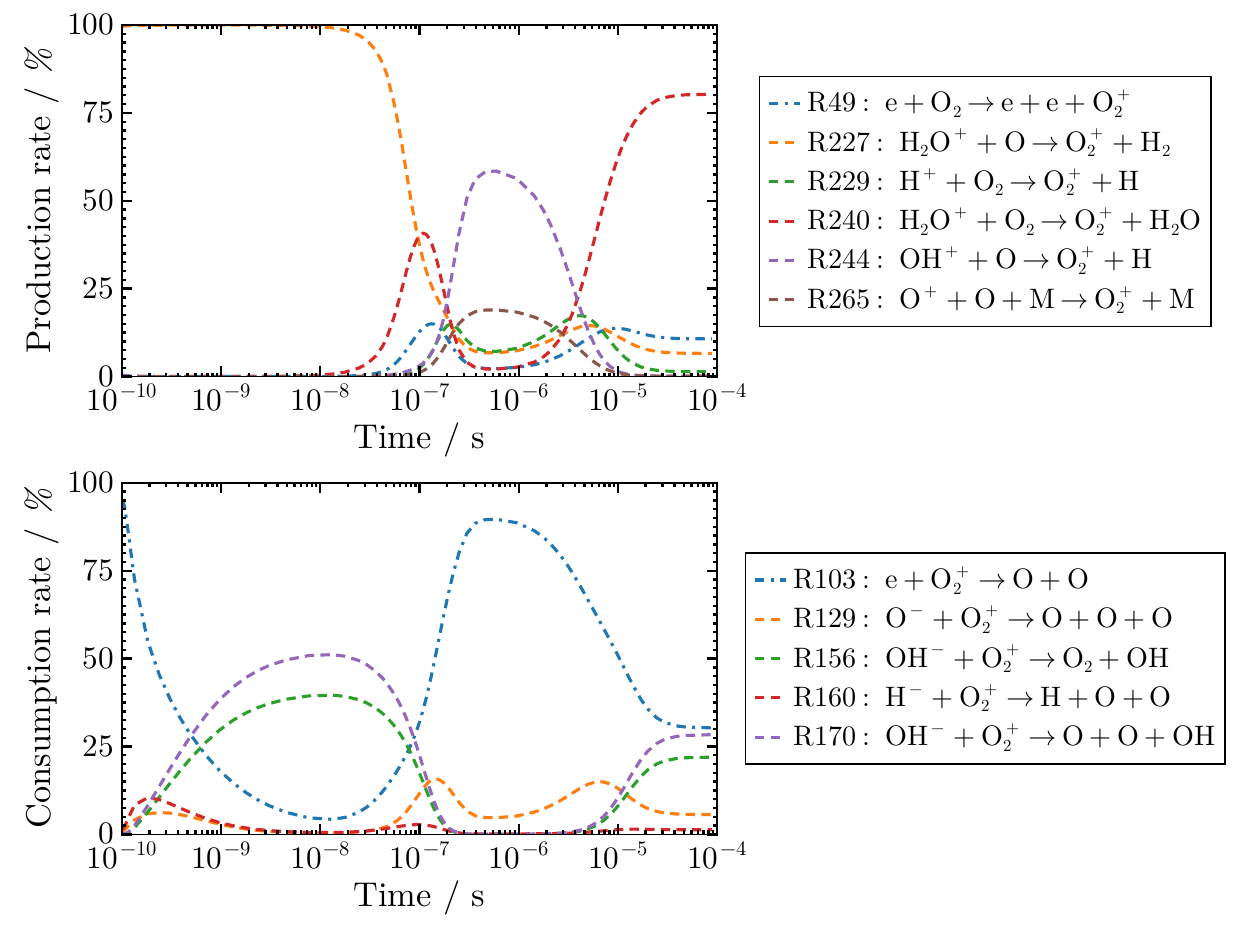}
    \caption{Production and consumption pathways of \ch{O2+} for the base case simulation.}
    \label{fig:base-case-prod_cons_O2+_film00_peak06_scaled}
\end{figure}

Lastly, the major production and consumption processes of the main positive ions, \ch{H2O+} and \ch{O2+}, are shown in Fig.~\ref{fig:base-case-prod_cons_H2O+_film00_peak06_scaled} and \ref{fig:base-case-prod_cons_O2+_film00_peak06_scaled}, respectively.
During Phase I, \ch{H2O+} is mainly produced via electron impact ionization (R47).
The consumption of \ch{H2O+} at early times is caused by a variety of electron-ion and ion-ion recombination reactions.
Electron-ion reactions dominate at the start of Phase I with collisions involving \ch{OH-} (R172) being particularly important starting around \qty{e-9}{\second}.
During Phase I, \ch{O2+} begins to form, mainly via charge exchange reactions of \ch{H2O+} with \ch{O} i.e. \ch{H2O+ + O -> O2+ + H2} (R227).
Similarly to \ch{H2O+}, consumption of \ch{O2+} at early times occurs via dissociative recombination with electrons, \ch{e + O2+ -> O + O} (R103), with ion-ion recombination reactions involving \ch{OH-} (R170) also playing an important role.
\\

Overall, the reaction pathways in Phase I are largely driven by electron impact collisions with \ch{H2O} to form its first dissociation (\ch{H}, \ch{OH}), ionization (\ch{H2O+}) and attachment (\ch{OH-}) products.
These products then undergo further electron impact collisions and participate in charge exchange, electron-ion and ion-ion recombination reactions, which define the gas composition at the start of Phase II.

\subsubsection{Phase II: power density peak}\label{subsubsec:base-case-power-dens-peak-phase}

During Phase II (\qtyrange{e-8}{e-5}{\second}), the power density reaches \qty{3.2e14}{\watt\per\cubic\metre}, significantly increasing dissociation and ionization of \ch{H2O}, and reducing its density by more than two orders of magnitude to \qty{1e23}{\per\cubic\metre} at its lowest, as shown in Fig.~\ref{fig:base-case-neutral_species_film00_peak06}.
Since the densities of electrons have substantially increased, electron impact collisions continue to be important during this phase.
These contribute significantly to the dissociation of all molecular species, as well as the formation of \ch{H2O+}, and also play a role in the detachment of electrons from \ch{OH-}.
In addition, the densities of highly reactive species such as \ch{H} and \ch{O} significantly increase during Phase II, and their reactions start to play a more important role in the overall chemistry.
\\

The high power input sustains a significant imbalance between atomic and molecular species.
For oxygen, the \ch{O}/\ch{O2} ratio peaks at approximately 35:1 around \qty{e-6}{\second}, driven by enhanced \ch{O2} dissociation.
The main \ch{O2} production pathway in this phase remains \ch{O + OH -> H + O2} (R294), as shown in Fig.~\ref{fig:base-case-prod_cons_O2_film00_peak06_scaled}, due to high \ch{OH} and \ch{O} densities, showing that the density of \ch{O2} is strongly dependent on that of \ch{OH}.
\ch{O2} consumption occurs via electron-impact dissociation (\ch{e + O2 -> e + O + O}, (R66, R67)), neutral-neutral reaction \ch{H + O2 -> O + OH} (R272), and charge exchange \ch{H2O+ + O2 -> O2+ + H2O} (R240), with absolute reaction rates rising by up to 8 orders of magnitude in this phase.
In contrast, Fig.~\ref{fig:base-case-prod_cons_O_film00_peak06_scaled} shows \ch{O} consumption occurs via \ch{O + OH -> H + O2} (R294), \ch{H2 + O -> H + OH} (R286) and charge exchange \ch{H+ + O -> O+ + H} (R228), replacing the \ch{H2O} dissociation reaction (R292) which dominated in Phase I.
At the same time, the production of \ch{O} shows a drastic shift across the second phase, with electron-impact reactions \ch{e + H2O+ -> H + H + O} (R95) and \ch{e + OH -> e + H + O} (R68) being mainly responsible at the beginning of the phase, a mix between neutral-neutral (R271, R273), charge exchange (R237) and electron-impact (R67, R68, R95, R103) reactions during the peak power, and neutral-neutral reactions (R271, R272, R273) winning out around the end of the phase.
\\

For hydrogen, the \ch{H}/\ch{H2} ratio stabilizes around 17:1 in Phase II.
Reaction rates for \ch{H2}, as presented in Fig.~\ref{fig:base-case-prod_cons_H2_film00_peak06_scaled}, show that \ch{H2} production is dominated by neutral-neutral dissociation of \ch{H} with \ch{OH} (R273) and \ch{H2O} (R288).
During the peak power timeframe, three-body recombination \ch{H + H + M -> H2 + M} (R305) shows an increasing contribution to \ch{H2} production (\qty{35}{\percent}) due the \ch{H} density reaching its maximum value and molecule densities being generally lower.
Additionally, charge exchange reactions like \ch{H2+ + H -> H+ + H2} (R233) become more important, due to the increase in positive ion densities.
Meanwhile, \ch{H2} consumption is primarily driven by \ch{H2 + O -> H + OH} (R286), which rapidly becomes dominant (\qtyrange{70}{80}{\percent}), while electron-impact dissociation (\ch{e + H2 -> e + H + H} (R54, R56)) and neutral-neutral reactions (\ch{H2 + OH -> H + H2O} (R285)) play minor roles.
\ch{H} production is sustained by electron-impact dissociation of \ch{H2O} (\ch{e + H2O -> e + H + OH} (R62)) at the beginning of Phase II, while neutral-neutral reactions \ch{H2 + O -> H + OH} (R286) and \ch{O + OH -> H + O2} (R294) play a major role for the remaining time.
Similar to the \ch{O} and \ch{H2} cases, during peak power deposition, charge exchange reactions show a non-negligible contribution to the production of \ch{H}.
\ch{H} consumption, shown in Fig.~\ref{fig:base-case-prod_cons_H_film00_peak06_scaled}, occurs primarily due to neutral-neutral reactions involving the other dominant neutral species \ch{H2O} (R288), \ch{OH} (R273) but also three-body recombination (\ch{H + H + M -> H2 + M} (R305)) and charge exchange \ch{O+ + H -> H+ + O} (R237)) reactions when atomic/molecular ratios and ion densities increase during the high power density peak at around \qty{e-6}{\second}.
\\

With molecules becoming increasingly dissociated by neutral-neutral reactions, as well as being ionized by charge exchange and, to a lesser extent, electron impact reactions, both atomic species densities and positive ion densities continue to increase, as shown in Fig.~\ref{fig:base-case-neutral_species_film00_peak06} and \ref{fig:base-case-positive_species_film00_peak06}.
Increasing ionization rates results in electron densities reaching their maximum value of around \qty{3e22}{\per\cubic\metre} (see Fig.~\ref{fig:base-case-ne_te_film00_peak06}) while the densities of the positive ions (see Fig.~\ref{fig:base-case-positive_species_film00_peak06}) \ch{H2+}, \ch{H3+}, \ch{O+}, \ch{OH+} and \ch{H+} grow to around \qtyrange{e20}{e21}{\per\cubic\metre}, and \ch{O2+} and \ch{H2O+} to around \qty{e22}{\per\cubic\metre}.
\ch{O2+} and \ch{H2O+} being the dominant positive ion species is consistent with their ionization potentials, with respect to the respective ground states (see Tab.~\ref{tab:ion-attributes}) being the lowest of the ions included in the reaction scheme at \qty{12.07}{\electronvolt} and \qty{12.61}{\electronvolt}, respectively.
This favours their production in charge exchange reactions.
At the time of the highest power density, the total ionization degree $\alpha \equiv n_{pos} / n_{total}$ of the discharge peaks at \qty{0.31}{\percent}.
Additionally, the previously growing negatively charged species are rapidly depleted, with the exception of \ch{O-}, as shown in Fig.~\ref{fig:base-case-negative_species_film00_peak06}, with \ch{OH-} showing a nearly four orders of magnitude reduction.
The negative charge deficit is accounted for by electron densities growing accordingly, balancing out the positive charge carriers.
The substantial depletion of \ch{OH-} is due to the multiple available pathways of recombination with \ch{H2O+} (e.g., (R158), (R172)) and associative detachment with \ch{H} in the form of reaction \ch{OH- + H -> e + H2O} (R259).
Since these consumption partners all have considerable densities during Phase II, consumption of \ch{OH-} occurs rapidly.
Fig.~\ref{fig:base-case-prod_cons_OH-_film00_peak06_scaled}, which illustrates the production and consumption reaction rates of \ch{OH-}, highlights the dominance of these consumption pathways during this phase.
Furthermore, the most significant production reactions for \ch{OH-} are reliant on charge transfer from other negative ions or on \ch{H2O2} in the reaction \ch{e + H2O2 -> OH- + OH} (R70), which are insufficient to offset the high consumption rates.
Meanwhile, the other negative ion species are supplied by attachment reactions involving \ch{H2O} and \ch{OH}, leading to their densities being proportionally less depleted, as their production mechanisms remain more active compared to those of \ch{OH-}.
\\

\subsubsection{Phase III: power density decline}\label{subsubsec:base-case-power-dens-decline-phase}

During the power density decline phase (\qtyrange{e-5}{e-4}{\second}), the reduced power input decreases dissociation and ionization rates, promoting the formation of stable molecular neutral species (\ch{H2O}, \ch{O2}, \ch{H2}) while atomic species (\ch{H}, \ch{O}) decline, though their concentrations remain significant.
\\

The shift toward molecule formation stabilizes the balance between atomic and molecular species.
For oxygen, the \ch{O}/\ch{O2} ratio stabilizes at 0.95:1, driven by \ch{O2} production via \ch{O + OH -> H + O2} (R294), presented in Fig.~\ref{fig:base-case-prod_cons_O2_film00_peak06_scaled}.
This contributes nearly \qty{100}{\percent} of \ch{O2} production, sustained by high densities of \ch{OH} and \ch{O}.
\ch{O2} consumption occurs primarily through two-body neutral-neutral reactions, such as \ch{H + O2 -> O + OH} (R272), with minor contributions (below \qty{5}{\percent}) from electron-impact dissociation (\ch{e + O2 -> e + O + O}, (R66), (R67)) and charge exchange \ch{H2O+ + O2 -> O2+ + H2O} (R240).
Atomic oxygen production similarly is driven by neutral-neutral reactions between radicals (e.g., \ch{OH + OH -> H2O + O} (R271), \ch{H + O2 -> O + OH} (R272) and \ch{H + OH -> H2 + O} (R273)), while electron dissociation pathways show a total contribution of below \qty{5}{\percent}.
\\

For hydrogen, the \ch{H}/\ch{H2} ratio stabilizes at 1.1:1, reflecting enhanced \ch{H2} formation, as shown in Fig.~\ref{fig:base-case-prod_cons_H_film00_peak06_scaled} and \ref{fig:base-case-prod_cons_H2_film00_peak06_scaled}.
\ch{H2} production is driven by two-body neutral-neutral reactions, including \ch{H + OH -> H2 + O} (R273) and \ch{H + H2O -> H2 + OH} (R288), with a small contribution of around \qty{1}{\percent} from three-body recombination \ch{H + H + M -> H2 + M} (R305).
\ch{H2} consumption occurs via \ch{H2 + OH -> H + H2O} (R285) and \ch{H2 + O -> H + OH} (R286, \qty{98}{\percent}), with electron-impact dissociation (\ch{e + H2 -> e + H + H}, R54, R56) contributing minimally (below \qty{2}{\percent}).
For \ch{H}, neutral-neutral reactions like \ch{H + O2 -> O + OH} (R272), \ch{H + OH -> H2 + O} (R273) and \ch{H + H2O -> H2 + OH} (R288) for consumption and \ch{H2 + OH -> H + H2O} (R285), \ch{H2 + O -> H + OH} (R286) and \ch{O + OH -> H + O2} (R294), for production, define the dominant pathways.
Electron- and ion-driven reactions only play minor roles, with \ch{e + H2O -> e + H + OH} (R62) being the most important of these reactions at just below \qty{5}{\percent}.
The \ch{OH} density (around \qty{4e23}{\per\cubic\metre}) together with \ch{H2O} (around \qty{2e24}{\per\cubic\metre}) contribute as secondary species in both the \ch{O}/\ch{O2} and \ch{H}/\ch{H2} balance.
For oxygen, \ch{OH} is critical for \ch{O2} production (R294, \qty{95}{\percent}) and \ch{O} production ((R271) and (R273), combined \qty{70}{\percent}), while also acting as a reactant in neutral-neutral collisions with \ch{O} (R294), in turn reducing the \ch{O} density.
\ch{OH} production is, in part, facilitated by reactions such as \ch{H2O + O -> OH + OH} (R292) and \ch{H + O2 -> O + OH} (R272), reinforcing its role in the oxygen cycle.
Similarly, \ch{H2O} takes part in \ch{O} consumption via reaction \ch{H2O + O -> OH + OH} (R292), acting as both a sink for \ch{O} species and a source for \ch{OH} species.
A similar behaviour can be seen for \ch{H} and \ch{H2}, where \ch{H2} production is facilitated by \ch{H2O} (R288) leading to \ch{OH} and \ch{H} production, while \ch{H2} consumption happens mainly due to \ch{OH} (R285) in turn producing \ch{H2O} and \ch{H}.
\\

The electron density stabilizes at \qty{7e20}{\per\cubic\metre} in Phase III, with the detachment reaction \ch{OH- + H -> e + H2O} (R259) contributing \qty{55}{\percent} to electron production, while dissociative attachment reactions with \ch{OH} (\ch{e + OH -> H- + O} (R86-R88)) and \ch{H2O} (\ch{e + H2O -> H- + OH} (R84)) dominate consumption.
The corresponding data is presented in Fig.~\ref{fig:base-case-prod_cons_ne_film00_peak06_scaled}.
Negative ion densities mostly follow the expected order stemming from their electron affinities, with \ch{O2-} being the exception, showing higher densities than \ch{H-} at \qty{8e18}{\per\cubic\metre} and \qty{3e18}{\per\cubic\metre}, respectively.
As for the electrons, positive ion densities decrease significantly due to the reduced power input, with \ch{H+}, \ch{O+}, \ch{OH+}, \ch{H2+}, and \ch{H3+} falling to \qty{4e17}{\per\cubic\metre} and below, while \ch{O2+} (\qty{7e20}{\per\cubic\metre}) and \ch{H2O+} (\qty{1e20}{\per\cubic\metre}) maintain relatively high densities, as shown in Fig.~\ref{fig:base-case-positive_species_film00_peak06}.
The total ionization degree drops to $\alpha = \qty{0.017}{\percent}$, reflecting the considerable reduction of ion densities in this phase.
\\

The base case simulation shows a shift in the gas phase chemistry across the three defined phases, with a clear change in the dominant reactions for each phase.
During Phase I, electron-driven processes, including ionization and dissociation of \ch{H2O}, enabled by the increasing power density, dominate, producing significant quantities of atomic species (\ch{H}, \ch{O}, \ch{OH}) and \ch{H2O+}.
Other reactions during this time are mostly driven by the abundance of \ch{H2O} and the growing densities of the reactive species.
In Phase II, in conjunction with peak power densities, the chemistry shifts towards a regime where both neutral and charged particle reactions are dominant.
Here, there are still significant roles for electron-driven reactions, charge exchange plays an important part in the interactions between neutrals and ions, and neutral reactions involving reactive species are gaining importance.
The latter are particularly relevant towards the end of Phase II and into Phase III, where they begin to supersede electron- and ion-driven reactions in most cases.
For Phase III, the reduced power input shifts the dominance of the neutral-neutral reactions towards those which produce more stable molecular species like \ch{H2O}, \ch{H2}, and \ch{O2}.
At this point, reactions driven by charged species are negligible, accounting for less than \qty{5}{\percent} for production and consumption reactions of the major neutral species.
\\

It can be seen that both electron-driven and neutral-neutral reactions shape the chemical evolution of the gas composition of the bubble, but with electron-driven reactions being mostly relevant for early stages of the bubble lifetime, while neutral-neutral reactions decide the final gas composition.
Many of the neutral reactions that are important are characterised by large temperature thresholds, such that their rate coefficients increase significantly with gas temperature.
For instance, \ch{H + H2O -> H2 + OH} (R288) and \ch{H2O + O -> OH + OH} (R292) are two of the most important reactions for \ch{H2O} consumption during Phase III.
These reactions have threshold temperatures close to \qty{10000}{\kelvin}, and are therefore strongly gas temperature dependent in the temperature range studied.
Taken together, both electron and high gas-temperature driven reactions have prominent roles in the base case microdischarge, with their contributions varying across the three phases.

\subsection{Gas temperature variation}\label{subsec:gas-temp-variation}

\begin{figure}
    \centering
    \includegraphics[width=\linewidth]{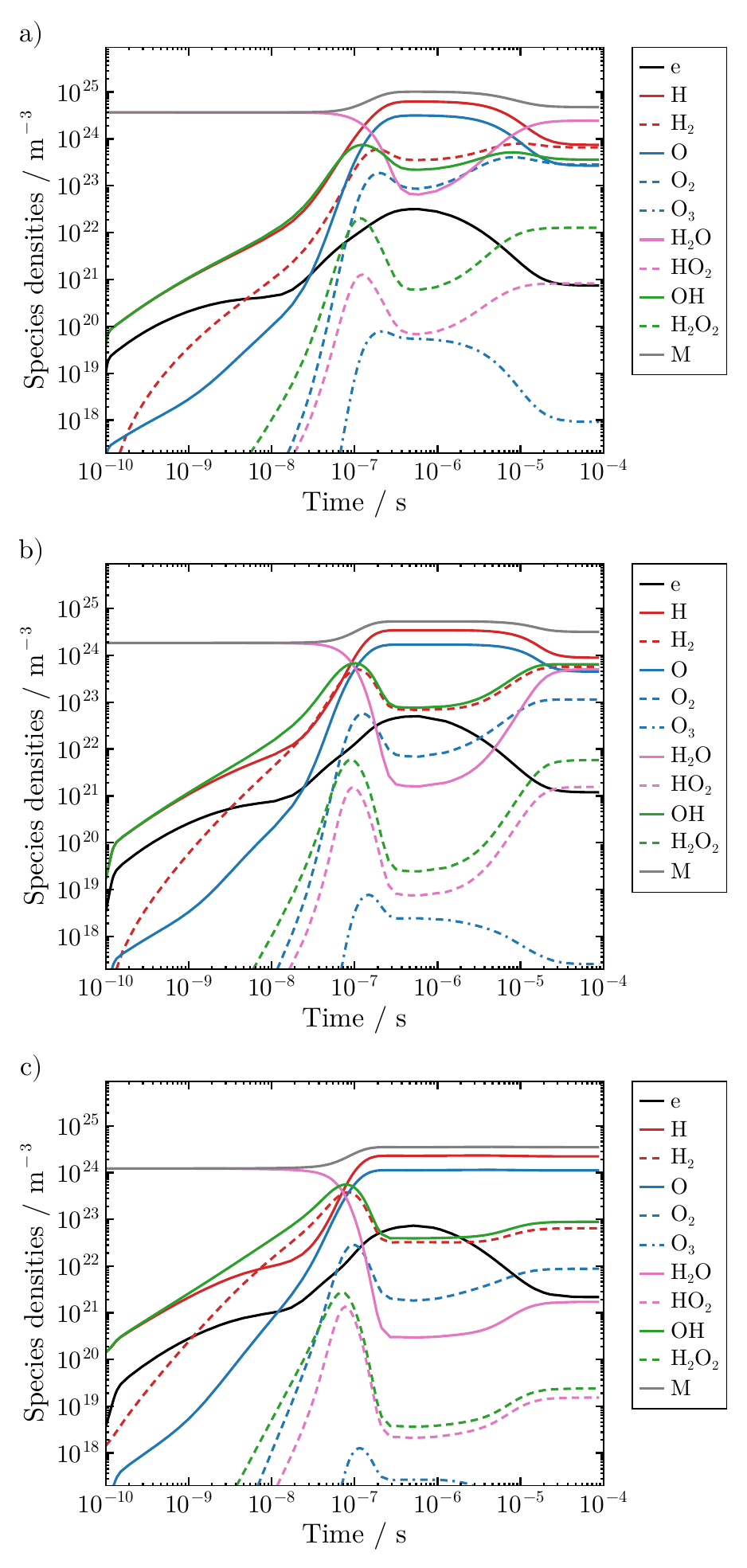}
    \caption{Neutral species densities at gas temperatures of a) \qty{2000}{\kelvin} i.e the base case simulation, b) \qty{4000}{\kelvin} and c) \qty{6000}{\kelvin}.}
     \label{fig:gas-temp-neutral_species}
\end{figure}

\begin{figure}
    \centering
    \includegraphics[width=\linewidth]{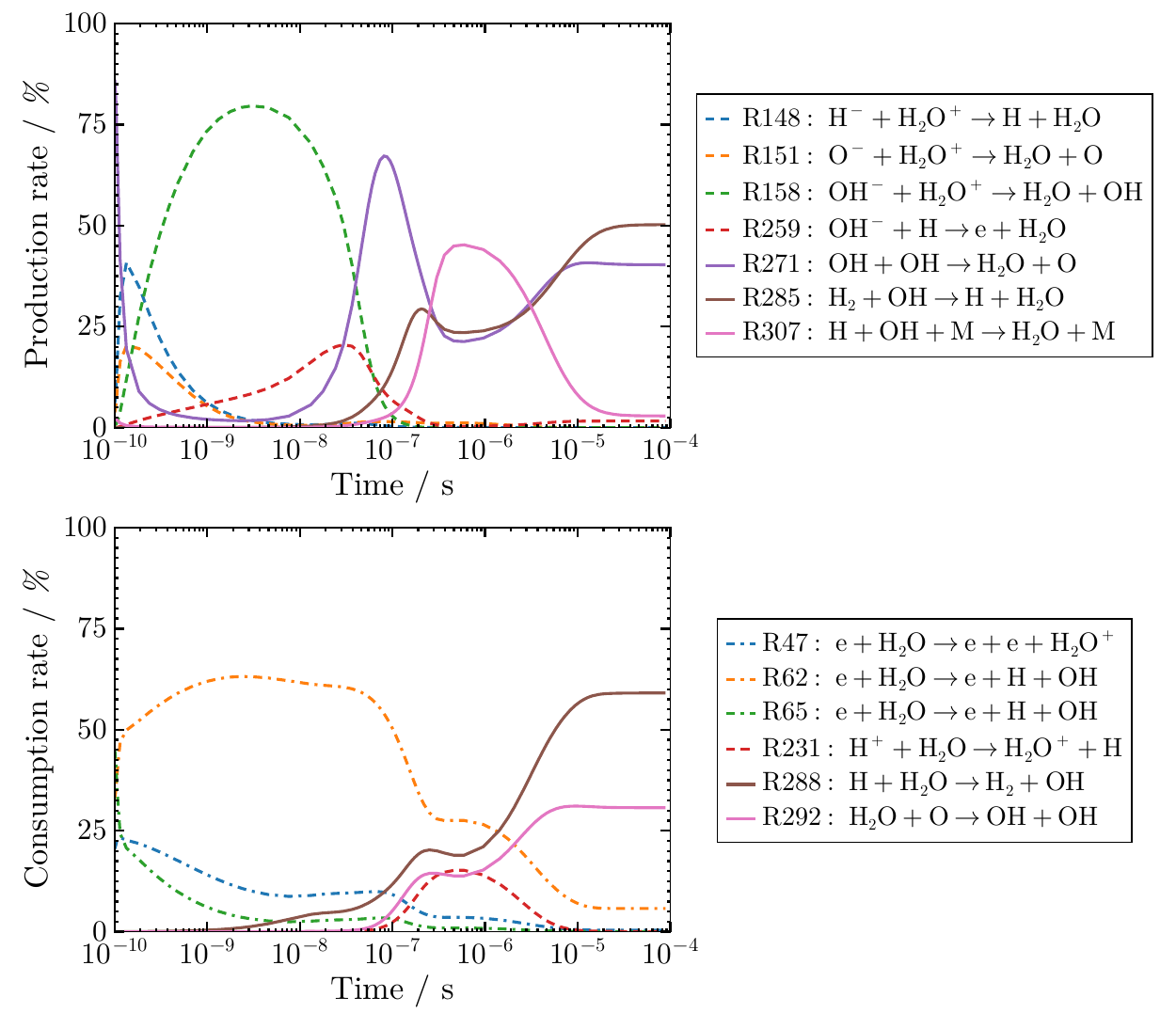}
    \caption{Production and consumption pathways of \ch{H2O} at a gas temperature of \qty{2000}{\kelvin} i.e. the base case simulation.}
    \label{fig:gas-temp-2000K-prod_cons_H2O_film00_peak06_scaled}
\end{figure}

\begin{figure}
    \centering
    \includegraphics[width=\linewidth]{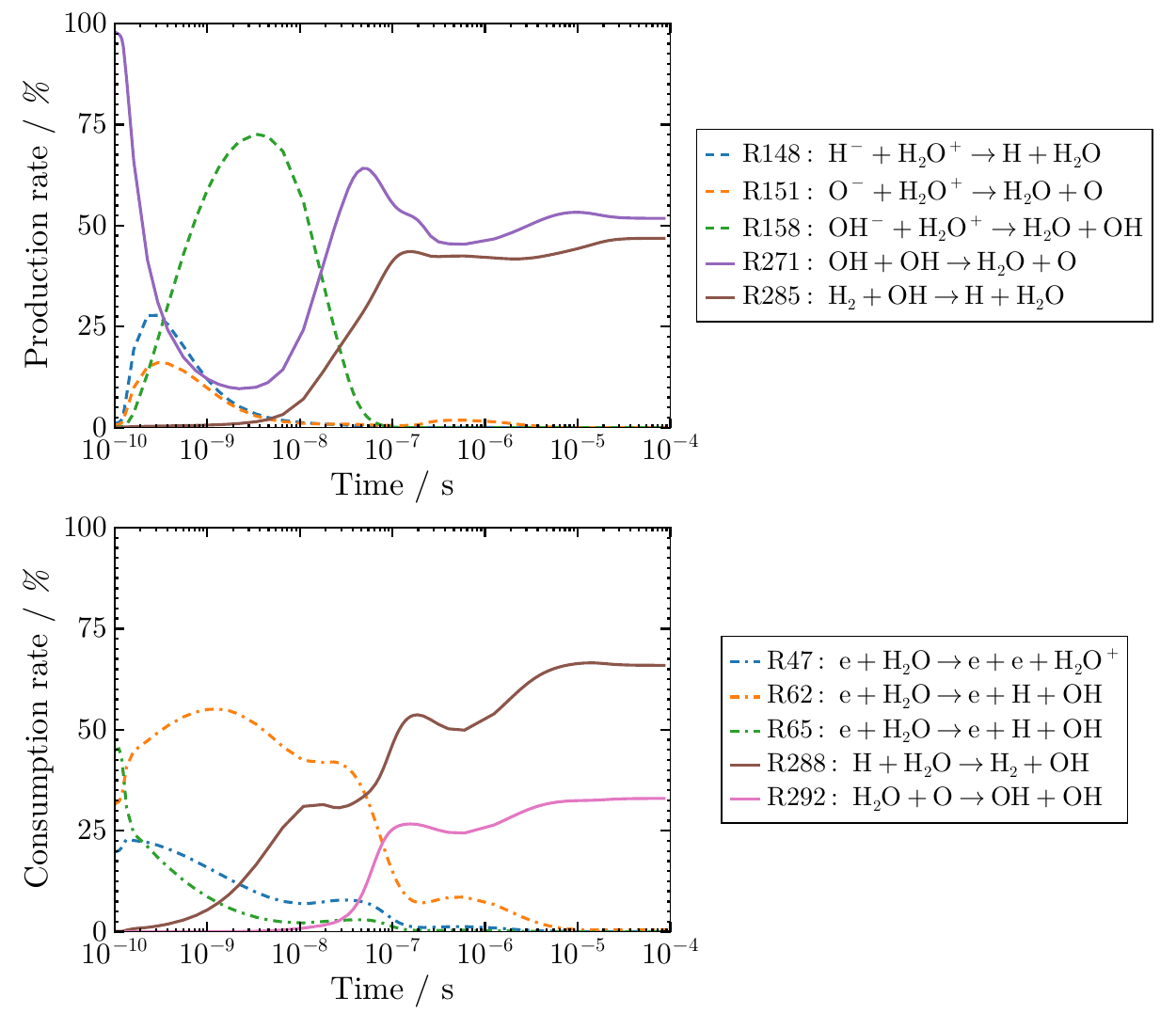}
    \caption{Production and consumption pathways of \ch{H2O} at a gas temperature of \qty{4000}{\kelvin}.}
    \label{fig:gas-temp-4000K-prod_cons_H2O_film00_peak06_scaled}
\end{figure}

\begin{figure}
    \centering
    \includegraphics[width=\linewidth]{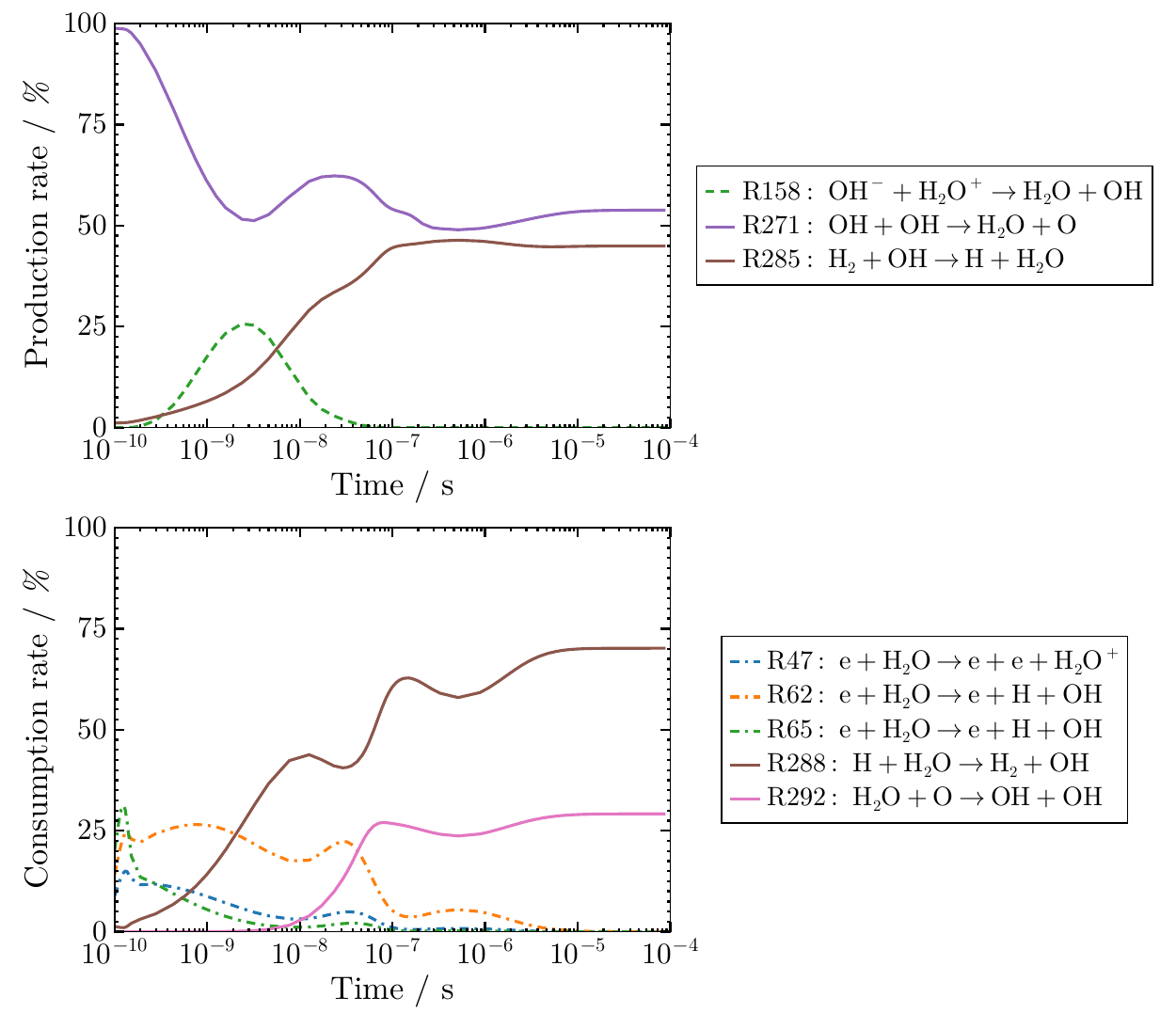}
    \caption{Production and consumption pathways of \ch{H2O} at a gas temperature of \qty{6000}{\kelvin}.}
    \label{fig:gas-temp-6000K-prod_cons_H2O_film00_peak06_scaled}
\end{figure}

To investigate the effects of varying gas temperature on reaction pathways and species densities, this section presents simulation results for the three power density phases introduced in Sec.~\ref{subsec:power-density} - Phase I (\qty{e-10}{\second} to \qty{e-8}{\second}), Phase II (\qty{e-8}{\second} to \qty{e-5}{\second}), and Phase III (\qty{e-5}{\second} to \qty{e-4}{\second}) - at gas temperatures of \qty{2000}{\kelvin}, \qty{4000}{\kelvin}, and \qty{6000}{\kelvin}.
These gas temperatures are chosen to cover the approximate span (\qty{2000}{\kelvin} and \qty{4000}{\kelvin}) of the experimental surface temperature measurements referenced in Sec.~\ref{sec:experiment}, as well as a higher temperature case where high-gas temperature chemistry may be expected to dominate the reaction pathways (\qty{6000}{\kelvin}).
The focus here is to analyze the differing impacts of thermal decomposition reactions (e.g., \ch{H2O + M -> H + OH + M} (R269)), electron impact reactions, and other neutral-neutral chemical reactions on \ch{H2O} reaction pathways at different gas temperatures, as well as the impact of the gas temperature on the evolution of the species densities themselves.
The evolution of the neutral species and electron densities are shown in Fig.~\ref{fig:gas-temp-neutral_species}a) ($T_{\mathrm{g}} = \qty{2000}{\kelvin}$), \ref{fig:gas-temp-neutral_species}b) ($T_{\mathrm{g}} = \qty{4000}{\kelvin}$) and \ref{fig:gas-temp-neutral_species}c) ($T_{\mathrm{g}} = \qty{6000}{\kelvin}$).
The corresponding production and consumption reactions for \ch{H2O} are shown in Fig.~\ref{fig:gas-temp-2000K-prod_cons_H2O_film00_peak06_scaled} for $T_{\mathrm{g}} = \qty{2000}{\kelvin}$, in Fig.~\ref{fig:gas-temp-4000K-prod_cons_H2O_film00_peak06_scaled} for $T_{\mathrm{g}} = \qty{4000}{\kelvin}$ and in Fig.~\ref{fig:gas-temp-6000K-prod_cons_H2O_film00_peak06_scaled} for $T_{\mathrm{g}} = \qty{6000}{\kelvin}$, to illustrate differences in some of the relevant reaction processes.
\\

As shown in Sec.~\ref{subsec:base-case-species-pathways}, during Phase I (\qty{e-10}{\second} to \qty{e-8}{\second}), the reaction dynamics are initially dominated by electron impact reactions at \qty{2000}{\kelvin}, particularly dissociation (\ch{e + H2O -> e + H + OH} (R62, R65)) and ionization (\ch{e + H2O -> e + e + H2O+} (R47)), which show significant contributions.
Thermal decomposition of \ch{H2O} is negligible at the lowest gas temperature (\qty{2000}{\kelvin}) as the large threshold temperature (almost \qty{50000}{\kelvin}) for this reaction means that its rate constant is low.
Other neutral-neutral reactions are also negligible during Phase I, as the densities of most neutral species remain too low for these reactions to play a role.
At \qty{4000}{\kelvin}, the higher gas temperature introduces a slight shift, with thermal decomposition (\ch{H2O + M -> H + OH + M} (R269)) contributing a small amount more to \ch{H2O} consumption than at \qty{2000}{\kelvin}.
This trend intensifies at \qty{6000}{\kelvin}, where thermal decomposition accelerates the production of dissociation products \ch{OH} and \ch{H}, influencing the reaction kinetics at early times.
\\

Higher densities of \ch{H} facilitate a larger contribution of \ch{H + H2O -> H2 + OH} (R288) towards \ch{H2O} dissociation as the gas temperature increases.
As noted above, this reaction is itself strongly gas temperature dependent, and so its rate constant also increases with gas temperature.
At \qty{e-8}{\second}, its contribution rises from \qty{5}{\percent} at \qty{2000}{\kelvin} to \qty{30}{\percent} at \qty{4000}{\kelvin} and \qty{45}{\percent} at \qty{6000}{\kelvin}.
Likewise, the reaction \ch{H2O + O -> OH + OH} (R292), the rate constant for which also scales with gas temperature, exhibits an earlier rise in its reaction rate, though its overall influence remains limited during this phase of the simulation.
This behavior is also tied to the \ch{O} density, which builds up through the dissociation of \ch{OH}, a process partially initiated by reaction (R269).
The contribution of (R269) to \ch{H2O} consumption increases with temperature, having a negligible contribution at \qty{2000}{\kelvin}, reaching a few percent at \qty{4000}{\kelvin}, and peaking at near \qty{50}{\percent} at the start of Phase I, before declining to \qty{25}{\percent} at \qty{e-8}{\second} for \qty{6000}{\kelvin}.
The increasing rate and contribution of this reaction at higher temperatures facilitates an earlier initiation of \ch{H2O} consumption via (R292).
\\

For electron impact reactions with \ch{H2O} at \qty{4000}{\kelvin} and \qty{6000}{\kelvin}, absolute reaction rates show only slight reductions compared to \qty{2000}{\kelvin}.
This can be explained by the initial \ch{H2O} density, calculated via the ideal gas law, decreasing with higher gas temperatures, leading to marginally lower rates for these reactions.
\\

In Phase II (\qty{1e-8}{\second} to \qty{1e-5}{\second}), electron impact reactions follow a similar trend to Phase I, decreasing in importance as neutral-neutral reactions, particularly \ch{H + H2O -> H2 + OH} (R288) and \ch{H2O + O -> OH + OH} (R292), become more significant.
This shift grows more pronounced with increasing gas temperature.
The absolute reaction rates for electron impact dissociation reactions drastically increase during this phase but then drop significantly, with the decline intensifying at higher temperatures.
For a gas temperature of \qty{2000}{\kelvin}, electron impact dissociation contributes around \qty{25}{\percent} to the dissociation of \ch{H2O} at \qty{1e-7}{\second}.
However, at \qty{4000}{\kelvin} this contribution decreases to below \qty{10}{\percent} at \qty{1e-7}{\second}, and at \qty{6000}{\kelvin} it only contributes around \qty{5}{\percent} at the same time point.
Overall, the consumption of \ch{H2O} during Phase II increases significantly, with the \ch{H2O} density decreasing by two orders of magnitude during this phase at a gas temperature of \qty{2000}{\kelvin}, three orders at \qty{4000}{\kelvin}, and three and a half orders at \qty{6000}{\kelvin}.
This shift in densities occurs at the time the power density input reaches its maximum, heavily increasing positive ion and electron densities, with ionization degrees of \qty{0.31}{\percent} at \qty{2000}{\kelvin}, \qty{0.95}{\percent} at \qty{4000}{\kelvin}, and \qty{2.1}{\percent} at \qty{6000}{\kelvin}.
These conditions favor reactive species production like \ch{OH}, which in turn shows a total contribution to the production of \ch{H2O} via reactions (R271) and (R285) of nearly \qty{100}{\percent} at \qty{6000}{\kelvin}.
It is notable that these reactions have comparably low temperature thresholds in comparison to other neutral-neutral reactions in the scheme at around \qty{200}{\kelvin} and \qty{2000}{\kelvin}, respectively.
\\

While the thermal decomposition reaction \ch{H2O + M -> H + OH + M} (R269) played an important role during the earlier Phase I, its contribution during Phase II diminishes, showing only negligible contributions even at \qty{6000}{\kelvin}.
\\

The consumption contributions lost by electron impact reactions are increasingly filled by collisions with H and O via (R288) and (R292), which drive \ch{H2O} consumption more effectively at higher temperatures.
This dominance is primarily due to the exponential increase in their rate coefficients with increasing temperature, due to their high activation energies (\qty{9270}{\kelvin} for (R288), \qty{8600}{\kelvin} for (R292)), outpacing the slower growth or decline of other pathways.
Meanwhile, reactions that produce \ch{H2O}, such as \ch{H + OH + M -> H2O + M} (R307), \ch{OH + OH -> H2O + O} (R271), and \ch{H2 + OH -> H + H2O} (R285), show decreasing contributions with increasing temperature.
This is driven by the inverse temperature dependence of the rate coefficient of (R307) ($k \propto T^{-2}$) and the increasing importance of charged-particle driven reactions for the formation of \ch{H2O} due to larger charged particle densities.
\\

In Phase III (\qty{1e-5}{\second} to \qty{1e-4}{\second}), the reaction dynamics exhibit distinct trends compared to Phases I and II.
Relative contributions for \ch{H2O} consumption and production rates remain largely consistent across different gas temperatures.
However, absolute reaction rates (not shown) do show significant temperature-dependent variations.
Rates of electron impact dissociation and ionization reactions decrease at higher gas temperatures, reflecting the reduced \ch{H2O} density.
Reactions \ch{H + H2O -> H2 + OH} (R288) and \ch{H2O + O -> OH + OH} (R292) are affected in a more complex way.
Despite the decreasing \ch{H2O} density, which should lower their reaction rates, the increased \ch{H} and \ch{O} densities at \qty{4000}{\kelvin} initially balance this effect, maintaining the rates.
However, \ch{H} and \ch{O} densities reach their maximum possible concentration in the gas phase due to the limited availability of precursor species (e.g., \ch{H2O}) for dissociation.
For instance, the dominant production reaction for \ch{H}, \ch{H2 + O -> H + OH} (R286), shows a rate of \qty{2e30}{\per\cubic\metre\per\second} at \qty{2000}{\kelvin}, rising to \qty{4e31}{\per\cubic\metre\per\second} at \qty{4000}{\kelvin} due to increased \ch{O} density balancing the drop in \ch{H2} density, and staying at a similar rate of \qty{4e31}{\per\cubic\metre\per\second} at \qty{6000}{\kelvin} as the \ch{H2} density continues to decline.
Another key \ch{H} production reaction, \ch{H2 + OH -> H + H2O} (R285), highlights the impact of decreasing molecular densities more starkly, as it depends on both \ch{H2} and \ch{OH}.
This reaction's contribution to total \ch{H} production decreases significantly across temperatures, from \qty{40}{\percent} at \qty{2000}{\kelvin} to \qty{25}{\percent} at \qty{4000}{\kelvin}, and below \qty{1}{\percent} at \qty{6000}{\kelvin}, driven by the combined reduction in \ch{H2} and \ch{OH} densities.
Consequently, at \qty{6000}{\kelvin}, the additional decrease in \ch{H2O} density (4.5 orders of magnitude compared to 4 orders at \qty{4000}{\kelvin}, as noted in Phase II) cannot be offset, resulting in lower reaction rates for (R288) and (R292) by approximately one order of magnitude.
\\

As observed in Phase II, \ch{H + OH + M -> H2O + M} (R307) continues its declining trend, becoming negligible at \qty{6000}{\kelvin} due to its inverse temperature dependence ($k \propto T^{-2}$).
Reactions \ch{OH + OH -> H2O + O} (R271) and \ch{H2 + OH -> H + H2O} (R285) show comparable relative contributions to \ch{H2O} production across different gas temperatures, mirroring the consistency seen in consumption reactions.
However, their absolute reaction rates reflect the temperature-dependent trends driven by reactant densities.
High \ch{H2} and \ch{OH} densities at \qty{2000}{\kelvin} result in elevated rates for both reactions.
Moving to \qty{4000}{\kelvin}, these rates remain within the same order of magnitude due to sustained \ch{H2} and \ch{OH} densities.
By \qty{6000}{\kelvin}, however, \ch{H2} and \ch{OH} densities drop by around one order of magnitude, causing the reaction rates to decrease by approximately one and a half orders of magnitude, from \qty{1e31}{\per\cubic\metre\per\second} to \qty{5e29}{\per\cubic\metre\per\second}.
Following the trend of Phase II, the thermal decomposition reaction \ch{H2O + M -> H + OH + M} (R269) is much less significant than neutral-neutral reactions involving radical species \ch{H} and \ch{O}.
Similarly, decreasing negative ion densities reduce \ch{H2O} production rates via the detachment reaction \ch{OH^- + H -> e + H2O} (R259), with absolute rates dropping by two and a half orders from the \qty{4000}{\kelvin} to \qty{6000}{\kelvin} case.
\\

The investigation of gas temperature effects across Phases I (\qty{1e-10}{\second} to \qty{1e-8}{\second}), II (\qty{1e-8}{\second} to \qty{1e-5}{\second}), and III (\qty{1e-5}{\second} to \qty{1e-4}{\second}) reveals a progressive shift from electron impact and thermal decomposition reactions to neutral-neutral reactions involving reactive species, with increasing temperature amplifying this transition.
At higher temperatures, notably \qty{6000}{\kelvin}, absolute reaction rates for dissociation, ionization, and production decline due to significant drops in molecular densities (e.g., \ch{H2O} by 4.5 orders), while reactions like (R288) and (R292) remain important through elevated radical densities and their highly temperature dependent rate coefficients, while being limited by reactant availability.
Overall, the pathways towards the densities of each neutral and charged particle in the microdischarges studied in this work are highly complex, depending on a range of reaction types, with electron and high gas temperature reactions both playing important roles.

%% file: 06_conclusion.tex
\section{Conclusion}\label{sec:conclusion}

In this work, the plasma chemistry of microdischarges formed in gas bubbles immersed in a liquid environment has been studied using a 0-D plasma chemical kinetics model. 
Microdischarges formed during plasma electrolytic oxidation of aluminium are used as a test case. 
Complementary experimental measurements have been performed, both to inform the power density, gas temperature, and gas pressure used as inputs to the model, as well as to validate simulated electron densities. 
In order to simulate the complex chemistry occurring in these discharges, a new reaction scheme is developed for microdischarges in water vapour, inclusive of the reaction kinetics expected to occur at both high electron densities, and high gas temperatures.
\\

In the experiment, the power density is found to vary strongly in time, peaking close to the start of the gas bubble expansion. 
Electron densities are measured in the experiment using the Stark broadening of the H$\alpha$ emission line. 
Fitting of the broadening is found to be improved by assuming two electron densities, which appears to be related to the varying electron density profile occurring during and between individual microdischarges. 
The absolute values of the simulated electron density profiles appear to be broadly consistent with these experimental measurements. 
Simulations also show that there is a strong variation of electron density during an individual microdischarge event, with electron densities following the profile of the power density. 
Experimentally measured power density profiles are found to be highly variable, and when used as input to the simulations also lead to significant variations in electron density values. 
These comparisons illustrate the challenges involved in comparing experiment and simulation for such a complex and stochastic experimental system.
\\

Having established a baseline comparison between experiment and simulation, simulations are then used to study the temporal profiles of neutrals and charged particles formed during a microdischarge event. 
For the base case, where a comparatively low gas temperature of \qty{2000}{\kelvin} is assumed, water vapour is found to be largely dissociated during the peak of the power density, with \ch{H} and \ch{O} forming the main constituents of the gas during this period. 
After the peak of the power density, much of these reactive species recombine to form \ch{H2O}, \ch{H2}, and \ch{O2} before the collapse of the gas bubble. 
At its peak, the ionization fraction in the microdischarge is found to be around \qty{0.31}{\percent}, while the electronegativity is generally low for most of the microdischarge lifetime, except at early stages. 
The main production and consumption reactions of the major charged and neutral species have been discussed, split into three distinct phases of the microdischarge, motivated by the different regimes of power density during the microdischarge lifetime. 
During the first two phases, electron impact reactions are found to be strong drivers of the overall chemical kinetics, leading to ionization and dissociation processes. 
In the second and third phases of the microdischarge, when electron impact dissociation processes have formed large densities of reactive species, high gas temperature reactions involving these species begin to contribute significantly towards the overall chemistry. 
For example, significant fractions of the dissociation of \ch{H2O} begins to occur via collisions with \ch{H} and \ch{O}, which have threshold temperatures of almost \qty{10000}{\kelvin}. 
Therefore, both electron- and gas-temperature driven chemistry are significant contributors to the chemical kinetics even at the lowest temperature studied.
\\

Lastly, the influence of increasing gas temperature on the chemical kinetics is investigated. 
As gas temperature increases to \qty{4000}{\kelvin} and \qty{6000}{\kelvin}, the degree of dissociation during the peak of the power density increases as well. 
Particularly at \qty{6000}{\kelvin}, the role of thermal dissociation of water vapour in collisions with non-reactive neutrals, with a threshold temperature of over \qty{50000}{\kelvin}, becomes significant. 
However, electron-driven processes still remain important, and dissociation of water in collisions with \ch{H} and \ch{O} still dominates its consumption after the power density peak, due to the large densities of those species present at that time.
\\

The aim of this work was to provide new insights into the gas-phase plasma chemistry occurring in microdischarges in bubbles at high temperatures. 
It has been demonstrated that an experimentally informed model of these systems can lead to electron densities in reasonable agreement with experimental measurements. 
From this, the model has been used to demonstrate the relative importance of electron- and gas temperature driven reactions in such discharges. 
While this study has focused on microdischarges in plasma electrolytic oxidation as a test case, the results and the plasma-chemical reaction scheme developed should be transferable to other bubble-based microdischarge systems, such as contact glow discharge electrolysis, for example.

%% file: 07_addendum.tex
\section{Appendix}\label{sec:appendix}

\begin{table*}[h!]
\footnotesize
\setlength{\tabcolsep}{8pt}
\caption{$T_\mathrm{e,eV}$: electron temperature in units of eV, $T_\mathrm{{e,K}}$: electron temperature in units of Kelvin. An energy threshold ($E_{\mathrm{thr}}$) given as $T_\mathrm{{e,eV}}$ means the threshold is taken to be the current value of $T_{e,eV}$. An $E_{\mathrm{thr}}$ of 0 does not imply that the process does not have a threshold, but instead that it is not included in the electron energy balance here. Rate coefficients are given in units of $\mathrm{m}^{3}\mathrm{s}^{-1}$ or $\mathrm{m}^{6}\mathrm{s}^{-1}$ depending on the order of the reaction.
A rate coefficient given as $f\left(T_\mathrm{{e,eV}}\right)$ is derived from cross section data. As described in section \ref{sec:simulation-framework}, excited states are not tracked in model. For clarity, reactions involving excited states are listed, along with how they are implemented here. Excited state notation follows the original source as far as possible.}
\begin{tabular}{p{1.5cm}p{5cm}p{1.5cm}p{5cm}p{1.5cm}}
no. & reaction                                                                     & $E_{\mathrm{thr}}$ / eV & rate coefficient                                                                      & reference                                     \\
\hline\multicolumn{5}{c}{\textbf{Elastic scattering and momentum transfer}} \\ \hline 
R1&$\mathrm{e} + \mathrm{H} \rightarrow \mathrm{e} + \mathrm{H}$&$0.0$ & $f(T_\mathrm{e,eV})$ & \ \cite{marques.2007,lxcat-istlisbon}\\ 
R2&$\mathrm{e} + \mathrm{H_2} \rightarrow \mathrm{e} + \mathrm{H_2}$&$0.0$ & $f(T_\mathrm{e,eV})$ & \ \cite{marques.2007,lxcat-istlisbon}\\ 
R3&$\mathrm{e} + \mathrm{H_{2}O} \rightarrow \mathrm{e} + \mathrm{H_{2}O}$&$0.0$ & $f(T_\mathrm{e,eV})$ & \ \cite{budde.2022}\\ 
R4&$\mathrm{e} + \mathrm{O} \rightarrow \mathrm{e} + \mathrm{O}$&$0.0$ & $f(T_\mathrm{e,eV})$ & \ \cite{alves.2016,lxcat-istlisbon-o}\\ 
R5&$\mathrm{e} + \mathrm{O_2} \rightarrow \mathrm{e} + \mathrm{O_2}$&$0.0$ & $f(T_\mathrm{e,eV})$ & \ \cite{itikawa.2009,lxcat-itikawa}\\ 
R6&$\mathrm{e} + \mathrm{OH} \rightarrow \mathrm{e} + \mathrm{OH}$&$0.0$ & $f(T_\mathrm{e,eV})$ & \ \cite{chakrabarti.2019}\\ 
\\ \multicolumn{5}{c}{\textbf{Electron impact excitation - electronic}} \\ \hline 
R7&$\mathrm{e} + \mathrm{H} \rightarrow \mathrm{e} + \mathrm{H(2p)}$ & $10.2043$ & $f(T_\mathrm{e,eV})$ & \ \cite{marques.2007,lxcat-istlisbon}\\ 
used as & $\mathrm{e} + \mathrm{H} \rightarrow \mathrm{e} + \mathrm{H}$& &  & \\[0.1cm] 
R8&$\mathrm{e} + \mathrm{H} \rightarrow \mathrm{e} + \mathrm{H(2s)}$ & $10.2043$ & $f(T_\mathrm{e,eV})$ & \ \cite{marques.2007,lxcat-istlisbon}\\ 
used as & $\mathrm{e} + \mathrm{H} \rightarrow \mathrm{e} + \mathrm{H}$& &  & \\[0.1cm] 
R9&$\mathrm{e} + \mathrm{H} \rightarrow \mathrm{e} + \mathrm{H(n{=}3)}$ & $12.0939$ & $f(T_\mathrm{e,eV})$ & \ \cite{marques.2007,lxcat-istlisbon}\\ 
used as & $\mathrm{e} + \mathrm{H} \rightarrow \mathrm{e} + \mathrm{H}$& &  & \\[0.1cm] 
R10&$\mathrm{e} + \mathrm{H} \rightarrow \mathrm{e} + \mathrm{H(n{=}4)}$ & $12.7553$ & $f(T_\mathrm{e,eV})$ & \ \cite{marques.2007,lxcat-istlisbon}\\ 
used as & $\mathrm{e} + \mathrm{H} \rightarrow \mathrm{e} + \mathrm{H}$& &  & \\[0.1cm] 
R11&$\mathrm{e} + \mathrm{H} \rightarrow \mathrm{e} + \mathrm{H(n{=}5)}$ & $13.0615$ & $f(T_\mathrm{e,eV})$ & \ \cite{marques.2007,lxcat-istlisbon}\\ 
used as & $\mathrm{e} + \mathrm{H} \rightarrow \mathrm{e} + \mathrm{H}$& &  & \\[0.1cm] 
R12&$\mathrm{e} + \mathrm{H_2} \rightarrow \mathrm{e} + \mathrm{H_2(B^1\Sigma^{+}_{u})}$ & $11.4$ & $f(T_\mathrm{e,eV})$ & \ \cite{marques.2007,lxcat-istlisbon}\\ 
used as & $\mathrm{e} + \mathrm{H_2} \rightarrow \mathrm{e} + \mathrm{H_2}$& &  & \\[0.1cm] 
R13&$\mathrm{e} + \mathrm{H_2} \rightarrow \mathrm{e} + \mathrm{H_2(C^1\Pi_{u})}$ & $12.4$ & $f(T_\mathrm{e,eV})$ & \ \cite{marques.2007,lxcat-istlisbon}\\ 
used as & $\mathrm{e} + \mathrm{H_2} \rightarrow \mathrm{e} + \mathrm{H_2}$& &  & \\[0.1cm] 
R14&$\mathrm{e} + \mathrm{H_2} \rightarrow \mathrm{e} + \mathrm{H_2(E^{1}\Sigma^{+}_{g}, F^1\Sigma^{+}_{g})}$ & $12.4$ & $f(T_\mathrm{e,eV})$ & \ \cite{marques.2007,lxcat-istlisbon}\\ 
used as & $\mathrm{e} + \mathrm{H_2} \rightarrow \mathrm{e} + \mathrm{H_2}$& &  & \\[0.1cm] 
R15&$\mathrm{e} + \mathrm{H_2} \rightarrow \mathrm{e} + \mathrm{H_2(e^3\Sigma^{+}_{u})}$ & $13.0$ & $f(T_\mathrm{e,eV})$ & \ \cite{marques.2007,lxcat-istlisbon}\\ 
used as & $\mathrm{e} + \mathrm{H_2} \rightarrow \mathrm{e} + \mathrm{H_2}$& &  & \\[0.1cm] 
R16&$\mathrm{e} + \mathrm{H_2} \rightarrow \mathrm{e} + \mathrm{H_2(B'^{1}\Sigma^{+}_{u})}$ & $13.8$ & $f(T_\mathrm{e,eV})$ & \ \cite{marques.2007,lxcat-istlisbon}\\ 
used as & $\mathrm{e} + \mathrm{H_2} \rightarrow \mathrm{e} + \mathrm{H_2}$& &  & \\[0.1cm] 
R17&$\mathrm{e} + \mathrm{H_2} \rightarrow \mathrm{e} + \mathrm{H_2(D^1\Pi_{u})}$ & $14.0$ & $f(T_\mathrm{e,eV})$ & \ \cite{marques.2007,lxcat-istlisbon}\\ 
used as & $\mathrm{e} + \mathrm{H_2} \rightarrow \mathrm{e} + \mathrm{H_2}$& &  & \\[0.1cm] 
R18&$\mathrm{e} + \mathrm{H_2} \rightarrow \mathrm{e} + \mathrm{H_2(B''^{1}\Sigma^{+}_{u})}$ & $14.6$ & $f(T_\mathrm{e,eV})$ & \ \cite{marques.2007,lxcat-istlisbon}\\ 
used as & $\mathrm{e} + \mathrm{H_2} \rightarrow \mathrm{e} + \mathrm{H_2}$& &  & \\[0.1cm] 
R19&$\mathrm{e} + \mathrm{H_2} \rightarrow \mathrm{e} + \mathrm{H_2(D'^{1}\Pi_{u})}$ & $14.6$ & $f(T_\mathrm{e,eV})$ & \ \cite{marques.2007,lxcat-istlisbon}\\ 
used as & $\mathrm{e} + \mathrm{H_2} \rightarrow \mathrm{e} + \mathrm{H_2}$& &  & \\[0.1cm] 
R20&$\mathrm{e} + \mathrm{H_2O} \rightarrow \mathrm{e} + \mathrm{H_2O(\tilde{A}^1B_1)}$ & $7.49$ & $f(T_\mathrm{e,eV})$ & \ \cite{budde.2022}\\ 
used as & $\mathrm{e} + \mathrm{H_{2}O} \rightarrow \mathrm{e} + \mathrm{H_{2}O}$& &  & \\[0.1cm] 
R21&$\mathrm{e} + \mathrm{H_2O} \rightarrow \mathrm{e} + \mathrm{H_2O(\tilde{a}^3B_1)}$ & $7.14$ & $f(T_\mathrm{e,eV})$ & \ \cite{budde.2022}\\ 
used as & $\mathrm{e} + \mathrm{H_{2}O} \rightarrow \mathrm{e} + \mathrm{H_{2}O}$& &  & \\[0.1cm] 
R22&$\mathrm{e} + \mathrm{O} \rightarrow \mathrm{e} + \mathrm{O(^1D)}$ & $1.96$ & $f(T_\mathrm{e,eV})$ & \ \cite{alves.2016,lxcat-istlisbon-o}\\ 
used as & $\mathrm{e} + \mathrm{O} \rightarrow \mathrm{e} + \mathrm{O}$& &  & \\[0.1cm] 
R23&$\mathrm{e} + \mathrm{O} \rightarrow \mathrm{e} + \mathrm{O(^1S)}$ & $4.18$ & $f(T_\mathrm{e,eV})$ & \ \cite{alves.2016,lxcat-istlisbon-o}\\ 
used as & $\mathrm{e} + \mathrm{O} \rightarrow \mathrm{e} + \mathrm{O}$& &  & \\[0.1cm] 
R24&$\mathrm{e} + \mathrm{O} \rightarrow \mathrm{e} + \mathrm{O(^4S^0)}$ & $9.2$ & $f(T_\mathrm{e,eV})$ & \ \cite{alves.2016,lxcat-istlisbon-o}\\ 
used as & $\mathrm{e} + \mathrm{O} \rightarrow \mathrm{e} + \mathrm{O}$& &  & \\[0.1cm] 
R25&$\mathrm{e} + \mathrm{O} \rightarrow \mathrm{e} + \mathrm{O(^2D^0)}$ & $12.5$ & $f(T_\mathrm{e,eV})$ & \ \cite{alves.2016,lxcat-istlisbon-o}\\ 
used as & $\mathrm{e} + \mathrm{O} \rightarrow \mathrm{e} + \mathrm{O}$& &  & \\[0.1cm]

\hline
\end{tabular}
\end{table*}

\begin{table*}[h!]
\footnotesize
\setlength{\tabcolsep}{8pt}
\caption{$T_\mathrm{e,eV}$: electron temperature in units of eV, $T_\mathrm{{e,K}}$: electron temperature in units of Kelvin. An energy threshold ($E_{\mathrm{thr}}$) given as $T_\mathrm{{e,eV}}$ means the threshold is taken to be the current value of $T_{e,eV}$. An $E_{\mathrm{thr}}$ of 0 does not imply that the process does not have a threshold, but instead that it is not included in the electron energy balance here. Rate coefficients are given in units of $\mathrm{m}^{3}\mathrm{s}^{-1}$ or $\mathrm{m}^{6}\mathrm{s}^{-1}$ depending on the order of the reaction.
A rate coefficient given as $f\left(T_\mathrm{{e,eV}}\right)$ is derived from cross section data. As described in section \ref{sec:simulation-framework}, excited states are not tracked in model. For clarity, reactions involving excited states are listed, along with how they are implemented here. Excited state notation follows the original source as far as possible.}
\begin{tabular}{p{1.5cm}p{5cm}p{1.5cm}p{5cm}p{1.5cm}}
no. & reaction                                                                     & $E_{\mathrm{thr}}$ / eV & rate coefficient                                                                      & reference                                     \\
\hline

R26&$\mathrm{e} + \mathrm{O} \rightarrow \mathrm{e} + \mathrm{O(^2P^0)}$ & $14.1$ & $f(T_\mathrm{e,eV})$ & \ \cite{alves.2016,lxcat-istlisbon-o}\\ 
used as & $\mathrm{e} + \mathrm{O} \rightarrow \mathrm{e} + \mathrm{O}$& &  & \\[0.1cm] 
R27&$\mathrm{e} + \mathrm{O} \rightarrow \mathrm{e} + \mathrm{O(^3P^0)}$ & $15.7$ & $f(T_\mathrm{e,eV})$ & \ \cite{alves.2016,lxcat-istlisbon-o}\\ 
used as & $\mathrm{e} + \mathrm{O} \rightarrow \mathrm{e} + \mathrm{O}$& &  & \\[0.1cm] 
R28&$\mathrm{e} + \mathrm{O_2} \rightarrow \mathrm{e} + \mathrm{O_2(a^1\Delta_g)}$ & $0.977$ & $f(T_\mathrm{e,eV})$ & \ \cite{gousset.1991,lxcat-istlisbon-o2}\\ 
used as & $\mathrm{e} + \mathrm{O_2} \rightarrow \mathrm{e} + \mathrm{O_2}$& &  & \\[0.1cm] 
R29&$\mathrm{e} + \mathrm{O_2} \rightarrow \mathrm{e} + \mathrm{O_2(b^1\Sigma_g^+)}$ & $1.627$ & $f(T_\mathrm{e,eV})$ & \ \cite{gousset.1991,lxcat-istlisbon-o2}\\ 
used as & $\mathrm{e} + \mathrm{O_2} \rightarrow \mathrm{e} + \mathrm{O_2}$& &  & \\[0.1cm] 
R30&$\mathrm{e} + \mathrm{O_2} \rightarrow \mathrm{e} + \mathrm{O_2(A^3\Sigma_u^+, C^3\Delta_u, c^1\Sigma_u^-)}$ & $4.5$ & $f(T_\mathrm{e,eV})$ & \ \cite{gousset.1991,lxcat-istlisbon-o2}\\ 
used as & $\mathrm{e} + \mathrm{O_2} \rightarrow \mathrm{e} + \mathrm{O_2}$& &  & \\[0.1cm] 
R31&$\mathrm{e} + \mathrm{O_2} \rightarrow \mathrm{e} + \mathrm{O_2}(9.97\,\mathrm{eV})$ & $9.97$ & $f(T_\mathrm{e,eV})$ & \ \cite{gousset.1991,lxcat-istlisbon-o2}\\ 
used as & $\mathrm{e} + \mathrm{O_2} \rightarrow \mathrm{e} + \mathrm{O_2}$& &  & \\[0.1cm] 
R32&$\mathrm{e} + \mathrm{O_2} \rightarrow \mathrm{e} + \mathrm{O_2}(14.7\,\mathrm{eV})$ & $14.7$ & $f(T_\mathrm{e,eV})$ & \ \cite{gousset.1991,lxcat-istlisbon-o2}\\ 
used as & $\mathrm{e} + \mathrm{O_2} \rightarrow \mathrm{e} + \mathrm{O_2}$& &  & \\[0.1cm] 
R33&$\mathrm{e} + \mathrm{OH} \rightarrow \mathrm{e} + \mathrm{OH(A^2\Sigma^+)}$ & $4.05$ & $f(T_\mathrm{e,eV})$ & \ \cite{chakrabarti.2019}\\ 
used as & $\mathrm{e} + \mathrm{OH} \rightarrow \mathrm{e} + \mathrm{OH}$& &  & \\[0.1cm] 
\\ \multicolumn{5}{c}{\textbf{Electron impact excitation - vibrational}} \\ \hline 
R34&$\mathrm{e} + \mathrm{H_2(}v{=}0) \rightarrow \mathrm{e} + \mathrm{H_2(}v{=}1)$ & $0.516$ & $f(T_\mathrm{e,eV})$ & \ \cite{marques.2007,lxcat-istlisbon}\\ 
used as & $\mathrm{e} + \mathrm{H_2} \rightarrow \mathrm{e} + \mathrm{H_2}$& &  & \\[0.1cm] 
R35&$\mathrm{e} + \mathrm{H_2(}v{=}0) \rightarrow \mathrm{e} + \mathrm{H_2(}v{=}2)$ & $1.0$ & $f(T_\mathrm{e,eV})$ & \ \cite{marques.2007,lxcat-istlisbon}\\ 
used as & $\mathrm{e} + \mathrm{H_2} \rightarrow \mathrm{e} + \mathrm{H_2}$& &  & \\[0.1cm] 
R36&$\mathrm{e} + \mathrm{H_2(}v{=}0) \rightarrow \mathrm{e} + \mathrm{H_2(}v{=}3)$ & $1.5$ & $f(T_\mathrm{e,eV})$ & \ \cite{marques.2007,lxcat-istlisbon}\\ 
used as & $\mathrm{e} + \mathrm{H_2} \rightarrow \mathrm{e} + \mathrm{H_2}$& &  & \\[0.1cm] 
R37&$\mathrm{e} + \mathrm{H_2O(}v{=}000) \rightarrow \mathrm{e} + \mathrm{H_2O(}v{=}010)$ & $0.198$ & $f(T_\mathrm{e,eV})$ & \ \cite{budde.2022}\\ 
used as & $\mathrm{e} + \mathrm{H_{2}O} \rightarrow \mathrm{e} + \mathrm{H_{2}O}$& &  & \\[0.1cm] 
R38&$\mathrm{e} + \mathrm{H_2O(}v{=}000) \rightarrow \mathrm{e} + \mathrm{H_2O(}v{=}100+001)$ & $0.453$ & $f(T_\mathrm{e,eV})$ & \ \cite{budde.2022}\\ 
used as & $\mathrm{e} + \mathrm{H_{2}O} \rightarrow \mathrm{e} + \mathrm{H_{2}O}$& &  & \\[0.1cm] 
R39&$\mathrm{e} + \mathrm{O_2(}v{=}0) \rightarrow \mathrm{e} + \mathrm{O_2(}v{=}1)$ & $0.19$ & $f(T_\mathrm{e,eV})$ & \ \cite{gousset.1991,lxcat-istlisbon-o2}\\ 
used as & $\mathrm{e} + \mathrm{O_2} \rightarrow \mathrm{e} + \mathrm{O_2}$& &  & \\[0.1cm] 
R40&$\mathrm{e} + \mathrm{O_2(}v{=}0) \rightarrow \mathrm{e} + \mathrm{O_2(}v{=}2)$ & $0.38$ & $f(T_\mathrm{e,eV})$ & \ \cite{gousset.1991,lxcat-istlisbon-o2}\\ 
used as & $\mathrm{e} + \mathrm{O_2} \rightarrow \mathrm{e} + \mathrm{O_2}$& &  & \\[0.1cm] 
R41&$\mathrm{e} + \mathrm{O_2(}v{=}0) \rightarrow \mathrm{e} + \mathrm{O_2(}v{=}3)$ & $0.6$ & $f(T_\mathrm{e,eV})$ & \ \cite{gousset.1991,lxcat-istlisbon-o2}\\ 
used as & $\mathrm{e} + \mathrm{O_2} \rightarrow \mathrm{e} + \mathrm{O_2}$& &  & \\[0.1cm] 
R42&$\mathrm{e} + \mathrm{O_2(}v{=}0) \rightarrow \mathrm{e} + \mathrm{O_2(}v{=}4)$ & $0.8$ & $f(T_\mathrm{e,eV})$ & \ \cite{gousset.1991,lxcat-istlisbon-o2}\\ 
used as & $\mathrm{e} + \mathrm{O_2} \rightarrow \mathrm{e} + \mathrm{O_2}$& &  & \\[0.1cm] 
\\ \multicolumn{5}{c}{\textbf{Electron impact excitation - rotational}} \\ \hline 
R43&$\mathrm{e} + \mathrm{H_2O}(v{=}000,J{=}000) \rightarrow \mathrm{e} + \mathrm{H_2O}(v{=}000,J{=}111)$ & $0.0046048737$ & $f(T_\mathrm{e,eV})$ & \ \cite{budde.2022}\\ 
used as & $\mathrm{e} + \mathrm{H_{2}O} \rightarrow \mathrm{e} + \mathrm{H_{2}O}$& &  & \\[0.1cm] 
\\ \multicolumn{5}{c}{\textbf{Electron impact ionization}} \\ \hline 
R44&$\mathrm{e} + \mathrm{OH} \rightarrow $2$\mathrm{e} + \mathrm{OH^{+}}$&$13.1$ & $\num{2.000e-16}\cdot T_\mathrm{e,eV}^{1.78}\cdot\exp(\frac{-14}{T_\mathrm{e,eV}})$ & \ \cite{liu.2010}\\ 
R45&$\mathrm{e} + \mathrm{H} \rightarrow $2$\mathrm{e} + \mathrm{H^{+}}$&$13.6057$ & $f(T_\mathrm{e,eV})$ & \ \cite{marques.2007,lxcat-istlisbon}\\ 
R46&$\mathrm{e} + \mathrm{H_2} \rightarrow $2$\mathrm{e} + \mathrm{H_2^{+}}$&$15.4$ & $f(T_\mathrm{e,eV})$ & \ \cite{marques.2007,lxcat-istlisbon}\\ 
R47&$\mathrm{e} + \mathrm{H_{2}O} \rightarrow $2$\mathrm{e} + \mathrm{H_{2}O^{+}}$&$13.5$ & $f(T_\mathrm{e,eV})$ & \ \cite{budde.2022}\\ 
R48&$\mathrm{e} + \mathrm{O} \rightarrow $2$\mathrm{e} + \mathrm{O^{+}}$&$13.6$ & $f(T_\mathrm{e,eV})$ & \ \cite{alves.2016,lxcat-istlisbon-o}\\ 
R49&$\mathrm{e} + \mathrm{O_2} \rightarrow $2$\mathrm{e} + \mathrm{O_2^{+}}$&$12.1$ & $f(T_\mathrm{e,eV})$ & \ \cite{gousset.1991,lxcat-istlisbon-o2}\\ 
\\ \multicolumn{5}{c}{\textbf{Electron impact dissociation}} \\ \hline 
R50&$\mathrm{e} + \mathrm{H_2O_2} \rightarrow \mathrm{e}$ + 2$\mathrm{OH}$&$0.0$ & $\num{2.360e-15}$ & \ \cite{tavant.2016,soloshenko.2009}\\ 
R51&$\mathrm{e} + \mathrm{H_2O_2} \rightarrow \mathrm{e} + \mathrm{H} + \mathrm{HO_2}$&$0.0$ & $\num{3.100e-17}$ & \ \cite{tavant.2016,soloshenko.2002}\\

\hline
\end{tabular}
\end{table*}

\begin{table*}[h!]
\footnotesize
\setlength{\tabcolsep}{8pt}
\caption{$T_\mathrm{e,eV}$: electron temperature in units of eV, $T_\mathrm{{e,K}}$: electron temperature in units of Kelvin. An energy threshold ($E_{\mathrm{thr}}$) given as $T_\mathrm{{e,eV}}$ means the threshold is taken to be the current value of $T_{e,eV}$. An $E_{\mathrm{thr}}$ of 0 does not imply that the process does not have a threshold, but instead that it is not included in the electron energy balance here. Rate coefficients are given in units of $\mathrm{m}^{3}\mathrm{s}^{-1}$ or $\mathrm{m}^{6}\mathrm{s}^{-1}$ depending on the order of the reaction.
A rate coefficient given as $f\left(T_\mathrm{{e,eV}}\right)$ is derived from cross section data. As described in section \ref{sec:simulation-framework}, excited states are not tracked in model. For clarity, reactions involving excited states are listed, along with how they are implemented here. Excited state notation follows the original source as far as possible.}
\begin{tabular}{p{1.5cm}p{5cm}p{1.5cm}p{5cm}p{1.5cm}}
no. & reaction                                                                     & $E_{\mathrm{thr}}$ / eV & rate coefficient                                                                      & reference                                     \\
\hline

R52&$\mathrm{e} + \mathrm{HO_2} \rightarrow \mathrm{e} + \mathrm{H} + \mathrm{O_2}$&$0.0$ & $\num{3.100e-15}$ & \ \cite{tavant.2016,soloshenko.2009}\\ 
R53&$\mathrm{e} + \mathrm{O_3} \rightarrow \mathrm{e} + \mathrm{O} + \mathrm{O_2}$&$0.0$ & $\num{5.880e-15}$ & \ \cite{tavant.2016,turner.2015,brisset.2021}\\ 
R54&$\mathrm{e} + \mathrm{H_2} \rightarrow \mathrm{e} + \mathrm{H_2(b^3\Sigma^{+}_u)} \rightarrow \mathrm{e} + 2\,\mathrm{H}$ & $7.93$ & $f(T_\mathrm{e,eV})$ & \ \cite{marques.2007,lxcat-istlisbon}\\ 
used as & $\mathrm{e} + \mathrm{H_2} \rightarrow \mathrm{e}$ + 2$\mathrm{H}$& &  & \\[0.1cm] 
R55&$\mathrm{e} + \mathrm{H_2} \rightarrow \mathrm{e} + \mathrm{H_2(a^3\Sigma^{+}_g)} \rightarrow \mathrm{e} + 2\,\mathrm{H}$ & $11.72$ & $f(T_\mathrm{e,eV})$ & \ \cite{marques.2007,lxcat-istlisbon}\\ 
used as & $\mathrm{e} + \mathrm{H_2} \rightarrow \mathrm{e}$ + 2$\mathrm{H}$& &  & \\[0.1cm] 
R56&$\mathrm{e} + \mathrm{H_2} \rightarrow \mathrm{e} + \mathrm{H_2(c^3\Pi_u)} \rightarrow \mathrm{e} + 2\,\mathrm{H}$ & $11.72$ & $f(T_\mathrm{e,eV})$ & \ \cite{marques.2007,lxcat-istlisbon}\\ 
used as & $\mathrm{e} + \mathrm{H_2} \rightarrow \mathrm{e}$ + 2$\mathrm{H}$& &  & \\[0.1cm] 
R57&$\mathrm{e} + \mathrm{H_2} \rightarrow \mathrm{e} + \mathrm{H} + \mathrm{H(2p)}$ & $14.68$ & $f(T_\mathrm{e,eV})$ & \ \cite{marques.2007,lxcat-istlisbon}\\ 
used as & $\mathrm{e} + \mathrm{H_2} \rightarrow \mathrm{e}$ + 2$\mathrm{H}$& &  & \\[0.1cm] 
R58&$\mathrm{e} + \mathrm{H_2} \rightarrow \mathrm{e} + \mathrm{H} + \mathrm{H(2s)}$ & $14.68$ & $f(T_\mathrm{e,eV})$ & \ \cite{marques.2007,lxcat-istlisbon}\\ 
used as & $\mathrm{e} + \mathrm{H_2} \rightarrow \mathrm{e}$ + 2$\mathrm{H}$& &  & \\[0.1cm] 
R59&$\mathrm{e} + \mathrm{H_2} \rightarrow \mathrm{e} + \mathrm{H} + \mathrm{H(n{=}3)}$ & $16.57$ & $f(T_\mathrm{e,eV})$ & \ \cite{marques.2007,lxcat-istlisbon}\\ 
used as & $\mathrm{e} + \mathrm{H_2} \rightarrow \mathrm{e}$ + 2$\mathrm{H}$& &  & \\[0.1cm] 
R60&$\mathrm{e} + \mathrm{H_2} \rightarrow \mathrm{e} + \mathrm{H} + \mathrm{H(n{=}4)}$ & $17.22$ & $f(T_\mathrm{e,eV})$ & \ \cite{marques.2007,lxcat-istlisbon}\\ 
used as & $\mathrm{e} + \mathrm{H_2} \rightarrow \mathrm{e}$ + 2$\mathrm{H}$& &  & \\[0.1cm] 
R61&$\mathrm{e} + \mathrm{H_2} \rightarrow \mathrm{e} + \mathrm{H} + \mathrm{H(n{=}5)}$ & $17.53$ & $f(T_\mathrm{e,eV})$ & \ \cite{marques.2007,lxcat-istlisbon}\\ 
used as & $\mathrm{e} + \mathrm{H_2} \rightarrow \mathrm{e}$ + 2$\mathrm{H}$& &  & \\[0.1cm] 
R62&$\mathrm{e} + \mathrm{H_{2}O} \rightarrow \mathrm{e} + \mathrm{H} + \mathrm{OH}$&$6.6$ & $f(T_\mathrm{e,eV})$ & \ \cite{budde.2022}\\ 
R63&$\mathrm{e} + \mathrm{H_2O} \rightarrow \mathrm{e} + \mathrm{OH(A)} + \mathrm{H}$ & $9.2$ & $f(T_\mathrm{e,eV})$ & \ \cite{budde.2022}\\ 
used as & $\mathrm{e} + \mathrm{H_{2}O} \rightarrow \mathrm{e} + \mathrm{H} + \mathrm{OH}$& &  & \\[0.1cm] 
R64&$\mathrm{e} + \mathrm{H_2O} \rightarrow \mathrm{e} + 2\,\mathrm{H} + \mathrm{O(^1S_0)}$ & $13.696$ & $f(T_\mathrm{e,eV})$ & \ \cite{budde.2022}\\ 
used as & $\mathrm{e} + \mathrm{H_{2}O} \rightarrow \mathrm{e}$ + 2$\mathrm{H} + \mathrm{O}$& &  & \\[0.1cm] 
R65&$\mathrm{e} + \mathrm{H_2O} \rightarrow \mathrm{e} + \mathrm{H_{2}O^{*}}$ & $18.0$ & $f(T_\mathrm{e,eV})$ & \ \cite{budde.2022}\\ 
used as & $\mathrm{e} + \mathrm{H_{2}O} \rightarrow \mathrm{e} + \mathrm{H} + \mathrm{OH}$& &  & \\[0.1cm] 
R66&$\mathrm{e} + \mathrm{O_2} \rightarrow \mathrm{e}$ + 2$\mathrm{O}$&$6.0$ & $f(T_\mathrm{e,eV})$ & \ \cite{gousset.1991,lxcat-istlisbon-o2}\\ 
R67&$\mathrm{e} + \mathrm{O_2} \rightarrow \mathrm{e} + \mathrm{O} + \mathrm{O(^1D)}$ & $8.4$ & $f(T_\mathrm{e,eV})$ & \ \cite{gousset.1991,lxcat-istlisbon-o2}\\ 
used as & $\mathrm{e} + \mathrm{O_2} \rightarrow \mathrm{e}$ + 2$\mathrm{O}$& &  & \\[0.1cm] 
R68&$\mathrm{e} + \mathrm{OH} \rightarrow \mathrm{e} + \mathrm{O} + \mathrm{H(2s)}$ & $7.3537$ & $f(T_\mathrm{e,eV})$ & \ \cite{chakrabarti.2019}\\ 
used as & $\mathrm{e} + \mathrm{OH} \rightarrow \mathrm{e} + \mathrm{H} + \mathrm{O}$& &  & \\[0.1cm] 
\\ \multicolumn{5}{c}{\textbf{(Dissociative) electron attachment}} \\ \hline 
R69&$\mathrm{e} + \mathrm{H_2O_2} \rightarrow \mathrm{O^{-}} + \mathrm{H_{2}O}$&$0.0$ & $\num{1.570e-16}\cdot T_\mathrm{e,eV}^{-0.55}$ & \ \cite{tavant.2016,nandi.2003}\\ 
R70&$\mathrm{e} + \mathrm{H_2O_2} \rightarrow \mathrm{OH^{-}} + \mathrm{OH}$&$0.0$ & $\num{2.700e-16}\cdot T_\mathrm{e,eV}^{-0.5}$ & \ \cite{tavant.2016,nandi.2003}\\ 
R71&$\mathrm{e} + \mathrm{O_3} \rightarrow \mathrm{O_2^{-}} + \mathrm{O}$&$0.0$ & $\num{1.000e-15}$ & \ \cite{tavant.2016,matejcik.1997}\\ 
R72&$\mathrm{e} + \mathrm{H_{2}O} + \mathrm{O_2} \rightarrow \mathrm{O_2^{-}} + \mathrm{H_{2}O}$&$0.0$ & $\num{1.400e-41}$ & \ \cite{aleksandrov.1988}\\ 
R73&$\mathrm{e} + \mathrm{O_2} + \mathrm{M} \rightarrow \mathrm{O_2^{-}} + \mathrm{M}$&$0.0$ & $\num{8.500e-44}$ & \ \cite{aleksandrov.1988}\\ 
R74&$\mathrm{e}$ + 2$\mathrm{O_2} \rightarrow \mathrm{O_2^{-}} + \mathrm{O_2}$&$0.0$ & $\num{2.000e-42}$ & \ \cite{aleksandrov.1988}\\ 
R75&$\mathrm{e} + \mathrm{H} + \mathrm{O_2} \rightarrow \mathrm{O_2^{-}} + \mathrm{H}$&$0.0$ & $\num{8.500e-44}$ & \ \cite{aleksandrov.1988}\\ 
R76&$\mathrm{e} + \mathrm{O} + \mathrm{O_2} \rightarrow \mathrm{O_2^{-}} + \mathrm{O}$&$0.0$ & $\num{8.500e-44}$ & \ \cite{aleksandrov.1988}\\ 
R77&$\mathrm{e} + \mathrm{H_2} + \mathrm{O_2} \rightarrow \mathrm{O_2^{-}} + \mathrm{H_2}$&$0.0$ & $\num{8.500e-44}$ & \ \cite{aleksandrov.1988}\\ 
R78&$\mathrm{e} + \mathrm{H_2O_2} + \mathrm{O_2} \rightarrow \mathrm{O_2^{-}} + \mathrm{H_2O_2}$&$0.0$ & $\num{8.500e-44}$ & \ \cite{aleksandrov.1988}\\ 
R79&$\mathrm{e} + \mathrm{HO_2} + \mathrm{O_2} \rightarrow \mathrm{O_2^{-}} + \mathrm{HO_2}$&$0.0$ & $\num{8.500e-44}$ & \ \cite{aleksandrov.1988}\\ 
R80&$\mathrm{e} + \mathrm{O_2} + \mathrm{O_3} \rightarrow \mathrm{O_2^{-}} + \mathrm{O_3}$&$0.0$ & $\num{8.500e-44}$ & \ \cite{aleksandrov.1988}\\ 
R81&$\mathrm{e} + \mathrm{O_2} + \mathrm{OH} \rightarrow \mathrm{O_2^{-}} + \mathrm{OH}$&$0.0$ & $\num{8.500e-44}$ & \ \cite{aleksandrov.1988}\\ 
R82&$\mathrm{e} + \mathrm{H_{2}O} \rightarrow \mathrm{OH^{-}} + \mathrm{H}$&$5.9$ & $f(T_\mathrm{e,eV})$ & \ \cite{budde.2022}\\ 
R83&$\mathrm{e} + \mathrm{H_{2}O} \rightarrow \mathrm{O^{-}} + \mathrm{H_2}$&$4.9$ & $f(T_\mathrm{e,eV})$ & \ \cite{budde.2022}\\ 
R84&$\mathrm{e} + \mathrm{H_{2}O} \rightarrow \mathrm{H^{-}} + \mathrm{OH}$&$5.7$ & $f(T_\mathrm{e,eV})$ & \ \cite{budde.2022}\\ 
R85&$\mathrm{e} + \mathrm{O_2} \rightarrow \mathrm{O^{-}} + \mathrm{O}$&$5.0$ & $f(T_\mathrm{e,eV})$ & \ \cite{gousset.1991,lxcat-istlisbon-o2}\\ 
R86&$\mathrm{e} + \mathrm{OH} \rightarrow \mathrm{O^-(A^1\Pi)} + \mathrm{H(2s)}$ & $2.53$ & $f(T_\mathrm{e,eV})$ & \ \cite{chakrabarti.2019}\\ 
used as & $\mathrm{e} + \mathrm{OH} \rightarrow \mathrm{O^{-}} + \mathrm{H}$& &  & \\[0.1cm] 
R87&$\mathrm{e} + \mathrm{OH} \rightarrow \mathrm{O^-(a^3\Pi)} + \mathrm{H(2s)}$ & $2.41$ & $f(T_\mathrm{e,eV})$ & \ \cite{chakrabarti.2019}\\ 
used as & $\mathrm{e} + \mathrm{OH} \rightarrow \mathrm{O^{-}} + \mathrm{H}$& &  & \\[0.1cm] 
R88&$\mathrm{e} + \mathrm{OH} \rightarrow \mathrm{O^-(1^3\Sigma^{+})} + \mathrm{H(2s)}$ & $1.56$ & $f(T_\mathrm{e,eV})$ & \ \cite{chakrabarti.2019}\\ 
used as & $\mathrm{e} + \mathrm{OH} \rightarrow \mathrm{O^{-}} + \mathrm{H}$& &  & \\[0.1cm]

\hline
\end{tabular}
\end{table*}

\begin{table*}[h!]
\footnotesize
\setlength{\tabcolsep}{8pt}
\caption{$T_\mathrm{e,eV}$: electron temperature in units of eV, $T_\mathrm{{e,K}}$: electron temperature in units of Kelvin. An energy threshold ($E_{\mathrm{thr}}$) given as $T_\mathrm{{e,eV}}$ means the threshold is taken to be the current value of $T_{e,eV}$. An $E_{\mathrm{thr}}$ of 0 does not imply that the process does not have a threshold, but instead that it is not included in the electron energy balance here. Rate coefficients are given in units of $\mathrm{m}^{3}\mathrm{s}^{-1}$ or $\mathrm{m}^{6}\mathrm{s}^{-1}$ depending on the order of the reaction.
A rate coefficient given as $f\left(T_\mathrm{{e,eV}}\right)$ is derived from cross section data. As described in section \ref{sec:simulation-framework}, excited states are not tracked in model. For clarity, reactions involving excited states are listed, along with how they are implemented here. Excited state notation follows the original source as far as possible.}
\begin{tabular}{p{1.5cm}p{5cm}p{1.5cm}p{5cm}p{1.5cm}}
no. & reaction                                                                     & $E_{\mathrm{thr}}$ / eV & rate coefficient                                                                      & reference                                     \\
\hline

\\ \multicolumn{5}{c}{\textbf{Electron detachment}} \\ \hline 
R89&$\mathrm{e} + \mathrm{O^{-}} \rightarrow $2$\mathrm{e} + \mathrm{O}$&$3.44$ & $\num{9.330e-14}\cdot T_\mathrm{e,eV}^{0.178}\cdot\exp(\frac{-3}{T_\mathrm{e,eV}})$ & \ \cite{turner.2015}\\ 
R90&$\mathrm{e} + \mathrm{H^{-}} \rightarrow $2$\mathrm{e} + \mathrm{H}$&$0.754$ & $f(T_\mathrm{e,eV})$ & \ \cite{deutsch.2003}\\ 
R91&$\mathrm{e} + \mathrm{O_2^{-}} \rightarrow $2$\mathrm{e} + \mathrm{O_2}$&$4.5$ & $f(T_\mathrm{e,eV})$ & \ \cite{deutsch.2003,pedersen.1999}\\ 
R92&$\mathrm{e} + \mathrm{OH^{-}} \rightarrow $2$\mathrm{e} + \mathrm{OH}$&$3.7$ & $f(T_\mathrm{e,eV})$ & \ \cite{pedersen.1999}\\ 
\\ \multicolumn{5}{c}{\textbf{Two-body electron-ion recombination}} \\ \hline 
R93&$\mathrm{e} + \mathrm{H^+} \rightarrow \mathrm{H} + h\nu$ & $T_\mathrm{e,eV}$ & $\num{3.500e-18}\cdot \left(\frac{T_\mathrm{e,K}}{300.0}\right)^{-0.75}$ & \ \cite{prasad.1980,umist,millar.2024}\\ 
used as & $\mathrm{e} + \mathrm{H^{+}} \rightarrow \mathrm{H}$& &  & \\[0.1cm] 
R94&$\mathrm{e} + \mathrm{H_2^+} \rightarrow \mathrm{H_{2}^{**}} \rightarrow \mathrm{H} + \mathrm{H^{*}}$ & $T_\mathrm{e,eV}$ & $\num{1.600e-14}\cdot \left(\frac{T_\mathrm{e,K}}{300.0}\right)^{-0.43}$ & \ \cite{brian.1990,umist,millar.2024}\\ 
used as & $\mathrm{e} + \mathrm{H_2^{+}} \rightarrow $2$\mathrm{H}$& &  & \\[0.1cm] 
R95&$\mathrm{e} + \mathrm{H_{2}O^{+}} \rightarrow $2$\mathrm{H} + \mathrm{O}$&$T_\mathrm{e,eV}$ & $\num{3.050e-13}\cdot \left(\frac{T_\mathrm{e,K}}{300.0}\right)^{-0.5}$ & \ \cite{rosen.2000,umist,millar.2024}\\ 
R96&$\mathrm{e} + \mathrm{H_{2}O^{+}} \rightarrow \mathrm{H} + \mathrm{OH}$&$T_\mathrm{e,eV}$ & $\num{8.600e-14}\cdot \left(\frac{T_\mathrm{e,K}}{300.0}\right)^{-0.5}$ & \ \cite{rosen.2000,umist,millar.2024}\\ 
R97&$\mathrm{e} + \mathrm{H_{2}O^{+}} \rightarrow \mathrm{H_2} + \mathrm{O}$&$T_\mathrm{e,eV}$ & $\num{3.900e-14}\cdot \left(\frac{T_\mathrm{e,K}}{300.0}\right)^{-0.5}$ & \ \cite{rosen.2000,umist,millar.2024}\\ 
R98&$\mathrm{e} + \mathrm{H_3^{+}} \rightarrow $3$\mathrm{H}$&$0.0$ & $\num{2.190e-15}\cdot T_\mathrm{e,eV}^{-0.8}$ & \ \cite{tavant.2016,hjartarson.2010}\\ 
R99&$\mathrm{e} + \mathrm{H_3^{+}} \rightarrow \mathrm{H^{-}} + \mathrm{H_2^{+}}$&$0.0$ & $\num{1.930e-16}\cdot T_\mathrm{e,eV}^{-1.07}\cdot\exp(\frac{-3}{T_\mathrm{e,eV}})$ & \ \cite{tavant.2016,hjartarson.2010}\\ 
R100&$\mathrm{e} + \mathrm{H_3^{+}} \rightarrow \mathrm{H^{-}} + \mathrm{H_2^{+}}$&$0.0$ & $\num{2.590e-15}\cdot T_\mathrm{e,eV}^{-1.27}\cdot\exp(\frac{-6}{T_\mathrm{e,eV}} + 0.1)$ & \ \cite{tavant.2016,hjartarson.2010}\\ 
R101&$\mathrm{e} + \mathrm{H_3^{+}} \rightarrow \mathrm{H} + \mathrm{H_2}$&$0.0$ & $\num{7.300e-16}\cdot T_\mathrm{e,eV}^{0.8}$ & \ \cite{tavant.2016,hjartarson.2010}\\ 
R102&$\mathrm{e} + \mathrm{O^+} \rightarrow \mathrm{O} + h\nu$ & $T_\mathrm{e,eV}$ & $\num{3.240e-18}\cdot \left(\frac{T_\mathrm{e,K}}{300.0}\right)^{-0.66}$ & \ \cite{nahar.1999,umist,millar.2024}\\ 
used as & $\mathrm{e} + \mathrm{O^{+}} \rightarrow \mathrm{O}$& &  & \\[0.1cm] 
R103&$\mathrm{e} + \mathrm{O_2^{+}} \rightarrow $2$\mathrm{O}$&$T_\mathrm{e,eV}$ & $\num{1.950e-13}\cdot \left(\frac{T_\mathrm{e,K}}{300.0}\right)^{-0.7}$ & \ \cite{alge.1983,umist,millar.2024}\\ 
R104&$\mathrm{e} + \mathrm{OH^{+}} \rightarrow \mathrm{H} + \mathrm{O}$&$T_\mathrm{e,eV}$ & $\num{3.750e-14}\cdot \left(\frac{T_\mathrm{e,K}}{300.0}\right)^{-0.5}$ & \ \cite{umist,brian.1990,millar.2024}\\ 
\\ \multicolumn{5}{c}{\textbf{Three-body electron-ion recombination}} \\ \hline 
R105&2$\mathrm{e} + \mathrm{H^{+}} \rightarrow \mathrm{e} + \mathrm{H}$&$T_\mathrm{e,eV}$ & $\num{1.000e-31}\cdot \left(\frac{T_\mathrm{e,K}}{300.0}\right)^{-4.5}$ & \ \cite{kossyi.1992}\\ 
R106&$\mathrm{e} + \mathrm{H^{+}} + \mathrm{M} \rightarrow \mathrm{H} + \mathrm{M}$&$T_\mathrm{e,eV}$ & $\num{6.000e-39}\cdot \left(\frac{T_\mathrm{e,K}}{300.0}\right)^{-1.5}$ & \ \cite{kossyi.1992}\\ 
R107&2$\mathrm{e} + \mathrm{H_2^{+}} \rightarrow \mathrm{e} + \mathrm{H_2}$&$T_\mathrm{e,eV}$ & $\num{1.000e-31}\cdot \left(\frac{T_\mathrm{e,K}}{300.0}\right)^{-4.5}$ & \ \cite{kossyi.1992}\\ 
R108&$\mathrm{e} + \mathrm{H_2^{+}} + \mathrm{M} \rightarrow \mathrm{H_2} + \mathrm{M}$&$T_\mathrm{e,eV}$ & $\num{6.000e-39}\cdot \left(\frac{T_\mathrm{e,K}}{300.0}\right)^{-1.5}$ & \ \cite{kossyi.1992}\\ 
R109&2$\mathrm{e} + \mathrm{H_{2}O^{+}} \rightarrow \mathrm{e} + \mathrm{H_{2}O}$&$T_\mathrm{e,eV}$ & $\num{1.000e-31}\cdot \left(\frac{T_\mathrm{e,K}}{300.0}\right)^{-4.5}$ & \ \cite{kossyi.1992}\\ 
R110&$\mathrm{e} + \mathrm{H_{2}O^{+}} + \mathrm{M} \rightarrow $2$\mathrm{H} + \mathrm{O} + \mathrm{M}$&$T_\mathrm{e,eV}$ & $\num{6.000e-39}\cdot \left(\frac{T_\mathrm{e,K}}{300.0}\right)^{-1.5}$ & \ \cite{kossyi.1992}\\ 
R111&2$\mathrm{e} + \mathrm{O^{+}} \rightarrow \mathrm{e} + \mathrm{O}$&$T_\mathrm{e,eV}$ & $\num{1.000e-31}\cdot \left(\frac{T_\mathrm{e,K}}{300.0}\right)^{-4.5}$ & \ \cite{kossyi.1992}\\ 
R112&$\mathrm{e} + \mathrm{O^{+}} + \mathrm{M} \rightarrow \mathrm{O} + \mathrm{M}$&$T_\mathrm{e,eV}$ & $\num{6.000e-39}\cdot \left(\frac{T_\mathrm{e,K}}{300.0}\right)^{-1.5}$ & \ \cite{kossyi.1992}\\ 
R113&2$\mathrm{e} + \mathrm{O_2^{+}} \rightarrow \mathrm{e} + \mathrm{O_2}$&$T_\mathrm{e,eV}$ & $\num{1.000e-31}\cdot \left(\frac{T_\mathrm{e,K}}{300.0}\right)^{-4.5}$ & \ \cite{kossyi.1992}\\ 
R114&$\mathrm{e} + \mathrm{O_2^{+}} + \mathrm{M} \rightarrow \mathrm{O_2} + \mathrm{M}$&$T_\mathrm{e,eV}$ & $\num{6.000e-39}\cdot \left(\frac{T_\mathrm{e,K}}{300.0}\right)^{-1.5}$ & \ \cite{kossyi.1992}\\ 
R115&2$\mathrm{e} + \mathrm{OH^{+}} \rightarrow \mathrm{e} + \mathrm{OH}$&$T_\mathrm{e,eV}$ & $\num{1.000e-31}\cdot \left(\frac{T_\mathrm{e,K}}{300.0}\right)^{-4.5}$ & \ \cite{kossyi.1992}\\ 
R116&$\mathrm{e} + \mathrm{OH^{+}} + \mathrm{M} \rightarrow \mathrm{OH} + \mathrm{M}$&$T_\mathrm{e,eV}$ & $\num{6.000e-39}\cdot \left(\frac{T_\mathrm{e,K}}{300.0}\right)^{-1.5}$ & \ \cite{kossyi.1992}\\

\hline
\end{tabular}
\end{table*}

\begin{table*}[h!]
\footnotesize
\setlength{\tabcolsep}{8pt}
\caption{$T_\mathrm{e,eV}$: electron temperature in units of eV, $T_\mathrm{{e,K}}$: electron temperature in units of Kelvin. An energy threshold ($E_{\mathrm{thr}}$) given as $T_\mathrm{{e,eV}}$ means the threshold is taken to be the current value of $T_{e,eV}$. An $E_{\mathrm{thr}}$ of 0 does not imply that the process does not have a threshold, but instead that it is not included in the electron energy balance here. Rate coefficients are given in units of $\mathrm{m}^{3}\mathrm{s}^{-1}$ or $\mathrm{m}^{6}\mathrm{s}^{-1}$ depending on the order of the reaction.
A rate coefficient given as $f\left(T_\mathrm{{e,eV}}\right)$ is derived from cross section data. As described in section \ref{sec:simulation-framework}, excited states are not tracked in model. For clarity, reactions involving excited states are listed, along with how they are implemented here. Excited state notation follows the original source as far as possible.}
\begin{tabular}{p{1.5cm}p{5cm}p{1.5cm}p{5cm}p{1.5cm}}
no. & reaction                                                                     & $E_{\mathrm{thr}}$ / eV & rate coefficient                                                                      & reference                                     \\
\hline

\\ \multicolumn{5}{c}{\textbf{Two-body collisions - negative ions}} \\ \hline 
R117&$\mathrm{H^-} + \mathrm{N_2} \rightarrow \mathrm{e} + \mathrm{H} + \mathrm{N_2}$ & $-$ & $\num{2.000e-18}\cdot \left(\frac{T_\mathrm{g}}{300.0}\right)^{0.5}\cdot\exp(\frac{-8751}{T_\mathrm{g}})$ & \ \cite{kossyi.1992}\\ 
used as & $\mathrm{H^{-}} + \mathrm{M} \rightarrow \mathrm{e} + \mathrm{H} + \mathrm{M}$& &  & \\[0.1cm] 
R118&$\mathrm{O_2^{-}} + \mathrm{O} \rightarrow \mathrm{e} + \mathrm{O_3}$&$-$ & $\num{8.500e-17}\cdot \left(\frac{T_\mathrm{g}}{300.0}\right)^{-1.8}$ & \ \cite{brisset.2021,ard.2013}\\ 
R119&$\mathrm{O_2^{-}} + \mathrm{O_2} \rightarrow \mathrm{e}$ + 2$\mathrm{O_2}$&$-$ & $\num{2.700e-16}\cdot \left(\frac{T_\mathrm{g}}{300.0}\right)^{0.5}\cdot\exp(\frac{-5590}{T_\mathrm{g}})$ & \ \cite{tavant.2016,bortner.1975}\\ 
R120&$\mathrm{O^-} + \mathrm{N_2} \rightarrow \mathrm{e} + \mathrm{O} + \mathrm{N_2}$ & $-$ & $\num{2.000e-18}\cdot \left(\frac{T_\mathrm{g}}{300.0}\right)^{0.5}\cdot\exp(\frac{-16945}{T_\mathrm{g}})$ & \ \cite{kossyi.1992}\\ 
used as & $\mathrm{O^{-}} + \mathrm{M} \rightarrow \mathrm{e} + \mathrm{O} + \mathrm{M}$& &  & \\[0.1cm] 
R121&$\mathrm{O^{-}} + \mathrm{O_2} \rightarrow \mathrm{e} + \mathrm{O_3}$&$-$ & $\num{1.000e-18}$ & \ \cite{brisset.2021}\\ 
R122&$\mathrm{O_2^{-}} + \mathrm{H} \rightarrow \mathrm{H^{-}} + \mathrm{O_2}$&$-$ & $\num{7.000e-16}$ & \ \cite{vanGaens.2013}\\ 
R123&$\mathrm{O_2^{-}} + \mathrm{H} \rightarrow \mathrm{e} + \mathrm{HO_2}$&$-$ & $\num{7.000e-16}$ & \ \cite{vanGaens.2013}\\ 
R124&$\mathrm{O_2^-} + \mathrm{N_2} \rightarrow \mathrm{e} + \mathrm{O_2} + \mathrm{N_2}$ & $-$ & $\num{2.000e-18}\cdot \left(\frac{T_\mathrm{g}}{300.0}\right)^{0.5}\cdot\exp(\frac{-5223}{T_\mathrm{g}})$ & \ \cite{kossyi.1992}\\ 
used as & $\mathrm{O_2^{-}} + \mathrm{M} \rightarrow \mathrm{e} + \mathrm{O_2} + \mathrm{M}$& &  & \\[0.1cm] 
R125&$\mathrm{OH^{-}} + \mathrm{O} \rightarrow \mathrm{e} + \mathrm{HO_2}$&$-$ & $\num{2.000e-16}$ & \ \cite{albritton.1978}\\ 
R126&$\mathrm{OH^{-}} + \mathrm{H_2} \rightarrow \mathrm{H^{-}} + \mathrm{H_{2}O}$&$-$ & $\num{5.000e-18}$ & \ \cite{albritton.1978}\\ 
R127&$\mathrm{OH^-} + \mathrm{N_2} \rightarrow \mathrm{e} + \mathrm{O_2} + \mathrm{N_2}$ & $-$ & $\num{2.000e-18}\cdot \left(\frac{T_\mathrm{g}}{300.0}\right)^{0.5}\cdot\exp(\frac{-21239}{T_\mathrm{g}})$ & \ \cite{kossyi.1992}\\ 
used as & $\mathrm{OH^{-}} + \mathrm{M} \rightarrow \mathrm{e} + \mathrm{OH} + \mathrm{M}$& &  & \\[0.1cm] 
\\ \multicolumn{5}{c}{\textbf{Two-body ion-ion recombination}} \\ \hline 
R128&$\mathrm{O^{-}} + \mathrm{O_2^{+}} \rightarrow \mathrm{O} + \mathrm{O_2}$&$-$ & $\num{1.500e-13}\cdot \left(\frac{T_\mathrm{g}}{300.0}\right)^{-0.5}$ & \ \cite{tavant.2016,vanGaens.2013,bogaerts.2009}\\ 
R129&$\mathrm{O^{-}} + \mathrm{O_2^{+}} \rightarrow $3$\mathrm{O}$&$-$ & $\num{1.000e-13}$ & \ \cite{tavant.2016,kossyi.1992}\\ 
R130&$\mathrm{O_2^{-}} + \mathrm{O_2^{+}} \rightarrow $2$\mathrm{O_2}$&$-$ & $\num{2.000e-13}\cdot \left(\frac{T_\mathrm{g}}{300.0}\right)^{-1.0}$ & \ \cite{tavant.2016,stafford.2004}\\ 
R131&$\mathrm{O_2^{-}} + \mathrm{O_2^{+}} \rightarrow $2$\mathrm{O} + \mathrm{O_2}$&$-$ & $\num{1.000e-13}$ & \ \cite{tavant.2016,stafford.2004}\\ 
R132&$\mathrm{H^{-}} + \mathrm{H_3^{+}} \rightarrow $4$\mathrm{H}$&$-$ & $\num{8.290e-13}\cdot \left(\frac{T_\mathrm{g}}{300.0}\right)^{-0.5}$ & \ \cite{tavant.2016}\\ 
R133&$\mathrm{H^{-}} + \mathrm{H_3^{+}} \rightarrow $2$\mathrm{H_2}$&$-$ & $\num{2.000e-13}\cdot \left(\frac{T_\mathrm{g}}{300.0}\right)^{-0.5}$ & \ \cite{liu.2010}\\ 
R134&$\mathrm{O^{-}} + \mathrm{H_3^{+}} \rightarrow \mathrm{H_2} + \mathrm{OH}$&$-$ & $\num{2.000e-13}\cdot \left(\frac{T_\mathrm{g}}{300.0}\right)^{-0.5}$ & \ \cite{liu.2010}\\ 
R135&$\mathrm{O_2^{-}} + \mathrm{H_3^{+}} \rightarrow \mathrm{H_{2}O} + \mathrm{OH}$&$-$ & $\num{2.000e-13}\cdot \left(\frac{T_\mathrm{g}}{300.0}\right)^{-0.5}$ & \ \cite{tavant.2016,liu.2010}\\ 
R136&$\mathrm{OH^{-}} + \mathrm{H_3^{+}} \rightarrow \mathrm{H_2} + \mathrm{H_{2}O}$&$-$ & $\num{2.000e-13}\cdot \left(\frac{T_\mathrm{g}}{300.0}\right)^{-0.5}$ & \ \cite{tavant.2016,liu.2010}\\ 
R137&$\mathrm{H^{-}} + \mathrm{H^{+}} \rightarrow $2$\mathrm{H}$&$-$ & $\num{7.510e-14}\cdot \left(\frac{T_\mathrm{g}}{300.0}\right)^{-0.5}$ & \ \cite{harada.2008,umist,millar.2024}\\ 
R138&$\mathrm{H^{-}} + \mathrm{H_2^{+}} \rightarrow \mathrm{H} + \mathrm{H_2}$&$-$ & $\num{7.510e-14}\cdot \left(\frac{T_\mathrm{g}}{300.0}\right)^{-0.5}$ & \ \cite{harada.2008,umist,millar.2024}\\ 
R139&$\mathrm{H^{-}} + \mathrm{O^{+}} \rightarrow \mathrm{H} + \mathrm{O}$&$-$ & $\num{7.510e-14}\cdot \left(\frac{T_\mathrm{g}}{300.0}\right)^{-0.5}$ & \ \cite{harada.2008,umist,millar.2024}\\ 
R140&$\mathrm{O^{-}} + \mathrm{H^{+}} \rightarrow \mathrm{H} + \mathrm{O}$&$-$ & $\num{7.510e-14}\cdot \left(\frac{T_\mathrm{g}}{300.0}\right)^{-0.5}$ & \ \cite{harada.2008,umist,millar.2024}\\ 
R141&$\mathrm{O^{-}} + \mathrm{O^{+}} \rightarrow $2$\mathrm{O}$&$-$ & $\num{7.510e-14}\cdot \left(\frac{T_\mathrm{g}}{300.0}\right)^{-0.5}$ & \ \cite{harada.2008,umist,millar.2024}\\ 
R142&$\mathrm{O_2^{-}} + \mathrm{H^{+}} \rightarrow \mathrm{H} + \mathrm{O_2}$&$-$ & $\num{7.510e-14}\cdot \left(\frac{T_\mathrm{g}}{300.0}\right)^{-0.5}$ & \ \cite{harada.2008,umist,millar.2024}\\ 
R143&$\mathrm{O_2^{-}} + \mathrm{O^{+}} \rightarrow \mathrm{O} + \mathrm{O_2}$&$-$ & $\num{7.510e-14}\cdot \left(\frac{T_\mathrm{g}}{300.0}\right)^{-0.5}$ & \ \cite{harada.2008,umist,millar.2024}\\ 
R144&$\mathrm{OH^{-}} + \mathrm{H^{+}} \rightarrow \mathrm{H} + \mathrm{OH}$&$-$ & $\num{7.510e-14}\cdot \left(\frac{T_\mathrm{g}}{300.0}\right)^{-0.5}$ & \ \cite{harada.2008,umist,millar.2024}\\ 
R145&$\mathrm{OH^{-}} + \mathrm{O^{+}} \rightarrow \mathrm{O} + \mathrm{OH}$&$-$ & $\num{7.510e-14}\cdot \left(\frac{T_\mathrm{g}}{300.0}\right)^{-0.5}$ & \ \cite{harada.2008,umist,millar.2024}\\ 
R146&$\mathrm{H^{-}} + \mathrm{O_2^{+}} \rightarrow \mathrm{H} + \mathrm{O_2}$&$-$ & $\num{2.000e-13}\cdot \left(\frac{T_\mathrm{g}}{300.0}\right)^{-0.5}$ & \ \cite{kossyi.1992}\\ 
R147&$\mathrm{H^{-}} + \mathrm{OH^{+}} \rightarrow \mathrm{H} + \mathrm{OH}$&$-$ & $\num{2.000e-13}\cdot \left(\frac{T_\mathrm{g}}{300.0}\right)^{-0.5}$ & \ \cite{kossyi.1992}\\ 
R148&$\mathrm{H^{-}} + \mathrm{H_{2}O^{+}} \rightarrow \mathrm{H} + \mathrm{H_{2}O}$&$-$ & $\num{2.000e-13}\cdot \left(\frac{T_\mathrm{g}}{300.0}\right)^{-0.5}$ & \ \cite{kossyi.1992}\\ 
R149&$\mathrm{O^{-}} + \mathrm{H_2^{+}} \rightarrow \mathrm{H_2} + \mathrm{O}$&$-$ & $\num{2.000e-13}\cdot \left(\frac{T_\mathrm{g}}{300.0}\right)^{-0.5}$ & \ \cite{kossyi.1992}\\ 
R150&$\mathrm{O^{-}} + \mathrm{OH^{+}} \rightarrow \mathrm{O} + \mathrm{OH}$&$-$ & $\num{2.000e-13}\cdot \left(\frac{T_\mathrm{g}}{300.0}\right)^{-0.5}$ & \ \cite{kossyi.1992}\\ 
R151&$\mathrm{O^{-}} + \mathrm{H_{2}O^{+}} \rightarrow \mathrm{H_{2}O} + \mathrm{O}$&$-$ & $\num{2.000e-13}\cdot \left(\frac{T_\mathrm{g}}{300.0}\right)^{-0.5}$ & \ \cite{kossyi.1992}\\

\hline
\end{tabular}
\end{table*}

\begin{table*}[h!]
\footnotesize
\setlength{\tabcolsep}{8pt}
\caption{$T_\mathrm{e,eV}$: electron temperature in units of eV, $T_\mathrm{{e,K}}$: electron temperature in units of Kelvin. An energy threshold ($E_{\mathrm{thr}}$) given as $T_\mathrm{{e,eV}}$ means the threshold is taken to be the current value of $T_{e,eV}$. An $E_{\mathrm{thr}}$ of 0 does not imply that the process does not have a threshold, but instead that it is not included in the electron energy balance here. Rate coefficients are given in units of $\mathrm{m}^{3}\mathrm{s}^{-1}$ or $\mathrm{m}^{6}\mathrm{s}^{-1}$ depending on the order of the reaction.
A rate coefficient given as $f\left(T_\mathrm{{e,eV}}\right)$ is derived from cross section data. As described in section \ref{sec:simulation-framework}, excited states are not tracked in model. For clarity, reactions involving excited states are listed, along with how they are implemented here. Excited state notation follows the original source as far as possible.}
\begin{tabular}{p{1.5cm}p{5cm}p{1.5cm}p{5cm}p{1.5cm}}
no. & reaction                                                                     & $E_{\mathrm{thr}}$ / eV & rate coefficient                                                                      & reference                                     \\
\hline

R152&$\mathrm{O_2^{-}} + \mathrm{H_2^{+}} \rightarrow \mathrm{H_2} + \mathrm{O_2}$&$-$ & $\num{2.000e-13}\cdot \left(\frac{T_\mathrm{g}}{300.0}\right)^{-0.5}$ & \ \cite{kossyi.1992}\\ 
R153&$\mathrm{O_2^{-}} + \mathrm{OH^{+}} \rightarrow \mathrm{O_2} + \mathrm{OH}$&$-$ & $\num{2.000e-13}\cdot \left(\frac{T_\mathrm{g}}{300.0}\right)^{-0.5}$ & \ \cite{kossyi.1992}\\ 
R154&$\mathrm{O_2^{-}} + \mathrm{H_{2}O^{+}} \rightarrow \mathrm{H_{2}O} + \mathrm{O_2}$&$-$ & $\num{2.000e-13}\cdot \left(\frac{T_\mathrm{g}}{300.0}\right)^{-0.5}$ & \ \cite{kossyi.1992}\\ 
R155&$\mathrm{OH^{-}} + \mathrm{H_2^{+}} \rightarrow \mathrm{H_2} + \mathrm{OH}$&$-$ & $\num{2.000e-13}\cdot \left(\frac{T_\mathrm{g}}{300.0}\right)^{-0.5}$ & \ \cite{kossyi.1992}\\ 
R156&$\mathrm{OH^{-}} + \mathrm{O_2^{+}} \rightarrow \mathrm{O_2} + \mathrm{OH}$&$-$ & $\num{2.000e-13}\cdot \left(\frac{T_\mathrm{g}}{300.0}\right)^{-0.5}$ & \ \cite{kossyi.1992}\\ 
R157&$\mathrm{OH^{-}} + \mathrm{OH^{+}} \rightarrow $2$\mathrm{OH}$&$-$ & $\num{2.000e-13}\cdot \left(\frac{T_\mathrm{g}}{300.0}\right)^{-0.5}$ & \ \cite{kossyi.1992}\\ 
R158&$\mathrm{OH^{-}} + \mathrm{H_{2}O^{+}} \rightarrow \mathrm{H_{2}O} + \mathrm{OH}$&$-$ & $\num{2.000e-13}\cdot \left(\frac{T_\mathrm{g}}{300.0}\right)^{-0.5}$ & \ \cite{kossyi.1992}\\ 
R159&$\mathrm{H^{-}} + \mathrm{H_2^{+}} \rightarrow $3$\mathrm{H}$&$-$ & $\num{1.000e-13}$ & \ \cite{kossyi.1992}\\ 
R160&$\mathrm{H^{-}} + \mathrm{O_2^{+}} \rightarrow \mathrm{H}$ + 2$\mathrm{O}$&$-$ & $\num{1.000e-13}$ & \ \cite{kossyi.1992}\\ 
R161&$\mathrm{H^{-}} + \mathrm{OH^{+}} \rightarrow $2$\mathrm{H} + \mathrm{O}$&$-$ & $\num{1.000e-13}$ & \ \cite{kossyi.1992}\\ 
R162&$\mathrm{H^{-}} + \mathrm{H_{2}O^{+}} \rightarrow $2$\mathrm{H} + \mathrm{OH}$&$-$ & $\num{1.000e-13}$ & \ \cite{kossyi.1992}\\ 
R163&$\mathrm{O^{-}} + \mathrm{H_2^{+}} \rightarrow $2$\mathrm{H} + \mathrm{O}$&$-$ & $\num{1.000e-13}$ & \ \cite{kossyi.1992}\\ 
R164&$\mathrm{O^{-}} + \mathrm{OH^{+}} \rightarrow \mathrm{H}$ + 2$\mathrm{O}$&$-$ & $\num{1.000e-13}$ & \ \cite{kossyi.1992}\\ 
R165&$\mathrm{O^{-}} + \mathrm{H_{2}O^{+}} \rightarrow \mathrm{H} + \mathrm{O} + \mathrm{OH}$&$-$ & $\num{1.000e-13}$ & \ \cite{kossyi.1992}\\ 
R166&$\mathrm{O_2^{-}} + \mathrm{H_2^{+}} \rightarrow $2$\mathrm{H} + \mathrm{O_2}$&$-$ & $\num{1.000e-13}$ & \ \cite{kossyi.1992}\\ 
R167&$\mathrm{O_2^{-}} + \mathrm{OH^{+}} \rightarrow \mathrm{H} + \mathrm{O} + \mathrm{O_2}$&$-$ & $\num{1.000e-13}$ & \ \cite{kossyi.1992}\\ 
R168&$\mathrm{O_2^{-}} + \mathrm{H_{2}O^{+}} \rightarrow \mathrm{H} + \mathrm{O_2} + \mathrm{OH}$&$-$ & $\num{1.000e-13}$ & \ \cite{kossyi.1992}\\ 
R169&$\mathrm{OH^{-}} + \mathrm{H_2^{+}} \rightarrow $2$\mathrm{H} + \mathrm{OH}$&$-$ & $\num{1.000e-13}$ & \ \cite{kossyi.1992}\\ 
R170&$\mathrm{OH^{-}} + \mathrm{O_2^{+}} \rightarrow $2$\mathrm{O} + \mathrm{OH}$&$-$ & $\num{1.000e-13}$ & \ \cite{kossyi.1992}\\ 
R171&$\mathrm{OH^{-}} + \mathrm{OH^{+}} \rightarrow \mathrm{H} + \mathrm{O} + \mathrm{OH}$&$-$ & $\num{1.000e-13}$ & \ \cite{kossyi.1992}\\ 
R172&$\mathrm{OH^{-}} + \mathrm{H_{2}O^{+}} \rightarrow \mathrm{H}$ + 2$\mathrm{OH}$&$-$ & $\num{1.000e-13}$ & \ \cite{kossyi.1992}\\ 
\\ \multicolumn{5}{c}{\textbf{Three-body ion-ion recombination}} \\ \hline 
R173&$\mathrm{H^{-}} + \mathrm{H^{+}} + \mathrm{M} \rightarrow $2$\mathrm{H} + \mathrm{M}$&$-$ & $\num{2.000e-37}\cdot \left(\frac{T_\mathrm{g}}{300.0}\right)^{-2.5}$ & \ \cite{kossyi.1992}\\ 
R174&$\mathrm{H^{-}} + \mathrm{H^{+}} + \mathrm{M} \rightarrow \mathrm{H_2} + \mathrm{M}$&$-$ & $\num{2.000e-37}\cdot \left(\frac{T_\mathrm{g}}{300.0}\right)^{-2.5}$ & \ \cite{kossyi.1992}\\ 
R175&$\mathrm{H^{-}} + \mathrm{H_2^{+}} + \mathrm{M} \rightarrow \mathrm{H} + \mathrm{H_2} + \mathrm{M}$&$-$ & $\num{2.000e-37}\cdot \left(\frac{T_\mathrm{g}}{300.0}\right)^{-2.5}$ & \ \cite{kossyi.1992}\\ 
R176&$\mathrm{H^{-}} + \mathrm{H_{2}O^{+}} + \mathrm{M} \rightarrow \mathrm{H} + \mathrm{H_{2}O} + \mathrm{M}$&$-$ & $\num{2.000e-37}\cdot \left(\frac{T_\mathrm{g}}{300.0}\right)^{-2.5}$ & \ \cite{kossyi.1992}\\ 
R177&$\mathrm{H^{-}} + \mathrm{H_3^{+}} + \mathrm{M} \rightarrow $2$\mathrm{H_2} + \mathrm{M}$&$-$ & $\num{2.000e-37}\cdot \left(\frac{T_\mathrm{g}}{300.0}\right)^{-2.5}$ & \ \cite{kossyi.1992}\\ 
R178&$\mathrm{H^{-}} + \mathrm{O^{+}} + \mathrm{M} \rightarrow \mathrm{H} + \mathrm{O} + \mathrm{M}$&$-$ & $\num{2.000e-37}\cdot \left(\frac{T_\mathrm{g}}{300.0}\right)^{-2.5}$ & \ \cite{kossyi.1992}\\ 
R179&$\mathrm{H^{-}} + \mathrm{O^{+}} + \mathrm{M} \rightarrow \mathrm{OH} + \mathrm{M}$&$-$ & $\num{2.000e-37}\cdot \left(\frac{T_\mathrm{g}}{300.0}\right)^{-2.5}$ & \ \cite{kossyi.1992}\\ 
R180&$\mathrm{H^{-}} + \mathrm{O_2^{+}} + \mathrm{M} \rightarrow \mathrm{H} + \mathrm{O_2} + \mathrm{M}$&$-$ & $\num{2.000e-37}\cdot \left(\frac{T_\mathrm{g}}{300.0}\right)^{-2.5}$ & \ \cite{kossyi.1992}\\ 
R181&$\mathrm{H^{-}} + \mathrm{O_2^{+}} + \mathrm{M} \rightarrow \mathrm{HO_2} + \mathrm{M}$&$-$ & $\num{2.000e-37}\cdot \left(\frac{T_\mathrm{g}}{300.0}\right)^{-2.5}$ & \ \cite{kossyi.1992}\\ 
R182&$\mathrm{H^{-}} + \mathrm{OH^{+}} + \mathrm{M} \rightarrow \mathrm{H} + \mathrm{OH} + \mathrm{M}$&$-$ & $\num{2.000e-37}\cdot \left(\frac{T_\mathrm{g}}{300.0}\right)^{-2.5}$ & \ \cite{kossyi.1992}\\ 
R183&$\mathrm{H^{-}} + \mathrm{OH^{+}} + \mathrm{M} \rightarrow \mathrm{H_{2}O} + \mathrm{M}$&$-$ & $\num{2.000e-37}\cdot \left(\frac{T_\mathrm{g}}{300.0}\right)^{-2.5}$ & \ \cite{kossyi.1992}\\ 
R184&$\mathrm{O^{-}} + \mathrm{H^{+}} + \mathrm{M} \rightarrow \mathrm{H} + \mathrm{O} + \mathrm{M}$&$-$ & $\num{2.000e-37}\cdot \left(\frac{T_\mathrm{g}}{300.0}\right)^{-2.5}$ & \ \cite{kossyi.1992}\\ 
R185&$\mathrm{O^{-}} + \mathrm{H^{+}} + \mathrm{M} \rightarrow \mathrm{OH} + \mathrm{M}$&$-$ & $\num{2.000e-37}\cdot \left(\frac{T_\mathrm{g}}{300.0}\right)^{-2.5}$ & \ \cite{kossyi.1992}\\ 
R186&$\mathrm{O^{-}} + \mathrm{H_2^{+}} + \mathrm{M} \rightarrow \mathrm{H_2} + \mathrm{O} + \mathrm{M}$&$-$ & $\num{2.000e-37}\cdot \left(\frac{T_\mathrm{g}}{300.0}\right)^{-2.5}$ & \ \cite{kossyi.1992}\\ 
R187&$\mathrm{O^{-}} + \mathrm{H_2^{+}} + \mathrm{M} \rightarrow \mathrm{H_{2}O} + \mathrm{M}$&$-$ & $\num{2.000e-37}\cdot \left(\frac{T_\mathrm{g}}{300.0}\right)^{-2.5}$ & \ \cite{kossyi.1992}\\ 
R188&$\mathrm{O^{-}} + \mathrm{H_{2}O^{+}} + \mathrm{M} \rightarrow \mathrm{H_{2}O} + \mathrm{O} + \mathrm{M}$&$-$ & $\num{2.000e-37}\cdot \left(\frac{T_\mathrm{g}}{300.0}\right)^{-2.5}$ & \ \cite{kossyi.1992}\\ 
R189&$\mathrm{O^{-}} + \mathrm{H_{2}O^{+}} + \mathrm{M} \rightarrow \mathrm{H_2O_2} + \mathrm{M}$&$-$ & $\num{2.000e-37}\cdot \left(\frac{T_\mathrm{g}}{300.0}\right)^{-2.5}$ & \ \cite{kossyi.1992}\\ 
R190&$\mathrm{O^{-}} + \mathrm{H_3^{+}} + \mathrm{M} \rightarrow \mathrm{H} + \mathrm{H_{2}O} + \mathrm{M}$&$-$ & $\num{2.000e-37}\cdot \left(\frac{T_\mathrm{g}}{300.0}\right)^{-2.5}$ & \ \cite{kossyi.1992}\\ 
R191&$\mathrm{O^{-}} + \mathrm{O^{+}} + \mathrm{M} \rightarrow $2$\mathrm{O} + \mathrm{M}$&$-$ & $\num{2.000e-37}\cdot \left(\frac{T_\mathrm{g}}{300.0}\right)^{-2.5}$ & \ \cite{kossyi.1992}\\ 
R192&$\mathrm{O^{-}} + \mathrm{O^{+}} + \mathrm{M} \rightarrow \mathrm{O_2} + \mathrm{M}$&$-$ & $\num{2.000e-37}\cdot \left(\frac{T_\mathrm{g}}{300.0}\right)^{-2.5}$ & \ \cite{kossyi.1992}\\ 
R193&$\mathrm{O^{-}} + \mathrm{O_2^{+}} + \mathrm{M} \rightarrow \mathrm{O} + \mathrm{O_2} + \mathrm{M}$&$-$ & $\num{2.000e-37}\cdot \left(\frac{T_\mathrm{g}}{300.0}\right)^{-2.5}$ & \ \cite{kossyi.1992}\\

\hline
\end{tabular}
\end{table*}

\begin{table*}[h!]
\footnotesize
\setlength{\tabcolsep}{8pt}
\caption{$T_\mathrm{e,eV}$: electron temperature in units of eV, $T_\mathrm{{e,K}}$: electron temperature in units of Kelvin. An energy threshold ($E_{\mathrm{thr}}$) given as $T_\mathrm{{e,eV}}$ means the threshold is taken to be the current value of $T_{e,eV}$. An $E_{\mathrm{thr}}$ of 0 does not imply that the process does not have a threshold, but instead that it is not included in the electron energy balance here. Rate coefficients are given in units of $\mathrm{m}^{3}\mathrm{s}^{-1}$ or $\mathrm{m}^{6}\mathrm{s}^{-1}$ depending on the order of the reaction.
A rate coefficient given as $f\left(T_\mathrm{{e,eV}}\right)$ is derived from cross section data. As described in section \ref{sec:simulation-framework}, excited states are not tracked in model. For clarity, reactions involving excited states are listed, along with how they are implemented here. Excited state notation follows the original source as far as possible.}
\begin{tabular}{p{1.5cm}p{5cm}p{1.5cm}p{5cm}p{1.5cm}}
no. & reaction                                                                     & $E_{\mathrm{thr}}$ / eV & rate coefficient                                                                      & reference                                     \\
\hline

R194&$\mathrm{O^{-}} + \mathrm{O_2^{+}} + \mathrm{M} \rightarrow \mathrm{O_3} + \mathrm{M}$&$-$ & $\num{2.000e-37}\cdot \left(\frac{T_\mathrm{g}}{300.0}\right)^{-2.5}$ & \ \cite{kossyi.1992}\\ 
R195&$\mathrm{O^{-}} + \mathrm{OH^{+}} + \mathrm{M} \rightarrow \mathrm{O} + \mathrm{OH} + \mathrm{M}$&$-$ & $\num{2.000e-37}\cdot \left(\frac{T_\mathrm{g}}{300.0}\right)^{-2.5}$ & \ \cite{kossyi.1992}\\ 
R196&$\mathrm{O^{-}} + \mathrm{OH^{+}} + \mathrm{M} \rightarrow \mathrm{HO_2} + \mathrm{M}$&$-$ & $\num{2.000e-37}\cdot \left(\frac{T_\mathrm{g}}{300.0}\right)^{-2.5}$ & \ \cite{kossyi.1992}\\ 
R197&$\mathrm{O_2^{-}} + \mathrm{H^{+}} + \mathrm{M} \rightarrow \mathrm{H} + \mathrm{O_2} + \mathrm{M}$&$-$ & $\num{2.000e-37}\cdot \left(\frac{T_\mathrm{g}}{300.0}\right)^{-2.5}$ & \ \cite{kossyi.1992}\\ 
R198&$\mathrm{O_2^{-}} + \mathrm{H^{+}} + \mathrm{M} \rightarrow \mathrm{HO_2} + \mathrm{M}$&$-$ & $\num{2.000e-37}\cdot \left(\frac{T_\mathrm{g}}{300.0}\right)^{-2.5}$ & \ \cite{kossyi.1992}\\ 
R199&$\mathrm{O_2^{-}} + \mathrm{H_2^{+}} + \mathrm{M} \rightarrow \mathrm{H_2} + \mathrm{O_2} + \mathrm{M}$&$-$ & $\num{2.000e-37}\cdot \left(\frac{T_\mathrm{g}}{300.0}\right)^{-2.5}$ & \ \cite{kossyi.1992}\\ 
R200&$\mathrm{O_2^{-}} + \mathrm{H_2^{+}} + \mathrm{M} \rightarrow \mathrm{H_2O_2} + \mathrm{M}$&$-$ & $\num{2.000e-37}\cdot \left(\frac{T_\mathrm{g}}{300.0}\right)^{-2.5}$ & \ \cite{kossyi.1992}\\ 
R201&$\mathrm{O_2^{-}} + \mathrm{H_{2}O^{+}} + \mathrm{M} \rightarrow \mathrm{H_{2}O} + \mathrm{O_2} + \mathrm{M}$&$-$ & $\num{2.000e-37}\cdot \left(\frac{T_\mathrm{g}}{300.0}\right)^{-2.5}$ & \ \cite{kossyi.1992}\\ 
R202&$\mathrm{O_2^{-}} + \mathrm{H_3^{+}} + \mathrm{M} \rightarrow \mathrm{H} + \mathrm{H_2} + \mathrm{O_2} + \mathrm{M}$&$-$ & $\num{2.000e-37}\cdot \left(\frac{T_\mathrm{g}}{300.0}\right)^{-2.5}$ & \ \cite{kossyi.1992}\\ 
R203&$\mathrm{O_2^{-}} + \mathrm{O^{+}} + \mathrm{M} \rightarrow \mathrm{O} + \mathrm{O_2} + \mathrm{M}$&$-$ & $\num{2.000e-37}\cdot \left(\frac{T_\mathrm{g}}{300.0}\right)^{-2.5}$ & \ \cite{kossyi.1992}\\ 
R204&$\mathrm{O_2^{-}} + \mathrm{O^{+}} + \mathrm{M} \rightarrow \mathrm{O_3} + \mathrm{M}$&$-$ & $\num{2.000e-37}\cdot \left(\frac{T_\mathrm{g}}{300.0}\right)^{-2.5}$ & \ \cite{kossyi.1992}\\ 
R205&$\mathrm{O_2^{-}} + \mathrm{O_2^{+}} + \mathrm{M} \rightarrow $2$\mathrm{O_2} + \mathrm{M}$&$-$ & $\num{2.000e-37}\cdot \left(\frac{T_\mathrm{g}}{300.0}\right)^{-2.5}$ & \ \cite{kossyi.1992}\\ 
R206&$\mathrm{O_2^{-}} + \mathrm{OH^{+}} + \mathrm{M} \rightarrow \mathrm{O_2} + \mathrm{OH} + \mathrm{M}$&$-$ & $\num{2.000e-37}\cdot \left(\frac{T_\mathrm{g}}{300.0}\right)^{-2.5}$ & \ \cite{kossyi.1992}\\ 
R207&$\mathrm{OH^{-}} + \mathrm{H^{+}} + \mathrm{M} \rightarrow \mathrm{H} + \mathrm{OH} + \mathrm{M}$&$-$ & $\num{2.000e-37}\cdot \left(\frac{T_\mathrm{g}}{300.0}\right)^{-2.5}$ & \ \cite{kossyi.1992}\\ 
R208&$\mathrm{OH^{-}} + \mathrm{H^{+}} + \mathrm{M} \rightarrow \mathrm{H_{2}O} + \mathrm{M}$&$-$ & $\num{2.000e-37}\cdot \left(\frac{T_\mathrm{g}}{300.0}\right)^{-2.5}$ & \ \cite{kossyi.1992}\\ 
R209&$\mathrm{OH^{-}} + \mathrm{H_2^{+}} + \mathrm{M} \rightarrow \mathrm{H_2} + \mathrm{OH} + \mathrm{M}$&$-$ & $\num{2.000e-37}\cdot \left(\frac{T_\mathrm{g}}{300.0}\right)^{-2.5}$ & \ \cite{kossyi.1992}\\ 
R210&$\mathrm{OH^{-}} + \mathrm{H_{2}O^{+}} + \mathrm{M} \rightarrow \mathrm{H_{2}O} + \mathrm{OH} + \mathrm{M}$&$-$ & $\num{2.000e-37}\cdot \left(\frac{T_\mathrm{g}}{300.0}\right)^{-2.5}$ & \ \cite{kossyi.1992}\\ 
R211&$\mathrm{OH^{-}} + \mathrm{H_3^{+}} + \mathrm{M} \rightarrow \mathrm{H_2} + \mathrm{H_{2}O} + \mathrm{M}$&$-$ & $\num{2.000e-37}\cdot \left(\frac{T_\mathrm{g}}{300.0}\right)^{-2.5}$ & \ \cite{kossyi.1992}\\ 
R212&$\mathrm{OH^{-}} + \mathrm{O^{+}} + \mathrm{M} \rightarrow \mathrm{O} + \mathrm{OH} + \mathrm{M}$&$-$ & $\num{2.000e-37}\cdot \left(\frac{T_\mathrm{g}}{300.0}\right)^{-2.5}$ & \ \cite{kossyi.1992}\\ 
R213&$\mathrm{OH^{-}} + \mathrm{O^{+}} + \mathrm{M} \rightarrow \mathrm{HO_2} + \mathrm{M}$&$-$ & $\num{2.000e-37}\cdot \left(\frac{T_\mathrm{g}}{300.0}\right)^{-2.5}$ & \ \cite{kossyi.1992}\\ 
R214&$\mathrm{OH^{-}} + \mathrm{O_2^{+}} + \mathrm{M} \rightarrow \mathrm{O_2} + \mathrm{OH} + \mathrm{M}$&$-$ & $\num{2.000e-37}\cdot \left(\frac{T_\mathrm{g}}{300.0}\right)^{-2.5}$ & \ \cite{kossyi.1992}\\ 
R215&$\mathrm{OH^{-}} + \mathrm{OH^{+}} + \mathrm{M} \rightarrow $2$\mathrm{OH} + \mathrm{M}$&$-$ & $\num{2.000e-37}\cdot \left(\frac{T_\mathrm{g}}{300.0}\right)^{-2.5}$ & \ \cite{kossyi.1992}\\ 
R216&$\mathrm{OH^{-}} + \mathrm{OH^{+}} + \mathrm{M} \rightarrow \mathrm{H_2O_2} + \mathrm{M}$&$-$ & $\num{2.000e-37}\cdot \left(\frac{T_\mathrm{g}}{300.0}\right)^{-2.5}$ & \ \cite{kossyi.1992}\\ 
\\ \multicolumn{5}{c}{\textbf{Two-body collisions - positive ions}} \\ \hline 
R217&$\mathrm{H_2^{+}} + \mathrm{O} \rightarrow \mathrm{OH^{+}} + \mathrm{H}$&$-$ & $\num{1.500e-15}$ & \ \cite{prasad.1980,umist,millar.2024}\\ 
R218&$\mathrm{H_2^{+}} + \mathrm{OH} \rightarrow \mathrm{H_{2}O^{+}} + \mathrm{H}$&$-$ & $\num{7.600e-16}\cdot \left(\frac{T_\mathrm{g}}{300.0}\right)^{-0.5}$ & \ \cite{prasad.1980,umist,millar.2024}\\ 
R219&$\mathrm{H_2^{+}} + \mathrm{H_{2}O} \rightarrow \mathrm{H} + \mathrm{H_{3}O^{+}}$ & $-$ & $\num{3.400e-15}\cdot \left(\frac{T_\mathrm{g}}{300.0}\right)^{-0.5}$ & \ \cite{kim.1975,umist,millar.2024}\\ 
used as & $\mathrm{H_2^{+}} + \mathrm{H_{2}O} \rightarrow \mathrm{H_{2}O^{+}} + \mathrm{H_2}$& &  & \\[0.1cm] 
R220&$\mathrm{H_2^{+}} + \mathrm{O_{2}} \rightarrow \mathrm{H} + \mathrm{HO_{2}^{+}}$ & $-$ & $\num{1.900e-15}$ & \ \cite{kim.1975,umist,millar.2024}\\ 
used as & $\mathrm{H_2^{+}} + \mathrm{O_2} \rightarrow \mathrm{O_2^{+}} + \mathrm{H_2}$& &  & \\[0.1cm] 
R221&$\mathrm{H_2^{+}} + \mathrm{H_2} \rightarrow \mathrm{H_3^{+}} + \mathrm{H}$&$-$ & $\num{2.080e-15}$ & \ \cite{theard.1974,umist,millar.2024}\\ 
R222&$\mathrm{O^{+}} + \mathrm{OH} \rightarrow \mathrm{O_2^{+}} + \mathrm{H}$&$-$ & $\num{3.600e-16}\cdot \left(\frac{T_\mathrm{g}}{300.0}\right)^{-0.5}$ & \ \cite{prasad.1980,umist,millar.2024}\\ 
R223&$\mathrm{H_{2}O} + \mathrm{OH^+} \rightarrow \mathrm{H_{3}O^{+}} + \mathrm{O}$ & $-$ & $\num{1.300e-15}\cdot \left(\frac{T_\mathrm{g}}{300.0}\right)^{-0.5}$ & \ \cite{huntress.1973,umist,millar.2024}\\ 
used as & $\mathrm{OH^{+}} + \mathrm{H_{2}O} \rightarrow \mathrm{H_{2}O^{+}} + \mathrm{OH}$& &  & \\[0.1cm] 
R224&$\mathrm{H_3^{+}} + \mathrm{O} \rightarrow \mathrm{H_{2}O^{+}} + \mathrm{H}$&$-$ & $\num{2.080e-16}\cdot \left(\frac{T_\mathrm{g}}{300.0}\right)^{-0.4}\cdot\exp(\frac{-5}{T_\mathrm{g}})$ & \ \cite{hillenbrand.2022,umist,millar.2024}\\ 
R225&$\mathrm{H_{2}O} + \mathrm{H_{3}^+} \rightarrow \mathrm{H_{2}} + \mathrm{H_{3}O^{+}}$ & $-$ & $\num{5.900e-15}\cdot \left(\frac{T_\mathrm{g}}{300.0}\right)^{-0.5}$ & \ \cite{anicich.1975,umist,millar.2024}\\ 
used as & $\mathrm{H_3^{+}} + \mathrm{H_{2}O} \rightarrow \mathrm{H_{2}O^{+}} + \frac{3}{2}\mathrm{H_2}$& &  & \\[0.1cm] 
R226&$\mathrm{H_{3}^{+}} + \mathrm{O_2} \rightarrow \mathrm{H_2} + \mathrm{HO_{2}^{+}}$ & $-$ & $\num{9.300e-16}\cdot\exp(\frac{-100}{T_\mathrm{g}})$ & \ \cite{adams.1984,umist,millar.2024}\\ 
used as & $\mathrm{H_3^{+}} + \mathrm{O_2} \rightarrow \mathrm{O_2^{+}} + \frac{3}{2}\mathrm{H_2}$& &  & \\[0.1cm] 
R227&$\mathrm{H_{2}O^{+}} + \mathrm{O} \rightarrow \mathrm{O_2^{+}} + \mathrm{H_2}$&$-$ & $\num{4.000e-17}$ & \ \cite{viggiano.1980,umist,millar.2024}\\

\hline
\end{tabular}
\end{table*}

\begin{table*}[h!]
\footnotesize
\setlength{\tabcolsep}{8pt}
\caption{$T_\mathrm{e,eV}$: electron temperature in units of eV, $T_\mathrm{{e,K}}$: electron temperature in units of Kelvin. An energy threshold ($E_{\mathrm{thr}}$) given as $T_\mathrm{{e,eV}}$ means the threshold is taken to be the current value of $T_{e,eV}$. An $E_{\mathrm{thr}}$ of 0 does not imply that the process does not have a threshold, but instead that it is not included in the electron energy balance here. Rate coefficients are given in units of $\mathrm{m}^{3}\mathrm{s}^{-1}$ or $\mathrm{m}^{6}\mathrm{s}^{-1}$ depending on the order of the reaction.
A rate coefficient given as $f\left(T_\mathrm{{e,eV}}\right)$ is derived from cross section data. As described in section \ref{sec:simulation-framework}, excited states are not tracked in model. For clarity, reactions involving excited states are listed, along with how they are implemented here. Excited state notation follows the original source as far as possible.}
\begin{tabular}{p{1.5cm}p{5cm}p{1.5cm}p{5cm}p{1.5cm}}
no. & reaction                                                                     & $E_{\mathrm{thr}}$ / eV & rate coefficient                                                                      & reference                                     \\
\hline

R228&$\mathrm{H^{+}} + \mathrm{O} \rightarrow \mathrm{O^{+}} + \mathrm{H}$&$-$ & $\num{6.860e-16}\cdot \left(\frac{T_\mathrm{g}}{300.0}\right)^{0.26}\cdot\exp(\frac{-224}{T_\mathrm{g}})$ & \ \cite{stancil.1999,umist,millar.2024}\\ 
R229&$\mathrm{H^{+}} + \mathrm{O_2} \rightarrow \mathrm{O_2^{+}} + \mathrm{H}$&$-$ & $\num{2.000e-15}$ & \ \cite{smith.1992,umist,millar.2024}\\ 
R230&$\mathrm{H^{+}} + \mathrm{OH} \rightarrow \mathrm{OH^{+}} + \mathrm{H}$&$-$ & $\num{2.100e-15}\cdot \left(\frac{T_\mathrm{g}}{300.0}\right)^{-0.5}$ & \ \cite{prasad.1980,umist,millar.2024}\\ 
R231&$\mathrm{H^{+}} + \mathrm{H_{2}O} \rightarrow \mathrm{H_{2}O^{+}} + \mathrm{H}$&$-$ & $\num{6.900e-15}\cdot \left(\frac{T_\mathrm{g}}{300.0}\right)^{-0.5}$ & \ \cite{smith.1992,umist,millar.2024}\\ 
R232&$\mathrm{H_2^{+}} + \mathrm{O_2} \rightarrow \mathrm{O_2^{+}} + \mathrm{H_2}$&$-$ & $\num{8.000e-16}$ & \ \cite{kim.1975,umist,millar.2024}\\ 
R233&$\mathrm{H_2^{+}} + \mathrm{H} \rightarrow \mathrm{H^{+}} + \mathrm{H_2}$&$-$ & $\num{6.400e-16}$ & \ \cite{anicich.1979,umist,millar.2024}\\ 
R234&$\mathrm{H_2^{+}} + \mathrm{OH} \rightarrow \mathrm{OH^{+}} + \mathrm{H_2}$&$-$ & $\num{7.600e-16}\cdot \left(\frac{T_\mathrm{g}}{300.0}\right)^{-0.5}$ & \ \cite{prasad.1980,umist,millar.2024}\\ 
R235&$\mathrm{H_2^{+}} + \mathrm{H_{2}O} \rightarrow \mathrm{H_{2}O^{+}} + \mathrm{H_2}$&$-$ & $\num{3.900e-15}\cdot \left(\frac{T_\mathrm{g}}{300.0}\right)^{-0.5}$ & \ \cite{kim.1975,umist,millar.2024}\\ 
R236&$\mathrm{O^{+}} + \mathrm{O_2} \rightarrow \mathrm{O_2^{+}} + \mathrm{O}$&$-$ & $\num{1.900e-17}$ & \ \cite{adams.1980,umist,millar.2024}\\ 
R237&$\mathrm{O^{+}} + \mathrm{H} \rightarrow \mathrm{H^{+}} + \mathrm{O}$&$-$ & $\num{5.660e-16}\cdot \left(\frac{T_\mathrm{g}}{300.0}\right)^{0.36}\cdot\exp(\frac{-9}{T_\mathrm{g}})$ & \ \cite{stancil.1999,umist,millar.2024}\\ 
R238&$\mathrm{O^{+}} + \mathrm{OH} \rightarrow \mathrm{OH^{+}} + \mathrm{O}$&$-$ & $\num{3.600e-16}\cdot \left(\frac{T_\mathrm{g}}{300.0}\right)^{-0.5}$ & \ \cite{prasad.1980,umist,millar.2024}\\ 
R239&$\mathrm{O^{+}} + \mathrm{H_{2}O} \rightarrow \mathrm{H_{2}O^{+}} + \mathrm{O}$&$-$ & $\num{3.200e-15}\cdot \left(\frac{T_\mathrm{g}}{300.0}\right)^{-0.5}$ & \ \cite{adams.1980,umist,millar.2024}\\ 
R240&$\mathrm{H_{2}O^{+}} + \mathrm{O_2} \rightarrow \mathrm{O_2^{+}} + \mathrm{H_{2}O}$&$-$ & $\num{4.600e-16}$ & \ \cite{rakshit.1980,umist,millar.2024}\\ 
R241&$\mathrm{OH^{+}} + \mathrm{O_2} \rightarrow \mathrm{O_2^{+}} + \mathrm{OH}$&$-$ & $\num{5.900e-16}$ & \ \cite{jones.1981,umist,millar.2024}\\ 
R242&$\mathrm{OH^{+}} + \mathrm{H_{2}O} \rightarrow \mathrm{H_{2}O^{+}} + \mathrm{OH}$&$-$ & $\num{1.590e-15}\cdot \left(\frac{T_\mathrm{g}}{300.0}\right)^{-0.5}$ & \ \cite{huntress.1973,umist,millar.2024}\\ 
R243&$\mathrm{OH^{+}} + \mathrm{H_2} \rightarrow \mathrm{H_{2}O^{+}} + \mathrm{H}$&$-$ & $\num{1.300e-15}$ & \ \cite{liu.2010}\\ 
R244&$\mathrm{OH^{+}} + \mathrm{O} \rightarrow \mathrm{O_2^{+}} + \mathrm{H}$&$-$ & $\num{7.100e-16}$ & \ \cite{liu.2010}\\ 
R245&$\mathrm{OH^{+}} + \mathrm{OH} \rightarrow \mathrm{H_{2}O^{+}} + \mathrm{O}$&$-$ & $\num{7.000e-16}$ & \ \cite{liu.2010}\\ 
R246&$\mathrm{H_3^{+}} + \mathrm{O} \rightarrow \mathrm{OH^{+}} + \mathrm{H_2}$&$-$ & $\num{4.650e-16}\cdot \left(\frac{T_\mathrm{g}}{300.0}\right)^{-0.14}\cdot\exp(\frac{-1}{T_\mathrm{g}})$ & \ \cite{tavant.2016,umist,millar.2024,hillenbrand.2022}\\ 
R247&$\mathrm{H_3^{+}} + \mathrm{OH} \rightarrow \mathrm{H_{2}O^{+}} + \mathrm{H_2}$&$-$ & $\num{1.300e-15}\cdot \left(\frac{T_\mathrm{g}}{300.0}\right)^{-0.5}$ & \ \cite{tavant.2016,umist,millar.2024,prasad.1980}\\ 
\\ \multicolumn{5}{c}{\textbf{Two-body collisions - negative ions}} \\ \hline 
R248&$\mathrm{O_2^{-}} + \mathrm{O} \rightarrow \mathrm{O^{-}} + \mathrm{O_2}$&$-$ & $\num{8.500e-17}\cdot \left(\frac{T_\mathrm{g}}{300.0}\right)^{-1.8}$ & \ \cite{tavant.2016,brisset.2021,ard.2013}\\ 
R249&$\mathrm{H^{-}} + \mathrm{H_{2}O} \rightarrow \mathrm{OH^{-}} + \mathrm{H_2}$&$-$ & $\num{4.800e-15}$ & \ \cite{brisset.2021,martinez.2010}\\ 
R250&$\mathrm{O^{-}} + \mathrm{H_{2}O} \rightarrow \mathrm{OH^{-}} + \mathrm{OH}$&$-$ & $\num{1.400e-15}$ & \ \cite{tavant.2016,brisset.2021,melton.1971}\\ 
R251&$\mathrm{H^{-}} + \mathrm{H} \rightarrow \mathrm{e} + \mathrm{H_2}$&$-$ & $\num{4.820e-15}\cdot \left(\frac{T_\mathrm{g}}{300.0}\right)^{0.02}\cdot\exp(\frac{-4}{T_\mathrm{g}})$ & \ \cite{bruhns.2010,umist,millar.2024}\\ 
R252&$\mathrm{O^{-}} + \mathrm{H} \rightarrow \mathrm{e} + \mathrm{OH}$&$-$ & $\num{5.000e-16}$ & \ \cite{prasad.1980,umist,millar.2024}\\ 
R253&$\mathrm{O^{-}} + \mathrm{H_2} \rightarrow \mathrm{e} + \mathrm{H_{2}O}$&$-$ & $\num{7.000e-16}$ & \ \cite{ferguson.1973,umist,millar.2024}\\ 
R254&$\mathrm{O^{-}} + \mathrm{H_2} \rightarrow \mathrm{OH^{-}} + \mathrm{H}$&$-$ & $\num{3.000e-17}$ & \ \cite{ferguson.1973,umist,millar.2024}\\ 
R255&$\mathrm{H^{-}} + \mathrm{O_2} \rightarrow \mathrm{e} + \mathrm{HO_2}$&$-$ & $\num{1.200e-15}$ & \ \cite{vanGaens.2013}\\ 
R256&$\mathrm{H^{-}} + \mathrm{O} \rightarrow \mathrm{e} + \mathrm{OH}$&$-$ & $\num{1.000e-15}$ & \ \cite{prasad.1980,umist,millar.2024}\\ 
R257&$\mathrm{O^{-}} + \mathrm{O} \rightarrow \mathrm{e} + \mathrm{O_2}$&$-$ & $\num{1.900e-16}$ & \ \cite{ferguson.1973,umist,millar.2024}\\ 
R258&$\mathrm{O^{-}} + \mathrm{O_2} \rightarrow \mathrm{O_2^{-}} + \mathrm{O}$&$-$ & $\num{7.300e-16}\cdot\exp(\frac{-890}{T_\mathrm{g}})$ & \ \cite{midey.2008,umist,millar.2024}\\ 
R259&$\mathrm{OH^{-}} + \mathrm{H} \rightarrow \mathrm{e} + \mathrm{H_{2}O}$&$-$ & $\num{1.400e-15}$ & \ \cite{ferguson.1973,umist,millar.2024}\\ 
R260&$\mathrm{H^{-}} + \mathrm{OH} \rightarrow \mathrm{e} + \mathrm{H_{2}O}$&$-$ & $\num{1.000e-16}$ & \ \cite{prasad.1980,umist,millar.2024}\\ 
\\ \multicolumn{5}{c}{\textbf{Three-body collisions - positive ions}} \\ \hline 
R261&$\mathrm{H^{+}} + \mathrm{H} + \mathrm{M} \rightarrow \mathrm{H_2^{+}} + \mathrm{M}$&$-$ & $\num{3.000e-41}$ & \ \cite{ikezoe.1987}\\ 
R262&$\mathrm{H^{+}} + \mathrm{H_2} + \mathrm{M} \rightarrow \mathrm{H_3^{+}} + \mathrm{M}$&$-$ & $\num{3.000e-41}$ & \ \cite{ikezoe.1987}\\ 
R263&$\mathrm{H^{+}} + \mathrm{O} + \mathrm{M} \rightarrow \mathrm{OH^{+}} + \mathrm{M}$&$-$ & $\num{3.000e-41}$ & \ \cite{ikezoe.1987}\\ 
R264&$\mathrm{H^{+}} + \mathrm{OH} + \mathrm{M} \rightarrow \mathrm{H_{2}O^{+}} + \mathrm{M}$&$-$ & $\num{3.000e-41}$ & \ \cite{ikezoe.1987}\\ 
R265&$\mathrm{O^{+}} + \mathrm{O} + \mathrm{M} \rightarrow \mathrm{O_2^{+}} + \mathrm{M}$&$-$ & $\num{3.000e-41}$ & \ \cite{ikezoe.1987}\\ 
R266&$\mathrm{O^{+}} + \mathrm{H} + \mathrm{M} \rightarrow \mathrm{OH^{+}} + \mathrm{M}$&$-$ & $\num{3.000e-41}$ & \ \cite{ikezoe.1987}\\ 
R267&$\mathrm{OH^{+}} + \mathrm{H} + \mathrm{M} \rightarrow \mathrm{H_{2}O^{+}} + \mathrm{M}$&$-$ & $\num{3.000e-41}$ & \ \cite{ikezoe.1987}\\ 
R268&$\mathrm{H^{+}}$ + 2$\mathrm{H_2} \rightarrow \mathrm{H_3^{+}} + \mathrm{H_2}$&$-$ & $\num{3.100e-41}\cdot \left(\frac{T_\mathrm{g}}{300.0}\right)^{-0.5}$ & \ \cite{tavant.2016,liu.2010}\\ 
\\ \multicolumn{5}{c}{\textbf{Two-body collisions}} \\ \hline 
R269&$\mathrm{H_{2}O} + \mathrm{M} \rightarrow \mathrm{H} + \mathrm{OH} + \mathrm{M}$&$-$ & $\num{5.800e-15}\cdot\exp(\frac{-52900}{T_\mathrm{g}})$ & \ \cite{locke.2012,manion.2015,baulch.1992}\\ 
R270&$\mathrm{H_2} + \mathrm{M} \rightarrow $2$\mathrm{H} + \mathrm{M}$&$-$ & $\num{1.500e-15}\cdot\exp(\frac{-48400}{T_\mathrm{g}})$ & \ \cite{locke.2012,manion.2015,baulch.1992}\\ 
R271&2$\mathrm{OH} \rightarrow \mathrm{H_{2}O} + \mathrm{O}$&$-$ & $\num{1.020e-18}\cdot \left(\frac{T_\mathrm{g}}{298.0}\right)^{1.4}\cdot\exp(\frac{-200}{T_\mathrm{g}})$ & \ \cite{locke.2012,manion.2015,tsang.1986}\\

\hline
\end{tabular}
\end{table*}

\begin{table*}[h!]
\footnotesize
\setlength{\tabcolsep}{8pt}
\caption{$T_\mathrm{e,eV}$: electron temperature in units of eV, $T_\mathrm{{e,K}}$: electron temperature in units of Kelvin. An energy threshold ($E_{\mathrm{thr}}$) given as $T_\mathrm{{e,eV}}$ means the threshold is taken to be the current value of $T_{e,eV}$. An $E_{\mathrm{thr}}$ of 0 does not imply that the process does not have a threshold, but instead that it is not included in the electron energy balance here. Rate coefficients are given in units of $\mathrm{m}^{3}\mathrm{s}^{-1}$ or $\mathrm{m}^{6}\mathrm{s}^{-1}$ depending on the order of the reaction.
A rate coefficient given as $f\left(T_\mathrm{{e,eV}}\right)$ is derived from cross section data. As described in section \ref{sec:simulation-framework}, excited states are not tracked in model. For clarity, reactions involving excited states are listed, along with how they are implemented here. Excited state notation follows the original source as far as possible.}
\begin{tabular}{p{1.5cm}p{5cm}p{1.5cm}p{5cm}p{1.5cm}}
no. & reaction                                                                     & $E_{\mathrm{thr}}$ / eV & rate coefficient                                                                      & reference                                     \\
\hline

R272&$\mathrm{H} + \mathrm{O_2} \rightarrow \mathrm{O} + \mathrm{OH}$&$-$ & $\num{3.300e-16}\cdot\exp(\frac{-8460}{T_\mathrm{g}})$ & \ \cite{manion.2015,baulch.1992}\\ 
R273&$\mathrm{H} + \mathrm{OH} \rightarrow \mathrm{H_2} + \mathrm{O}$&$-$ & $\num{6.860e-20}\cdot \left(\frac{T_\mathrm{g}}{298.0}\right)^{2.8}\cdot\exp(\frac{-1950}{T_\mathrm{g}})$ & \ \cite{locke.2012,manion.2015,schofield.1973}\\ 
R274&$\mathrm{H_2O_2} + \mathrm{OH} \rightarrow \mathrm{H_{2}O} + \mathrm{HO_2}$&$-$ & $\num{2.910e-18}\cdot\exp(\frac{-160}{T_\mathrm{g}})$ & \ \cite{locke.2012,manion.2015,tsang.1986}\\ 
R275&$\mathrm{O_2} + \mathrm{OH} \rightarrow \mathrm{HO_2} + \mathrm{O}$&$-$ & $\num{3.700e-17}\cdot\exp(\frac{-26500}{T_\mathrm{g}})$ & \ \cite{locke.2012,manion.2015,tsang.1986}\\ 
R276&2$\mathrm{OH} \rightarrow $2$\mathrm{H}$ + 2$\mathrm{O}$&$-$ & $\num{4.090e-15}\cdot\exp(\frac{-50033}{T_\mathrm{g}})$ & \ \cite{locke.2012}\\ 
R277&$\mathrm{H_2O_2} + \mathrm{O} \rightarrow \mathrm{HO_2} + \mathrm{OH}$&$-$ & $\num{1.420e-18}\cdot \left(\frac{T_\mathrm{g}}{298.0}\right)^{2.0}\cdot\exp(\frac{-2000}{T_\mathrm{g}})$ & \ \cite{locke.2012,manion.2015,tsang.1986}\\ 
R278&$\mathrm{H_2O_2} + \mathrm{O_2} \rightarrow $2$\mathrm{HO_2}$&$-$ & $\num{9.000e-17}\cdot\exp(\frac{-20000}{T_\mathrm{g}})$ & \ \cite{locke.2012,manion.2015,tsang.1986}\\ 
R279&$\mathrm{HO_2} + \mathrm{OH} \rightarrow \mathrm{H_{2}O} + \mathrm{O_2}$&$-$ & $\num{4.810e-17}\cdot\exp(\frac{-250}{T_\mathrm{g}})$ & \ \cite{locke.2012,manion.2015,baulch.1992}\\ 
R280&$\mathrm{HO_2} + \mathrm{O} \rightarrow \mathrm{O_2} + \mathrm{OH}$&$-$ & $\num{2.910e-17}\cdot\exp(\frac{-200}{T_\mathrm{g}})$ & \ \cite{locke.2012,manion.2015,tsang.1986}\\ 
R281&$\mathrm{H} + \mathrm{HO_2} \rightarrow $2$\mathrm{OH}$&$-$ & $\num{2.810e-16}\cdot\exp(\frac{-440}{T_\mathrm{g}})$ & \ \cite{manion.2015,baulch.1992}\\ 
R282&$\mathrm{H} + \mathrm{HO_2} \rightarrow \mathrm{H_2} + \mathrm{O_2}$&$-$ & $\num{7.110e-17}\cdot\exp(\frac{-710}{T_\mathrm{g}})$ & \ \cite{manion.2015,baulch.1992}\\ 
R283&2$\mathrm{HO_2} \rightarrow \mathrm{H_2O_2} + \mathrm{O_2}$&$-$ & $\num{3.010e-18}$ & \ \cite{locke.2012,manion.2015,tsang.1986}\\ 
R284&$\mathrm{H_2} + \mathrm{HO_2} \rightarrow \mathrm{H} + \mathrm{H_2O_2}$&$-$ & $\num{5.000e-17}\cdot\exp(\frac{-13100}{T_\mathrm{g}})$ & \ \cite{locke.2012,manion.2015,tsang.1986}\\ 
R285&$\mathrm{H_2} + \mathrm{OH} \rightarrow \mathrm{H} + \mathrm{H_{2}O}$&$-$ & $\num{2.970e-18}\cdot \left(\frac{T_\mathrm{g}}{298.0}\right)^{1.21}\cdot\exp(\frac{-2371}{T_\mathrm{g}})$ & \ \cite{locke.2012,manion.2015,isaacson.1997}\\ 
R286&$\mathrm{H_2} + \mathrm{O} \rightarrow \mathrm{H} + \mathrm{OH}$&$-$ & $\num{3.440e-19}\cdot \left(\frac{T_\mathrm{g}}{298.0}\right)^{2.67}\cdot\exp(\frac{-3160}{T_\mathrm{g}})$ & \ \cite{locke.2012,manion.2015,baulch.1992}\\ 
R287&$\mathrm{H} + \mathrm{O_3} \rightarrow \mathrm{O_2} + \mathrm{OH}$&$-$ & $\num{2.710e-17}\cdot \left(\frac{T_\mathrm{g}}{300.0}\right)^{0.75}$ & \ \cite{burkholder.2015,manion.2015,shaw.1977}\\ 
R288&$\mathrm{H} + \mathrm{H_{2}O} \rightarrow \mathrm{H_2} + \mathrm{OH}$&$-$ & $\num{6.820e-18}\cdot \left(\frac{T_\mathrm{g}}{298.0}\right)^{1.6}\cdot\exp(\frac{-9270}{T_\mathrm{g}})$ & \ \cite{manion.2015,baulch.1992}\\ 
R289&$\mathrm{H} + \mathrm{H_2O_2} \rightarrow \mathrm{H_{2}O} + \mathrm{OH}$&$-$ & $\num{4.000e-17}\cdot\exp(\frac{-2000}{T_\mathrm{g}})$ & \ \cite{manion.2015,tsang.1986}\\ 
R290&$\mathrm{H} + \mathrm{H_2O_2} \rightarrow \mathrm{H_2} + \mathrm{HO_2}$&$-$ & $\num{8.000e-17}\cdot\exp(\frac{-4000}{T_\mathrm{g}})$ & \ \cite{manion.2015,tsang.1986}\\ 
R291&$\mathrm{H} + \mathrm{HO_2} \rightarrow \mathrm{H_{2}O} + \mathrm{O}$&$-$ & $\num{5.000e-17}\cdot\exp(\frac{-866}{T_\mathrm{g}})$ & \ \cite{manion.2015,baulch.1992}\\ 
R292&$\mathrm{H_{2}O} + \mathrm{O} \rightarrow $2$\mathrm{OH}$&$-$ & $\num{1.250e-17}\cdot \left(\frac{T_\mathrm{g}}{298.0}\right)^{1.3}\cdot\exp(\frac{-8600}{T_\mathrm{g}})$ & \ \cite{manion.2015,tsang.1986}\\ 
R293&$\mathrm{O} + \mathrm{O_3} \rightarrow $2$\mathrm{O_2}$&$-$ & $\num{1.830e-17}\cdot\exp(\frac{-2164}{T_\mathrm{g}})$ & \ \cite{manion.2015,jian.2022}\\ 
R294&$\mathrm{O} + \mathrm{OH} \rightarrow \mathrm{H} + \mathrm{O_2}$&$-$ & $\num{4.550e-18}\cdot \left(\frac{T_\mathrm{g}}{298.0}\right)^{0.4}\cdot\exp(\frac{-372}{T_\mathrm{g}})$ & \ \cite{manion.2015,miller.1997}\\ 
R295&$\mathrm{H_{2}O} + \mathrm{HO_2} \rightarrow \mathrm{H_2O_2} + \mathrm{OH}$&$-$ & $\num{4.650e-17}\cdot\exp(\frac{-16500}{T_\mathrm{g}})$ & \ \cite{manion.2015,lloyd.1974}\\ 
R296&$\mathrm{H_{2}O} + \mathrm{O_2} \rightarrow \mathrm{HO_2} + \mathrm{OH}$&$-$ & $\num{7.720e-18}\cdot\exp(\frac{-37300}{T_\mathrm{g}})$ & \ \cite{manion.2015,mayer.1968}\\ 
R297&$\mathrm{HO_2} + \mathrm{O_3} \rightarrow $2$\mathrm{O_2} + \mathrm{OH}$&$-$ & $\num{1.970e-22}\cdot \left(\frac{T_\mathrm{g}}{298.0}\right)^{4.57}\cdot\exp(\frac{-693}{T_\mathrm{g}})$ & \ \cite{manion.2015,atkinson.2004}\\ 
R298&$\mathrm{O_3} + \mathrm{OH} \rightarrow \mathrm{HO_2} + \mathrm{O_2}$&$-$ & $\num{3.760e-19}\cdot \left(\frac{T_\mathrm{g}}{298.0}\right)^{1.99}\cdot\exp(\frac{-604}{T_\mathrm{g}})$ & \ \cite{manion.2015,ju.2007}\\ 
R299&$\mathrm{H_2O_2} + \mathrm{M} \rightarrow $2$\mathrm{OH} + \mathrm{M}$&$-$ & $\num{2.030e-09}\cdot \left(\frac{T_\mathrm{g}}{298.0}\right)^{-4.86}\cdot\exp(\frac{-26800}{T_\mathrm{g}})$ & \ \cite{manion.2015,tsang.1986}\\ 
R300&$\mathrm{HO_2} + \mathrm{M} \rightarrow \mathrm{H} + \mathrm{O_2} + \mathrm{M}$&$-$ & $\num{2.410e-14}\cdot \left(\frac{T_\mathrm{g}}{298.0}\right)^{-1.18}\cdot\exp(\frac{-24400}{T_\mathrm{g}})$ & \ \cite{manion.2015,tsang.1986}\\ 
R301&$\mathrm{O_2} + \mathrm{M} \rightarrow $2$\mathrm{O} + \mathrm{M}$&$-$ & $\num{1.990e-16}\cdot\exp(\frac{-54200}{T_\mathrm{g}})$ & \ \cite{manion.2015,warnatz.1984}\\ 
R302&$\mathrm{O_3} + \mathrm{M} \rightarrow \mathrm{O} + \mathrm{O_2} + \mathrm{M}$&$-$ & $\num{7.160e-16}\cdot\exp(\frac{-11200}{T_\mathrm{g}})$ & \ \cite{manion.2015,heimerl.1979}\\ 
R303&$\mathrm{OH} + \mathrm{M} \rightarrow \mathrm{H} + \mathrm{O} + \mathrm{M}$&$-$ & $\num{4.000e-15}\cdot\exp(\frac{-50000}{T_\mathrm{g}})$ & \ \cite{manion.2015,tsang.1986}\\ 
\\ \multicolumn{5}{c}{\textbf{Three-body collisions}} \\ \hline 
R304&$\mathrm{H} + \mathrm{O} + \mathrm{M} \rightarrow \mathrm{OH} + \mathrm{M}$&$-$ & $\num{2.000e-44}$ & \ \cite{locke.2012,manion.2015,schofield.1973}\\ 
R305&2$\mathrm{H} + \mathrm{M} \rightarrow \mathrm{H_2} + \mathrm{M}$&$-$ & $\num{1.680e-44}$ & \ \cite{locke.2012,manion.2015,hurle.1969}\\ 
R306&$\mathrm{H} + \mathrm{O_2} + \mathrm{M} \rightarrow \mathrm{HO_2} + \mathrm{M}$&$-$ & $\num{4.520e-44}\cdot \left(\frac{T_\mathrm{g}}{298.0}\right)^{-0.88}$ & \ \cite{manion.2015,baulch.1992}\\ 
R307&$\mathrm{H} + \mathrm{OH} + \mathrm{M} \rightarrow \mathrm{H_{2}O} + \mathrm{M}$&$-$ & $\num{4.380e-42}\cdot \left(\frac{T_\mathrm{g}}{298.0}\right)^{-2.0}$ & \ \cite{manion.2015,baulch.1992}\\ 
R308&2$\mathrm{O} + \mathrm{M} \rightarrow \mathrm{O_2} + \mathrm{M}$&$-$ & $\num{5.210e-47}\cdot\exp(\frac{-900}{T_\mathrm{g}})$ & \ \cite{manion.2015,tsang.1986}\\ 
R309&$\mathrm{O} + \mathrm{O_2} + \mathrm{M} \rightarrow \mathrm{O_3} + \mathrm{M}$&$-$ & $\num{1.340e-46}\cdot \left(\frac{T_\mathrm{g}}{298.0}\right)^{-1.0}$ & \ \cite{manion.2015,hippler.1990}\\ 
R310&2$\mathrm{OH} + \mathrm{M} \rightarrow \mathrm{H_2O_2} + \mathrm{M}$&$-$ & presented in Table~\ref{tab:hydroxyl-ratecoeff} & \ \cite{baulch.1992}\\

\hline
\end{tabular}
\end{table*}